\documentclass[twoside,reqno,11pt]{amsproc}
 \usepackage{amssymb}

\usepackage{enumerate}

\usepackage[english]{babel}

\newtheorem{thm}{\noindent \bf Theorem}[section]
\newtheorem{prop}[thm]{\noindent \bf Proposition}
\newtheorem{defin}[thm]{\noindent \bf Definition}
\newtheorem{lemma}[thm]{\noindent \bf Lemma}
\newtheorem{corol}[thm]{\noindent \bf Corollary}
\newtheorem{coro}[thm]{\noindent \bf Corollary}

\usepackage[dvipsone]{color}
\usepackage[dvips]{graphicx}

\definecolor{blue2}{rgb}{0,0,0.6}
\definecolor{turck}{rgb}{0,0.5,0.5}
\definecolor{sturck}{rgb}{0,0.9,0.8} 
\definecolor{orange}{rgb}{1,0.4,0}
\definecolor{bord}{rgb}{1,0.3,0.3}
\definecolor{violet}{rgb}{0.4,0.2,0.8}
\definecolor{sblue}{rgb}{0.2,0.4,1.}
\definecolor{yellowa}{rgb}{1.,0.8,0}
\definecolor{greena}{rgb}{0.2,0.8,0}
\definecolor{greenb}{rgb}{0,0.5,0.2}
\definecolor{greenc}{rgb}{0.1,0.7,0.3}
\definecolor{greend}{cmyk}{0.7,0,0.3,0}
\definecolor{greene}{cmyk}{0.7,0,0.3,0.1}
\definecolor{greya}{cmyk}{0,0,0,0.1}
\definecolor{greyb}{cmyk}{0,0,0,0.2}
\definecolor{greyc}{cmyk}{0,0,0,0.3}
\definecolor{greyd}{cmyk}{0,0,0,0.4}
\definecolor{greye}{cmyk}{0,0,0,0.5}
\definecolor{greyf}{cmyk}{0,0,0,0.6}
\definecolor{greyg}{cmyk}{0,0,0,0.7}
\definecolor{greyh}{cmyk}{0,0,0,0.8}
\definecolor{greyi}{cmyk}{0,0,0,0.9}
\newcommand{\red}{\textcolor{red}}

\newcommand{\blue}{\textcolor{blue}}

\newcommand{\linetwo}[2]{{\substack{#1 \\ #2}}} 
\newcommand{\inftwo}[2]{\inf_{\substack{#1 \\ #2}}} 
\newcommand{\infthree}[3]{\inf_{\substack{#1 \\ #2 \\ #3}}} 
\newcommand{\sumtwo}[2]{\sum_{\substack{#1 \\ #2}}} 

\newcommand{\nn}{\nonumber}
\newcommand{\st}{\scriptstyle}

\newcommand{\dis}{\displaystyle}
\newcommand{\al}{\alpha}
\newcommand{\eps}{\epsilon}

\newcommand{\La}{\Lambda}
\newcommand{\la}{\lambda}
\newcommand{\ga}{\gamma}
\newcommand{\Ga}{\Gamma}
\newcommand{\Om}{\Omega}
\newcommand{\om}{\omega}
\newcommand{\si}{\sigma}
\newcommand{\Si}{\Sigma}

\newcommand{\Zz}{\mathbb{Z}}

\newcommand{\Rr}{\mathbb{R}}
\newcommand{\Nn}{\mathbb{N}}

\newcommand{\und}{\underline}
\newcommand{\cA}{\mathcal{A}}
\newcommand{\cAA}{\mathcal{A}^{\pm}}
\newcommand{\cB}{\mathcal{B}}

\newcommand{\cD}{\mathcal{D}}
\newcommand{\cE}{\mathcal{E}}
\newcommand{\cF}{\mathcal{F}}

\newcommand{\cH}{\mathcal{H}}
\newcommand{\cI}{\mathcal{I}}
\newcommand{\cJ}{\mathcal{J}}
\newcommand{\cL}{\mathcal{L}}
\newcommand{\cM}{\mathcal{M}}

\newcommand{\cR}{\mathcal{R}}
\newcommand{\cS}{\mathcal{S}}

\newcommand{\cX}{\mathcal{X}}

\newcommand{\hG}{{G}}

\newcommand{\fH}{\mathfrak{H}}
\newcommand{\fJ}{\mathfrak{J}}

\newcommand{\fK}{{\mathfrak{K}}}
\newcommand{\fM}{\mathfrak{M}}
\newcommand{\fZ}{\mathfrak{Z}}

\newcommand{\fA}{\mathfrak{A}}

\newcommand{\Ii}{\text{\bf 1}}

\newcommand{\loc}{\mathrm{\rm loc}}
\newcommand{\out}{\mathrm{\rm out}}
\newcommand{\ins}{\mathrm{\rm in}}
\newcommand{\mf}{\mathrm{mf}}  
\newcommand{\ssp}{\mathrm{sp}} 
\newcommand{\Int}{\mathrm{Int}} 
\newcommand{\Ext}{\mathrm{Ext}} 

\newcommand{\dist}{\mathrm{dist}}
\newcommand{\abs}{\mathrm{abs}}
\newcommand{\eff}{\mathrm{eff}}
\newcommand{\q}{\hat q}
\newcommand{\p}{\hat p}

\newcommand{\bcg}{\beta_c(\gamma)}
\newcommand{\bcmf}{\beta^{\mf}_c}
\newcommand{\ds}{-1}
\newcommand{\mmmintone}[1]{{\dis{\int\kern -.43cm
-}}_{\kern-.21cm\substack{#1}}\;\;}
\newcommand{\mmmintwo}[2]{{\dis{\int\kern -.43cm
-}}_{\kern-.21cm\substack{#1}}^{\substack{#2}}\;\;}
\newcommand{\submint}{{\scriptstyle{\int\kern -.66em -}}}
\newcommand{\submintone}[1]{{\scriptstyle{\int\kern -.66em
-}}_{\tiny{\kern-.21em\linetwo{}{\substack{#1}}}}}
\newcommand{\fracmint}{{\textstyle{\int\kern -.88em -}}}
\newcommand{\fracmintone}[1]{{\textstyle{\int\kern -.88em
-}}_{\tiny{\kern-.34em\substack{#1}}}\;}

\newcommand{\ba}{\begin{array}}
\newcommand{\ea}{\end{array}}
\newcommand{\bq}{\begin{eqnarray}}
\newcommand{\eq}{\end{eqnarray}}
\newcommand{\bqw}{\begin{eqnarray*}}
\newcommand{\eqw}{\end{eqnarray*}}
\newcommand{\lng}{\langle}
\newcommand{\rng}{\rangle}
\newcommand{\vw}{\vec \xi}
\newcommand{\vvw}{\vec {\delta \xi}}
\newcommand{\w}{\xi}

\newcommand{\vL}{\vec \cL}

\newcommand{\cXx}{\Om}
\title[Finite-range Potts Models]
{First-Order Phase Transition in Potts Models with finite-range
interactions}

\author{T. Gobron}
\address{Thierry Gobron, CNRS-UMR 8089, Laboratoire
de physique th\'eorique et mod\'elisation, Universit\'e de
Cergy-Pontoise,
France}
\email{gobron@ptm.u-cergy.fr}

\author{I. Merola}
\address{Immacolata Merola,
Dipartimento di matematica pura e applicata, Universit\`a
dell'Aquila,
 Italy}
\email{merola@univaq.it}
\begin{document}

\vskip .5cm \noindent\vskip .5cm \noindent\vskip .5cm
\noindent

\begin{abstract}
We consider the $Q$-state  Potts model on $\mathbb Z^d$, $Q\ge 3$,
$d\ge 2$, with Kac ferromagnetic interactions and scaling parameter
$\ga$. We prove  the existence of a first order phase transition for
large but finite potential ranges. More precisely we prove that for
$\ga$ small enough there is a value of the temperature at which
coexist $Q+1$ Gibbs states. The proof is obtained by a perturbation
around mean-field using Pirogov-Sinai theory. The result is valid in
particular for $d=2$, $Q=3$,
 in contrast with the case of nearest-neighbor interactions  for which available results
indicate a second order phase transition.
Putting both results
together provides an example of
a system which undergoes a transition from second to first order phase
transition
by changing only the finite range of the interaction.

\keywords{Kac potentials, Lebowitz-Penrose limit,
Pirogov-Sinai theory}
\end{abstract}
\maketitle
\section{Introduction}
The Potts model is one of the most studied systems in Statistical
Mechanics not to mention its interest in other areas of mathematics
and computer sciences. Since its original description by Potts as a
simplified version of the clock model \cite{Potts}, it has become an
ever growing source of interest, in particular in the field of phase
transition. Originally introduced as the simplest generalization of
the Ising Model (classical spins with $Q$ values interacting through
alike/unlike interactions), it acquired a further significance
through the Fortuin-Kasteleyn representation
\cite{FK}, 
which allows for both a straightforward generalization to
any real positive value of the parameter $Q$ (random
cluster model \cite{Grimmett}-\cite{Grimm-a}), and a
direct connection between its partition function and the
Tutte dichromatic polynomial \cite{Sokal}-\cite{WM},
which have a central meaning in large areas of graph
theory. The FK random cluster
representation, gave also rise to important connections with
percolation theory
($Q\to 1$) and resistor networks ($Q\to 0$). 

Though the Potts model is in general not solvable, it has been
regarded since the original work by Potts as a simple example of an
order-disorder phase transition. A lot of work has been dedicated to
a rigorous study of the critical properties of the model and their
dependence on the number $Q$ of spin values and the dimension $d$ of
the lattice. The are exact computations \cite{baxter73},
\cite{baxter78} which show that the transition is first order for
nearest neighbor interactions in two dimensions when $Q> 4$ while it
is continuous for $Q\le 4$, (see \cite{Wu} for a review), but a
complete proof is still missing. A thorough analysis is however
available for the mean field version of the model where the
transition is continuous for $Q\le 2$ and first order for $Q\ge 3$,
independently of the dimensions.

There have been several attempts in various directions to
weaken the mean field hypothesis. The idea is to regard the Potts model as a
perturbation of its mean field version and this has been
achieved in three different regimes: large number of
dimensions $d$, \cite{KS}, large number of components $Q$,
\cite{BKL}, \cite{KotS}, and long range interactions,
\cite{ACCN}, \cite{Biskup-Chayes}, \cite{b-chays-c} and the
present paper. We also mention that early attempts led to
an heuristic determination of a value $Q_c(d)$
beyond which the transition becomes mean-field like, and in
particular to a few exact results $Q_c(2)=4$, $Q_c(4)=2$,
$Q_c(6)=1$ \cite{Wu}.

Indeed the most natural way to approximate mean field is to use long
range interactions  as in  \cite{ACCN} where the occurrence of a
phase transition is proved in one dimension with an interaction
which decays as $1/r^2$. Numerical results, \cite{RD}, indicate that
a transition from continuous to first order occurs for interactions
with a power law decay $1/r^s$ when $s$ varies across some
$Q$-dependent critical value. More recently \cite{b-chays-c} have
proved for long-range interactions the existence at a special value
of the temperature of $Q+1$ distinct DLR states, where $Q$ of them
describe ordered phases, each one with a dominant spin, while the
last one describes a disordered phase where all spins have same
average value. The result applies to special interactions (they
should be ``reflection positive") and requires slow power law decay
in low dimension, $d< 3$, while for $ d\ge 3$ some exponentially
decaying potentials can be also considered.

As already mentioned the present paper is also based on
approximating mean field by using long range interactions, but in a
sense and with a methodology different from the above papers.  We
follow the approach proposed by Kac, in particular in its
implementation by Lebowitz and Penrose \cite{LP}.  Calling $\ga$ the
scaling parameter of the Kac potential (so that mean field is
recovered in the limit $\ga\to 0$) we will prove that a mean field
behavior is observed also at finite $\ga$'s, i.e.\ without taking
$\ga\to 0$, which, for the potentials that we consider, means that
the range of the interaction is strictly finite. More precisely we
will show that the coexistence in the mean-field model of $Q+1$
phases at the inverse critical temperature $\beta_c^{\rm mf}$
implies that the same occurs also for finite (and suitably small)
values of $\ga$ but at an inverse temperature $\beta_c=\beta_c(\ga)$
which is close to but not necessarily equal to $\beta_c^{\rm mf}$.
Moreover in a paper still in preparation we show that the
present techniques allow to determine also the structure of
the phase diagram around $\beta_c(\ga)$: there are $\delta$
and $\ga_\delta$ both positive so that for any $\ga\le
\ga_\delta$ the following holds. When $\beta\in
(\beta_c(\ga)-\delta, \beta_c(\ga))$ there is a unique
extremal, translational invariant [disordered] DLR state
while for $\beta\in (\beta_c(\ga),\beta_c(\ga)+\delta)$
there are exactly $Q$ extremal, translational invariant
[ordered] DLR states.
This shows that the transition is first order for $Q\ge 3$ and all
$d\ge 2$ provided the interaction has a ``sufficiently long range'',
but recall its range is strictly finite.

In some respect, the above results are quite surprising
 and contradict
some deeply rooted beliefs and in particular:

\vskip .5cm

\begin{itemize}
  \item
All finite range models with the same symmetries and the same
dimensions behave the same, in particular nearest neighbors
and finite range ferromagnetic bounded spin systems are in
a same universality class.

  \item
The Pirogov Sinai theory applies away from the critical
point.

\end{itemize}

\vskip .5cm

$\bullet$
Though a rigorous analysis is still missing, the
available results for the ($Q=3, \ d=2$) Potts model with
nearest neighbor interaction strongly indicate a second
order phase transition  \cite{baxter}. In
contradistinction,  our results show that for
 a finite but long enough range of interactions, the transition is
first order, which in turn suggests that there is a
critical interaction range where the transition changes nature
from first to second order.
If this was to be the case, we would get an example of a modification of
the qualitative behavior induced by changing the [finite] range of the
interaction and  the first item above would
be proven false.

$\bullet$ The second statement about Pirogov-Sinai applies
to our context because we use extensively the Pirogov-Sinai
techniques by ``perturbing the mean field ground states''.
The perturbation is on the inverse interaction range which
is 0 in mean field and $\ga>0$ for the true system. To our
knowledge this is the first example where the Pirogov-Sinai
theory works in a range which includes the critical point.

The idea of perturbing mean field with Kac potentials is
clearly contained in the original papers by Lebowitz and
Penrose \cite{LP}, who introduced a coarse grained
description of the model which then plays a fundamental
role in the proofs. Using this approach not only as a tool
to derive the limit $\ga\to 0$ but also in order to study
rigorously phase transitions for fixed  (small) values of
$\ga>0$  is much more recent, \cite{CP} and \cite{BZ1}-\cite{BZ2}.  The
above papers deal with a ferromagnetic Ising systems with
Kac potentials and the spin flip symmetry allows to avoid
Pirogov-Sinai.
Such a symmetry is absent in the models considered in
\cite{lmp}, \cite{BZ1} and \cite{bkmp1} and a
Pirogov-Sinai approach \cite{ps} is required, as well as in the Potts
model we are considering here.  Unfortunately the idea that
it is sufficient to take $\ga$ (instead of the temperature)
as the small parameter to get the classical Pirogov-Sinai
theory working as well for Kac potentials  is a little
naive and there is no paper, we believe, in the huge
literature on Pirogov-Sinai which covers our case.  We thus
have to enter into the theory itself and not only check
that our model verifies a list of general conditions.

A discussion on the differences with the classical
Pirogov-Sinai theory and the techniques used to overcome
the corresponding problems has a rather technical nature
and does not fit well in an introductory section,  so we
postpone it to Section 4 where we also outline the scheme
of the proofs.  We just mention here that an output of the
Pirogov-Sinai theory is a control of the local structure of
the phase diagram.  The work however becomes much simpler
if less ambitiously we restrict to the problem of finding a
temperature at which the $Q+1$ phases coexist.  This does
not require to determine the landscape of the metastable
free energies of the $Q+1$ phases, but only the existence
of a temperature where they are all equal.  This is what we
do here, the result as it stands, is indeed compatible with
the existence of many other nearby temperatures where the
same happens.
In the forthcomming paper mentioned previously, we will
exclude such possibilities with an argument which extends
\cite{bmpz} avoiding the analysis of metastable free
energies landscape and allows to characterize all the
ergodic DLR states.

\vskip .5cm

 \noindent
The paper is organized as follows: In
Section 2, we define the model and state our main result.
In section 3, we introduce the scales
which appear in the problem and define the contours we will deal with.
In Section 4, we explain the strategy of the proofs and introduce
abstract contours models which are at the core of the
Pirogov-Sinai theory in the Zahradn\`\i k approach, \cite{Z}. 
In section 5, we define the
coarse-grained configurations and prove a ``Lebowitz
Penrose theorem", introducing a mean field functional.
In section 6 we identify the value of the inverse
temperature $\beta$ at which the order-disorder phase
transition takes place. Section 7 contains an estimate for
the finite volume corrections to the pressure, which
requires both a control on the decay of correlation and a
small deviation estimate. In Section 8, we prove an approximate factorization theorem 
for the contour weights stated in section 4 and derive the large
deviation estimate which provides the Peierls constant. 

Six appendices are added at the end of the paper: Appendix
A contains a short review of the mean field theory for the
Potts model as well as the derivation of the properties on
which we rely in the rest of the paper. Appendix B discuss
the local equilibrium properties which are used in various
parts of the paper. In Appendix C, we prove the existence
of the pressure for the abstract models introduced in
section 6. In Appendix D and E, we prove two lemmas needed in
section 7.  Finally, in Appendix E, we give the proof of theorem \ref{thmz4.1}.

\vskip 2.5cm \noindent
\section{Model and main
{results.}}
{\bf Two equivalent representations:}
The $Q$-state Potts model on $\Zz^d$, $Q>2$ an integer,
 may be equivalently regarded as a system of classical spins which
take $Q$ values, called ``colors'', or else as a system of $Q$
species of particles with the constraint that at each site there is
one and only one particle.  In the sequel, we will rather stick to
the second interpretation since its implementation  fits better both
with the coarse-graining we  need to consider, and with the
mean-field free energy functional to be introduced later.

In the first interpretation we call $\si(i)$ the spin at site $i\in
\Zz^d$,  {${\tilde\cXx}_o:=\{a_1,\dots a_Q\}$, the set of ``colors''
the spins take value in, ${\tilde\cXx} :={\tilde\cXx}_o^{\Zz^d}$,
 the configuration
space, and ${\tilde\cXx}_\La:={\tilde\cXx}_o^\La$
its restriction to a finite subset $\La$ of $\Zz^d$.

In the second interpretation $\w_q(i)$ denotes the occupation number
at site $i\in \Zz^d$ of the species $q\in Q$. Let $\Omega_o:= \{\vec
u_1,\cdots,\vec u_Q\}$ be the set of unit vectors in $\Rr^Q$ with
components $u_{q,k}=\delta_{q,k}$. Due to the constraint $\dis{\sum
_{q\in Q} \w_q(i)=1}$, the collection of all occupation numbers at
site $i$ can be written as a density vector,
$\dis{\vw(i)=(\w_q(i))_{q\in Q}}$, taking value in $\Omega_o$. We
denote the configuration space as $\Omega :=\Omega_o^{\Zz^d}$.

There is obviously a one to one correspondence between  $\Omega$ and
$\tilde\Omega$, defined by associating to each element $\si$ of
$\tilde\Omega$, a vector configuration $\vw$ of $\Omega$ as
        \begin{eqnarray}
\vw(i)= \vec u_q \Longleftrightarrow \si(i)=a_q
       \end{eqnarray}

{\bf Kac potentials:} Denoting by $\ga>0$  a ``scaling
parameter'', let $J_\ga$ be the kernel defined on
$\Rr^d\times \Rr^d$ as
     \begin{eqnarray}
       \label{def:J}
{J_\ga(x,y)}=\ga^{d}\mathcal J(\ga(x-y))
     \end{eqnarray}
where $\mathcal J(r)$ is a spherically symmetric
probability density supported by the unit
ball and differentiable with bounded derivative.

Then the Potts-Kac energy in a finite region $\La$  with
boundary conditions $\si_{\La^c}\in {\tilde\cXx}_{\La^c}$
is

    \begin{eqnarray}
    \label{def:Kac-Ham-b}
{H_{\ga,\La}}(\si_\La|\si_{\La^c}):=-\frac{1}{2}\sumtwo{i,j\in
\La}{i\neq j}{J_\ga(i,j)}\Ii_{\{\si_\La(i)=\si_\La(j)\}}
-\sumtwo{i\in \La}{j\in
\La^c}{J_\ga(i,j)}\Ii_{\{\si_\La(i)=\si_{\La^c}(j)\}}
    \end{eqnarray}
which, in the particle representation, reads  

   \begin{eqnarray}
        \label{def:Kac-Ham-c}
{H_{\ga,\La}}(\vw_\La|\vec s_{\La^c}) = -\frac{1}{2}\sumtwo{i,j\in
\La}{i\neq j}{J_\ga(i,j)} \vw_\La(i)\cdot\vw_\La(j) -\sumtwo{i\in
\La}{j\in \La^c}{J_\ga(i,j)} \vw_\La(i)\cdot\vec s_{\La^c}(j)
       \end{eqnarray}
where the characteristic functions in \eqref{def:Kac-Ham-b} have
been substituted by a scalar product between density vectors,
$\dis{\vw \cdot \vw' :=\sum_{q=1}^{Q} \w_q \w_q'}$. This
representation allows in particular to extend the definition
\eqref{def:Kac-Ham-c} to a wider set of boundary conditions $\vec
s_{\La^c}$, where $\vec s$ is taken in $ L^{\infty}(\Zz^{d},S_{Q})$,
with $S_Q$ is the set of all density vectors in $\Rr^Q$,
\begin{eqnarray}
\label{def:Kac-Ham-c-2}
S_Q = \bigl\{\vec\rho\in\Rr^Q_+,\sum_q \rho_q =1\bigr\}
\end{eqnarray}
 }
The set $S_Q$ is the convex set in $\Rr^Q$ which extremal points
identify with the elements of $\Omega_o$: \vskip .5cm \noindent

  The finite-volume Gibbs
specifications are then the probability measures
  \begin{eqnarray}
\label{eq:gibbskac} \mu_{\ga,
\beta,\La}(\vw_\La|\vec s_{\La^{c}})= \frac{e^{-\beta
{H_{\ga,\La}}(\vw_\La|\vec s_{\La^{c}})}}{Z_{\ga,\beta,\La}(\vec s_{\La^{c}})}
   \end{eqnarray}
where $Z_{\ga,\beta,\La}(\vec s_{\La^{c}})$ is the partition
function
\begin{eqnarray*}
Z_{\ga,\beta,\La}(\vec s_{\La^c}):=\sum_{\vw_\La}e^{-\beta
{H_{\ga,\La}}(\vw_\La|\vec s_{\La^{c}})}
\end{eqnarray*}

   \vskip .5cm

{\bf Mean field:} The mean field free energy density is
    \begin{eqnarray}
   \label{def:mean-field-free-energy.a}
\phi^{\mf}_\beta(\vec\rho) = -\frac{1}{2}\sum_{q=1}^{Q}
\rho_q^2 +\frac{1}{\beta} \sum_{q=1}^{Q}\rho_q \ln
(\rho_q), \quad \vec\rho=(\rho_q)_{ q\in Q} \in S_Q
    \end{eqnarray}
The $q$-th component $\rho_q$ is interpreted as the density of particles of
species $q$, the first term in
\eqref{def:mean-field-free-energy.a} is then (at leading
order in the number of particles)
the energy density
supposing that each particle interacts equally with all the
others, and the second term is the entropy. Referring to Appendix
\ref{app:meanfield} for details, we recall that in the mean field theory,
for each value of $Q$,
there is a critical inverse temperature $\bcmf$ such that:

 for all $\beta<\bcmf$,  $\phi^{\mf}_\beta$ has a unique
minimizer denoted by $\vec\rho^{\,-1}_{\beta}$;

for all $\beta>\bcmf$ there are $Q$ minimizers $\vec\rho_\beta^{\,
p}$, $p\in\{1,\cdots,Q\}$;

for $\beta=\bcmf$ there are $Q+1$ minimizers $\vec\rho^{\,\p}_{c}$,
$\p\in\{-1,1,\cdots,Q\}$.

In the above result and in the sequel, we label with $\p$, $\p \in
\{-1,1,\cdots, Q\}$, the $Q+1$ mean field minimizers at the critical
temperature, $\p = -1$ referring to the disordered phase, and
$\p=q$, $q>0$ the ordered one in which the color $q$ dominates.

\vskip.5cm {\bf Main result:}  For the finite range Kac-Potts models
with $\ga$ small enough, a situation similar to the mean-field
results holds. Calling a set of DLR measures {\it mutually
independent } if none of them is a convex combination of the others,
we will prove in the sequel the following:

\vskip.5cm

     \begin{thm}
      \label{thm:main-p}
For any $d\ge 2$ and $Q\ge 3$, there exists $\bar\ga>0$ such that
for any $\ga\in(0,\bar\ga)$, there is a value $\beta=\bcg$ at which
there are $Q+1$   mutually independent DLR measures with Gibbs
specifications \eqref{eq:gibbskac}, $\mu^{\,\p}_{\ga,\bcg}$,
$\p=-1,1, \cdots,Q$.
        \end{thm}

\vskip.5cm

In the course of the proof, we will characterize quite explicitly
the support properties of the DLR measures $\mu^{\,\p}_{\ga,\bcg}$
which will make evident closeness to mean field, in particular we
will see that for a suitable constant $c$
  \[
|\bcg-\beta_c^{\mf}|< c \ga^{1/2}
  \]
and prove that with large probability in $\mu^{\,\p}_{\ga,\bcg}$ the
empirical average of $\vw(i)$ over suitably large blocks is close to
the (critical) mean field value $ \vec\rho^{\,\p}_{c}$.

In a forthcoming paper we will also prove that the DLR measures
$\mu^{\,\p}_{\ga,\bcg}$ are translational invariant and have trivial
$\si$-algebra at infinity; moreover any other translational
invariant DLR measure is a convex combination of the
$\mu^{\,\p}_{\ga,\bcg}$ which are then the only ergodic DLR
measures.

\vskip 1.5cm

\section{ Scales, Phase Indicators and Contours}
   \label{sec:scales}
section3
Coarse graining is the master word in the proof of Theorem
\ref{thm:main-p}. We will need to define three scale lengths
$\ell_0$, $\ell_{-,\ga}$ and $\ell_{+,\ga}$, depending on the scale
parameter $\ga$. The first one, the shortest, will be used to
estimate partition functions \`a la Lebowitz-Penrose.  $\ell_0$ is
much shorter than the interaction range $\ga^{-1}$, yet much larger
than the lattice spacing, set equal to 1. $\ell_{-,\ga}$ is the
scale at which one ``recognizes a phase'': the empirical average of
the spins in boxes of side $\ell_{-,\ga}$ will be used as an
indicator of the local state of the system. When compared to the
mean field equilibrium value it will allow us to check whether the
system is locally close to an equilibrium. $\ell_{-,\ga}$ will be
chosen much larger than  $\ell_0$, yet still much smaller than the
range $\ga^{-1}$ so that the indicator can be regarded as a truly
local estimator. Finally $\ell_{+,\ga}$ is chosen much larger than
$\ga^{-1}$, such that if in a box of side $\ell_{+,\ga}$ the phase
indicator constantly indicates the same equilibrium, then the
behavior of the spins inside the box but far from the boundaries is
almost uncorrelated to the outside.

A possible choice for $\ell_0$, $\ell_{-,\ga}$ and $\ell_{+,\ga}$ is
to have them scale as $\ga^{-\frac{1}{2}}$, $\ga^{-(1-\alpha)}$ and
$\ga^{-(1+\alpha)}$, with $\alpha$ positive and small enough.
More precisely we set
these three lengthes  as the closest powers of 2 to these values
   \begin{eqnarray}
\ell_0=2^{\left[\frac{1}{2}\frac{\ln\ga^{-1}}{\ln
2}\right]},\quad
\ell_{\pm,\ga}=2^{\left[(1\pm\alpha)\frac{\ln\ga^{-1}}{\ln
2}\right]}
   \end{eqnarray}
($[\cdot]$ is the integer part of $\cdot$), so that the ratios
$\frac{\ell_{+,\ga}}{\ell_{-,\ga}}$ and
$\frac{\ell_{-,\ga}}{\ell_0}$ are integers. We can then construct
three partitions $\cD^{\ell}$ of $\Rr^d$ in cubes of size $\ell$,
 $\ell=\ell_0,\ell_{-,\ga},\ell_{+,\ga}$  which are one coarser than the other (if $\ga$ is small enough).
In order to define our local phase indicator,we need to define an
accuracy parameter $a$. In the course of the proof, various
restrictions on the possible choices of $\alpha$ and $a$ will
appear, none of which critical, nor necessarily optimal. We write
them here for the reader's convenience, but it is somewhat simpler
to keep in mind the choice $\alpha << 1$ and $a << \alpha$. In the
sequel, we will require:
\begin{eqnarray*}
\alpha &<& \frac{1}{16 d} \\
a &<& \min(\frac{1}{4},\frac{\alpha}{2})\\
-d (1 - \alpha) + 2 a &<& -2 d \alpha
\end{eqnarray*}

We define a local phase indicator as
    \begin{eqnarray}
\label{def:eta} \eta_x(\vec \xi):=
  \begin{cases}
    \p & \text{if }~
    \dis{\|\vw^{\,\ell_{-,\ga}}(x)-
   {\vec \rho_{c}^{\,\p}}\|_{\star}< \ga^{a}}, \quad
    {\p\in\{{\ds},1,\dots, Q\}}\\
    \\
    0 & \text{otherwise}.
  \end{cases}
    \end{eqnarray}
where  $\|\cdot\|_{\star}$ is the sup norm, $\vec \rho_{c}^{\,\p}$
is the $\p$-th minimizer of the mean field free energy functional at
$\beta=\bcmf$ and $\vw^{\,\ell}(x)$ is the empirical average of
$\vw$ over $C_x^{\ell}$, the cube of the partition $\cD^{\ell}$
which contains $x$:
    \begin{eqnarray}
    \label{def:p-ell0}
\vw^{\,\ell}(x) = \ell^{-d} \sum_{i\in C_x^{\ell}} {\vw}(i)
\end{eqnarray}

We also define a phase indicator at scale $\ell_{+,\ga}$,
  \begin{eqnarray}
  \label{def:Theta}
\Theta_x(\vec \xi)=  \begin{cases}
    \eta_x(\vec \xi) & \text{if }\eta_y(\vec \xi)= \p \quad\forall y
    : {C_y^{\ell_{+,\ga}}}\sim {C_x^{\ell_{+,\ga}}}, \\
    0 & \text{otherwise}.
  \end{cases}
   \end{eqnarray}

The set $\{y: C^{\ell_{+,\ga}}_y\sim C^{\ell_{+,\ga}}_x\}$ is the
set of cubes in $\cD^{\ell_{+,\ga}}$ $*$-connected with
$C^{\ell_{+,\ga}}_x$, i.e. :
$\overline{C^{\ell_{+,\ga}}_y}\sqcap\overline{C^{\ell_{+,\ga}}_x}\neq
\emptyset$.

By definition, for any $\p \ne \q$, the distance between any regions
$\{x:\Theta_x(\vec s)=a_{\p}\}$ and $\{x:\Theta_x(\vec s)=a_{\q}\}$
is at least $2\ell_{+,\ga}$. The interspace between these regions
will be the support for the {\em contours} which we now define with
respect to a configurations $\vw \in \Omega$. A similar definition
will also hold with respect to a continuous profile $\vec \rho\in
L^\infty(\Rr^d, S_Q)$.

\begin{defin}
\label{def:contours} A contour $\Ga\equiv(\ssp(\Ga),{\eta_\Ga})$ for
a configuration $\vw$ in $\Om$ , is specified by a couple
$(\ssp(\Ga),{\eta_\Ga})$ where $\ssp(\Ga)$ is one of the maximal
connected component of the subset $\{x:\Theta_x(\vw)=0\}$  and
{$\eta_{\Ga}$} is the coarse grained configuration on $\ssp(\Ga)$ at
scale $\ell_{\ga,-}$: {$\eta_{\Ga}\equiv\{\eta_x(\vw)\}_{x\in
\ssp(\Ga)}$.}
\end{defin}

\vskip .5cm \noindent

We now define the weight of these contours in the following way:
\vskip .5cm \noindent

We denote by $|\Ga|\equiv|\ssp(\Ga)|$ the volume of the region
$\ssp(\Ga)$, and by $N_\Ga:= \frac{|\Ga|}{|C_0^{\ell_{+,\ga}}|}$ the
number of $\cD^{\ell_{+,\ga}}$-cubes  in $\ssp(\Ga)$. For any
bounded contour ($|\Ga|<\infty$), we denote by $\Ext(\Ga)$ the
(unique) unbounded connected component of $\ssp(\Ga)^{c}$ and  by
$\{\Int_{i}(\Ga)\}$, $i\in I$ the collection of its bounded
connected components.

For any set $B\sqsubset \Rr^{d}$ $B\sqsubset \Rr^{d}$, $r\in \Rr$,
we define
\begin{eqnarray}
\delta_{\ins}^r[B]&:=&\{x\in B: \dist(x,B^c)\leq r\}\\
\delta_{\out}^r[B]&:=&\{x\in B^{c}: \dist(x,B)\leq r\}
\end{eqnarray}
where $\dist(x,B)=\inf_{y\in B} \dist(x,y)$ and
$\dist(x,y)=\sup_{1\le k\le d} |x_k-y_k|$.

Now, for any point $x$ in $\delta_{\out}^{\ell_{+,\ga}}[\ssp(\Ga)]$,
$\Theta_x\ne 0$ and its value is fixed by $\eta_\Ga$. We define
\begin{eqnarray}
A^{\q}\equiv A^{\q}(\Ga):=\{x\in
\delta_{\out}^{\ell_{+,\ga}}[\ssp(\Ga)]: \Theta_x= a_q \}
\qquad ; \qquad A(\Ga):= \bigsqcup_{\q} A^{\q} =
\delta_{\out}^{\ell_{+,\ga}}[\ssp(\Ga)]
\end{eqnarray}
\begin{eqnarray}
I^{\q}(\Ga):=\{i: \Int_i(\Ga)\sqcap A^{\q}\neq \emptyset;
\} \qquad ; \qquad \Int^{\q}(\Ga):= \bigsqcup_{i\in
I^{\q}(\Ga)}\Int_i(\Ga)
\end{eqnarray}
We call $\Ga$ a ``$\p$-contour'' if $\Ext(\Ga)\sqcap
A^{\p}\neq \emptyset$
and define
\begin{eqnarray}
c(\Ga):=\ssp(\Ga)\sqcup_{i\in I}\Int_{i}(\Ga)
\end{eqnarray}
\vskip .5cm \noindent

\vskip .5cm \noindent Let denote by $\cE(\Ga,\p)$ the event
that $\Ga$ is a $\p$-contour:
\begin{eqnarray*}
& &\cE(\Ga,\p):= \left\{\vw: \eta_x(\vw)=\eta_\Ga
~~\forall~ x\in \ssp(\Ga)~~; ~~
\linetwo{\Theta_x(\vw)=a_{\q} ~~\forall~  x\in
A^{\q}(\Ga)~\forall \q\in \{{\ds},1,\dots,
Q\}}{\Theta_x(\vw)=a_{\p} ~~\forall~  x\in A(\Ga)\sqcap
\Ext(\Ga)}\right\}
\end{eqnarray*}
and by $\cE(\not\Ga,\p)$ the event that the phase $\p$
extends on $\ssp(\Ga)\sqcup A(\Ga)$
\begin{eqnarray*}
& &\cE(\not\Ga,\p):= \{\vw: \Theta_x(\vw)=a_{\p} ~~\forall~
x\in \ssp(\Ga)\sqcup A(\Ga)\}
\end{eqnarray*}

\vskip .5cm \noindent We then define the weight of a $\p$-contour as
the ratio:
\begin{eqnarray}
\label{weight} w^{\p}_{\ga,\beta}(\Ga;\vw_{A^{\p}}):=
\frac{\mu_{\ga,\beta,c(\Ga)\setminus
\Int^{\p}(\Ga)}(\cE(\Ga,\p)|{\vw_{A^{\p}}})}{\mu_{\ga,\beta,c(\Ga)\setminus
\Int^{\p}(\Ga)}(\cE(\not\Ga,\p)|{\vw_{A^{\p}}})}
\end{eqnarray}

\vskip .5cm \noindent Let $\cX^{\p}$ be the set of ``correct"
$\p$-configurations:

   \begin{eqnarray}
    &&
    \cX^{\p}:=\{\vec s: \Theta_x(\vec s)=a_{\p}~\forall x\in
\Rr^{d}\}
       \label{cX}
       \end{eqnarray}
where, depending on the context,  $\vec s$ will be either in $\Om$
or in $L^{\infty}(\Zz^d,S_Q)$

Using an iteration procedure, the partition function with boundary
conditions in $\cX^{\p}$ can be rewritten in terms of $\p$-contours
as:
\begin{eqnarray}
\label{part-fun} Z_{\ga,\beta,\La}^{\p}(\vec s_{\La^c})=
\sum_{\und\Ga\in\cB^{\p}_{\La}}
\sum_{\vw_{\La}\in\cX^{\p}_{\La}}\prod_{\Ga\in \und\Ga
}w^{\p}_{\ga,\beta}(\Ga;\vw_{A^{\p}})e^{-\beta
H_{\ga,\La}(\vw_{\La}|\vec s_{\La^c})}
\end{eqnarray}
where $\und \Ga$ is a configuration of compatible $p$-contours and
$\cB^{\p}_{\La}$ is a the set of all possible configurations of
compatible $\p$-contours with support inside $\La$.

\vskip .5cm \noindent

The expression \eqref{part-fun} already shows some of
  the main differences we incounter here with respect to the classical Pirogov-Sinai
  theory at low temperature.

\begin{itemize}
  \item
  The ``ground configurations" $\cX^{\p}, \p=-1,1,\dots,Q$
are not fixed configurations, but ensemble of configurations.
 The study of these  configurations and of their perturbations is then more
 complicated and
  involve the study of variational problem for a non local functional.
 \item
  The partition function cannot be express in terms of p-compatible contours
  because it persists a weak interaction between
  contours
\end{itemize}

\vskip .5cm \noindent

A similar situation appear in \cite{BZ1}-\cite{BZ2}, where Kac
models are considered  at low temperature. For such models, the
references configurations cannot be chosen as the ground states of
the energy even at low temperatures,
 because the direct interaction between two spin is too weak, an one
 needs to take into account the local entropy. The partition function is
then expressed in terms of interacting contours.

Nevertheless the techniques developed in \cite{BZ1}-\cite{BZ2}, cannot be
immediately applied here, since they are based on a cluster
expansion at low temperatures in order to extend the classical PS
theory to the case of weak interactions, giving a result uniform in
the range of the interaction.

Here, we will follow the extension of Pirogov-Sinai techniques to
high temperatures used in \cite{LMP} and \cite{errico-leip} which also deal with
perturbation of a mean-field theory. \vskip .5cm \noindent

\vskip .7cm \noindent

\setcounter{equation}{0} \vskip 2.5cm \noindent
\section{Main steps of the proof}

 In this section we state, postponing the proofs, the main
 steps that
lead to the proof of the
  Theorem \ref{thm:main-p}.

A very preliminary step is an approximate factorization of the
contours weights \eqref{weight}, which relies on properties of the
following mean-field free energy functional
 defined on the
functions in $\vec \rho \in L^\infty(\La,S_Q)$ as follows:
\begin{eqnarray}
\label{def:freeenergy}
F_{{\ga,\beta,\La}}(\vec\rho_\La|\vec\rho_{\La^c}) =
V_{\ga,\La}(\vec\rho_\La|\vec\rho_{\La^c}) -\frac{1}{\beta}
I(\vec\rho_\La)
\end{eqnarray}
where $\vec\rho_{\La^c}\in L^\infty(\La^c,S_Q)$ defines the boundary
conditions. The two functionals
$V_{\ga,\La}(\vec\rho_\La|\vec\rho_{\La^c})$ and $I(\vec\rho_\La)$
are respectively the energy and the entropy of $\vec\rho_\La$,
\begin{eqnarray*}
V_{{\ga,\La}}(\vec\rho_\La|\vec\rho_{\La^c}) =
-\frac{1}{2}\int_\La\int_\La {J_\ga(x,y)} \bigl(
\vec\rho_\La(x)\cdot\vec\rho_\La(y) \bigr) dx dy
-\int_\La\int_{\La^c} {J_\ga(x,y)} \bigl(
\vec\rho_\La(x)\cdot\vec\rho_{\La^c}(y) \bigr) dx dy
\end{eqnarray*}
\begin{eqnarray*}
I(\vec\rho_\La) = - \int_\La \sum_{q=1}^Q \rho_{\La,q}(x) \ln
(\rho_{\La,q}(x))   dx
\end{eqnarray*}

These functionals comes out naturally after a coarse graining
procedure on the scale $\ell_0$. In particular, we will prove the
following
\begin{prop}
There exists a constant $c>0$ such that for all $\ga>0$ and all
bounded $\cD^{\ell_0}$-measurable regions $\La$ in $\Rr^d$,
\begin{eqnarray*}
\big|\log Z_{\ga,\beta,\La} (\vw_{\La^c}) +\beta \inf_{\vec\rho_\La}
 F_{{\ga,\beta,\La}}(\vec\rho_\La|\vw^{\,\ell_0}_{\La^c}) \big| \le \beta c
 \ga^\frac{1}{2}|\La|
\end{eqnarray*}
\end{prop}
This result together with a local stability result for the
functional around its minimizers allows us to set $\vec
\rho_c^{\,\q}$ (respectively $\vec \rho_c^{\,\p}$) on $A^{\q}$,
$\q\not=\p$, in the expression of the numerator (respectively
denominator) of the left hand side of \eqref{weight} at the price of
a small error and get the following bound:

\begin{thm}
   \label{thm:Peierls-factoriz-b}
  There are $\bar \ga, \bar b$ and a constant $c$ such that for all $\ga<\bar \ga,
\linebreak |\beta-\bcmf|<\bar b $
:
       \begin{eqnarray}
       \label{eq:Peierls-factoriz-b}
       w^{\p}_{\ga,\beta}(\Ga;\vw_{A^{\p}})&\le &
       \prod_{\q\ne \p} \frac{Z^{\q}_{\ga,\beta,\Int^{\q}(\Ga)\setminus A^{\q}}
       ({\vec\rho^{\q}})}{
        Z^{\p}_{\ga,\beta,\Int^{\q}(\Ga)\setminus A^{\q}}(\vec\rho^{\p})}
\exp\{-\beta |A^{\q}|
(\phi^\mf_\beta(\vec\rho^{\,\q})-\phi^\mf_\beta(\vec\rho^{\,\p}))\}\\
       &\times&
\frac{Z_{\ga,\beta,sp(\Ga)}(\cE(\Ga,\p)|{\vw_{A^{\p}}},\{\vec\rho^{\q}_{A^{\q}},
\q\not= \p\})}
{Z_{\ga,\beta,sp(\Ga)}(\cE(\not\Ga,\p)|{\vw_{A^{\p}}},\{\vec\rho^{\p}_{A^{\q}},\q\not=\p\})}\nn\\
& \times&\exp\{c\ga^{\frac{1}{2}}|\Ga|\}
       \nn
       \end{eqnarray}
\end{thm}

Provided a good control on the ratios of partition functions in the
first line of \eqref{eq:Peierls-factoriz-b}, the factor in the
second line will provide the Peierls bound, using a large deviation
result for the functional $F_{{\ga,\beta}}$ on $\ssp(\Ga)$ and paying
again a price of order $\exp\{c\ga^{\frac{1}{2}}|\Ga|\}$.
This will finally lead to the the following Theorem:

\begin{thm}
   \label{thm:Peierls-0}
There is $\bar \ga $ such that for any $\ga<\bar\ga$ there exists $\bcg$, such that $\forall \q$:
       \begin{equation}
       \label{eq:key}
       w^{\q}_{\ga,\bcg}(\Ga;\vw_{A^{\q}})\le \exp\{-\fK_\ga |N_\Ga|\}\hskip1cm
  \text{with }~~    \fK_\ga= c_f \ga^{2a}
|C^{\ell_{-,\ga}}|
       \end{equation}
      with $c_f$ a constant depending only on $d$ and $Q$,
\begin{equation}
c_f=\frac{1}{3^{d+1}}(Q-\bcmf) > 0
\end{equation}

\end{thm}

 The proof of the Theorem \ref{thm:main-p} will follow immediately from the above,
 using the well known Peierls argument that will be omitted here.
\vskip 1.5cm \noindent

Now the main work is to obtain a good control on the ratios
\begin{eqnarray}
\label{ratio}
\frac{Z^{\q}_{\ga,\beta,\Delta}
       ({\vec\rho^{\q}})}{Z^{\p}_{\ga,\beta,\Delta}(\vec\rho^{\p})}
\end{eqnarray}
Of course, whenever both $\p$ and $\q$ refer to ordered states, the ratio
is equal to one by symmetry (by permutation of the colors), and we need essentially to consider
the case when $\p$ or $\q$ equals $-1$. Here we need the Pirogov-Sinai theory and we follow the Zahradn\`\i k's approach.
We introduce $Q+1$ ``abstract contour models'' defined on
the product spaces: $\cX^{\p}\times \cB^{\p}$. The
partition function of the $\p$-th abstract model and the weights of the contours are defined recursively as:

\begin{eqnarray}\label{z4.7}
Z_{\abs,\beta,\La}^{\p}(\vec s_{\La^c})=
\sum_{\und\Ga\in\cB^{\p}_{\La}}
\sum_{\vw_{\La}\in\cX^{\p}_{\La}}\prod_{\Ga\in \und\Ga
}W^{\p}(\Ga;\vw)e^{-\beta H_{\ga,\La}(\vw_{\La}|\vec s_{\La^c})}
\end{eqnarray}
and
 \begin{eqnarray}
 \label{cutoff-weight-b}
\hat W^{\p}_\ga(\Ga;\vec s):=\min
\{w^{\p}_{\abs,\beta}(\Ga;\vec s), e^{-\frac{\fK_{\ga}}{2}
N_\Ga}\}
 \end{eqnarray}
 where $\fK_\ga$ is given by \eqref{eq:key}
 and   $w^{\p}_{\abs,\beta}(\Ga;\vec s)$ is given by

\begin{equation}
\label{def:w-abs} w^{\p}_{\abs,\beta}(\Ga;\vec
s_{A^{\p}}):= \frac{ \dis{\sum_{\vw_{\hG^{\p}}\in
\Om_{\hG^{\p}}}e^{-\beta
H_{\ga,\hG^{\p}}(\vw_{\hG^{\p}}|\vec s_{A^{\p}})}
\Ii_{\eta(\vw_{\ssp(\Ga)})=\eta_\Ga}\prod_{\q\ne\p}\Ii_{\eta_{\vw_{A^{\q}}}=a_{\q}}
 Z_{\abs,\beta,\Int^{\q}(\Ga)\setminus A^{\q}}^{\q}(\vw_{A^{\q}})}}
{\dis{\sum_{\vw_{\hG^{\p}}\in \Om_{\hG^{\p}}}e^{-\beta
H_{\ga,\hG^{\p}}(\vw_{\hG^{\p}}|\vec s_{A^{\p}})}
\Ii_{\eta(\vw_{\hG^{\p}})=a_{\p}}\prod_{\q\ne\p}
 Z_{\abs,\beta,\Int^{\q}(\Ga)\setminus A^{\q}}^{\q}(\vw_{A^{\q}})}}
\end{equation}
where $\hG^{\p}=\ssp(\Ga)\bigsqcup_{\q\ne \p} A^{\q}$.

\vskip .5cm \noindent

 We stress   that the elements of a
pair $(\vw,\und \Ga)$ in an abstract model are totally
unrelated, in fact the configuration $\vw$ is in $\mathcal
X^{\p}$ and therefore has no contours. The sum
$\sum_{\vw_{\hG^{\p}}\in \Om_{\hG^{\p}}}$ appearing in the
definition \eqref{def:w-abs} enters only as definition of
the weights, without any relations with  the configurations
$\vw\in\cX^{\p}$ of the abstract model that we are
considering.

\vskip .5cm \noindent

For any bounded $\mathcal D^{\ell_{+,\ga}}$ measurable region $\La$ and any $\vec s_{\La^c}$ in $L^{\infty}(\Zz^{d},S_{Q})$,
we define the ``dilute'', finite volume Gibbs measures on
$\cX^{\p}_{\La} \times \mathcal B^{\p}_\La$ as,

       \begin{equation}
         \label{z4.6}
\mu_{\abs,\La}^{\p}(\vw',\und \Ga|\vec s_{\La^{c}}) :=
\frac{{\bf 1}_{\vw' \in
\cX^{\p}}}{Z_{\abs,\beta,\La}^{\p}(\vec s_{\La^{c}})}
\prod_{\Ga \in \und\Ga} W^{\p}(\Ga,\vw') e^{-H_{\ga,\La}(
\vw'_\La | \vec s_{\La^c})}
        \end{equation}

\vskip .5cm \noindent

The following theorem states the relation between the true
model and the abstract ones.  Its proof can be easily obtained by induction on the volume,
but we omit it here since it is a standard result in Pirogov Sinai theory.
\vskip .5cm \noindent
\begin{thm}
   \label{thm:ab-true}
If for any $\p$ and any $\p$-contour $\Ga$ the weights
$\hat W^{\p}_\ga(\Ga;\vw)$, defined by
\eqref{cutoff-weight-b},
satisfy 
       \begin{equation}
       \label{eq:ab-true1-p}
       \hat W^{\p}_\ga(\Ga;\vec s)< e^{-\frac{\fK_\ga}{2} N_\Ga}
       \end{equation}

then 
\begin{equation}
       \label{eq:ab-true2-p-p}
W^{\p}_\ga(\Ga;\vec s)= w^{\p}_{\ga,\beta}(\Ga;\vec s
)\hskip1cm ;\hskip1cm
       Z_{\abs,\beta,\La}^{\p}( \vec s_{\La^{c}})=
       Z_{\beta,\La}^{\p}( \vec s_{\La^{c}})
       \end{equation}
\end{thm}

\vskip .5cm \noindent

\vskip .5cm \noindent

 Let denote by $P^{\p}_{\abs,\La,\ga,\beta}(\und\vw_{\La^c})$
 the ``finite volume pressure" of the $\p$-th abstract model:

\begin{equation*}
P^{\p}_{\abs,\La,\ga,\beta}(\und\vw_{\La^c}):=\frac{1}{\beta
|\La|}\ln Z_{\abs,\beta,\La}^{\p}(\und\vw_{\La^c})
\end{equation*}

The following theorem characterizes the infinite volume limit of these pressures
\begin{thm}
   \label{thm:equalitypresure-b}
   {Let $\{\La_n\}$ an increasing  sequence of sets in $\Rr^d$ of side $2^n\ell_{+,\ga}$.
  For each $\p$ in $\{-1,1,\cdots,Q\}$, there exists the limit:
\begin{eqnarray}
\label{def:P-abs} P^{\p}_{\abs,\ga,\beta} :=
\lim_{n\to\infty}\frac{1}{\beta |\La_n|}\ln
P^{\p}_{\abs,\La_n,\ga,\beta}(\vec\rho^{\p}_{\beta})
\end{eqnarray}
which is continuous in $\beta$. Moreover there are
constants} $c_b$, $\bar \ga$ such that for any $\ga<\bar \ga$
there is a value of $\beta$, noted  $\bcg$, with $|\bcmf-\bcg|<c_b\ga^{1/2}$
such that all the pressures are equal
       \begin{equation}
       \label{eq:abs1}
P^{\p}_{\abs,\ga,\bcg}=P^{\q}_{\abs,\ga,\bcg}\hskip1cm \forall \p,\q
    \end{equation}

\end{thm}

The existence of the
limits follows by general arguments for regular
interactions and the existence of a (non necessarily unique) value  of $\beta$ at
which they are equal follows, for $\ga$ small enough,
using a continuity argument
and the fact that the mean field pressures
are crossing at $\beta=\bcmf$.
Fixing $\beta=\bcg$ \eqref{eq:abs1} holds and
the ratios:

\begin{eqnarray}
\label{ratios}
\frac{Z^{\q}_{\abs,\beta,\La}
       ({\vec\rho^{\q}})}{Z^{\p}_{\abs,\beta,\La}(\vec\rho^{\p})}
\end{eqnarray}

converge to $1$ in the limit $|\La|\to\infty$ but the
control of the finite volume corrections requires an extra analysis
with respect to the standard low
temperature case, where the reference configurations are
singletons. Here we use a partial cluster expansion to sum over the contours using the measure \eqref{z4.6}
and write its marginal $\mu_{\abs,\La}^{\p}(.|\vec s_{\La^{c}})$ on $\cX^{\p}_\La$.
We characterize then the marginals $\mu_{\abs,\La}^{\p}$ using a generalized Dobrushin argument.

Following Dobrushin, we
introduce an interpolation Hamiltonian as follows. Let $u\in
[0,1]$ and :

\begin{eqnarray}
\label{def:h-u}
\hat h^{\p}_u(\vw_{\La}|\vw_{\La^c}):=u H_{\ga,\La}(\vw_{\La}|\vw_{\La^c}) +(1-u)\fH_{\ga,\La}^{\p}(\vw)
\end{eqnarray}
where $\fH^{\p}$ are the one body ``mean field"
Hamiltonians:
\begin{eqnarray}
\label{def:H0-b}\fH_{\ga,\La}^{\p}(\vw):=\sum_{i\in \La}\sum_{j\ne i} J_{\ga}(i,j)
((\vw-\vec\rho^{\p})\cdot \vec\rho^{\p})+ H_{\ga,\La}(\vec\rho^{\p}|\vec\rho^{\p})
\end{eqnarray}

We denote by
$\mu^{\pm}_{\abs,\La;u}(\vw_i|\und\vw)$
the [finite volume] Gibbs measure with hamiltonian $\hat
h^{\pm}_u(\vw)$ on  $\cX^{\pm}_\La$ and by
$Z_{\abs,\beta,\La;u}^{\pm}$ the associated partition function.
The finite volume pressures of the abstract model
can be written in terms of correlations
as:
     \begin{eqnarray}
P^{\p}_{\abs,\ga,\La }
=
\frac{1}{\beta|\La|}\ln  Z_{\abs,\beta,\La;0}^{\p}
- \frac{1}{|\La|}\int_0^1 du\; \left\langle \tilde H^{\p}_\La( \vw'_\La|\vec\rho^{\p})-
 \fH^{\p}_{\La}( \vw'_\La) \right\rangle_{\tilde \mu^{\p}_{\abs,\La;u}}
          \label{interpol}
    \end{eqnarray}

The estimates for the finite volume corrections to the pressure will
follow essentially from the proof of exponentially decay of
correlations for the measures
$\mu^{\p}_{\abs,\La;u}(\vw_i|\und\vw)$, which
 is based mainly on  the
following two results:
\begin{thm}
   \label{thm:firstassumption}

For any $i\in \Zz^d$ there is a measurable set
$G^{\p}_i\sqsubset \cX^{\p}$ depending only on
$\{\und\vw_j,\; j\in C^{\ell_{-,\ga}}_i\setminus i\}$, such
that there exists $b(i,j)$ with the following properties:
\begin{eqnarray}
\label{firstassumption}
&&b(i,j)\ge 0 \;\; ;\;\; b(i,i)=0 \nonumber\\
&&\sup_{i\in\Zz^d}\sum_{j\in C^{\ell_{-,\ga}}_i} b(i,j)<\delta< 1 \\
&&
\cR(\mu^{\p}_{\abs,i;u}(\vw_i|\und\vw),\mu^{\p}_{\abs,i;u}(\vw_i|\und\vw'))\le
\sum_j b(i,j)\; \dist(\und\vw_j,\und\vw_j') \hskip1.6cm
\text{for any $\und\vw,\und\vw'\in G_i$ and any $u\in (0,1)$} \nonumber
\end{eqnarray}
where
$\cR(\mu^{\p}_{\abs,i;{u}}(\vw_i|\und\vw),\mu^{\p}_{\abs,i;{u}}(\vw_i|\und\vw'))$
is the  Vaserstein distance associated to the metric on the
configuration space defined by:

\begin{eqnarray}
\label{def:dist2} \dist(\vw_i,\vw'_i):=
\frac{1}{2}\sum_i|\w_i(i)-\w'_i(i)|
\end{eqnarray}

\end{thm}
 We will prove that  the theorem holds with  $G^{\p}_i$ defined as :
\begin{eqnarray}
\label{Gx} G^{\p}_i:=\{\vw\in \cX^{({\p})}:\vw^{(i,q)}\in
\cX^{({\p})} \forall q\}
\end{eqnarray}
 where we have denoted
\begin{eqnarray}
\hskip1cm \vw^{(i,q)}_j=
  \begin{cases}
    \vw_j & \text{$j\ne i$}, \\
    \vec u_q & \text{$j=i$}.
  \end{cases}
\end{eqnarray}

$G^{\p}_i$ is then the set of configurations $\vw$ which
belong to $\cX^{({\p})}$ independently of the value of
$\vw_i$ and it is measurable on $\vw_{i^c}$.

Notice that when $\und\vw,\und\vw'$ are not in $G^{\p}_i$,
the probability measures for $\vw_i$ have support on a
strict subset of $\Om$ and the statement of the
theorem \ref{thm:firstassumption}
would not hold in general.

\vskip .75cm \noindent
 The other result that allows
  to prove exponentially decay of the correlations 
is a bound on the probability of the ``bad set" of
configurations,  ${G_i^{\p}}^c$:

\begin{eqnarray}
\label{assumpt2b} \sup_{ \vw^*\in
\cX^{\p}}\mu^{({\p})}_{\abs,D,u}
({G_i^{\p}}^{c}|\vw^{~*})\le
 e^{-c\ga^{2a}\ell^{d}_{-,\ga}}
\end{eqnarray}
$c$ a positive constant and $D=C^{\ell_{-,\ga}}_i$.

Together with the exponential decay of the correlations, we need also to have a control on
the contribution of the parts close to the boundary. This
control follows from the next theorem that, in words,
states that well inside a ``correct region" (i.e. $\Theta=
\q$, $\q\ne 0 $), the typical configurations becomes ``very
close" (i.e. on the small scale $\ell_0$) to the
corresponding mean field value.

\begin{thm}
Let $A$ a finite subset of $\Zz^{d}$, $\hat A:= A\sqcup
\delta_{\out}^{\ell_{+,\ga}}[A]$, $\hat
h_u(\vw_{\hat A}|s_{\hat A^c})$
as in \eqref{def:hu} and
$\hat \mu^{\p}_{\abs;\hat A,u}$ is the measure on
$\cX^{\p}_{\hat A}$ associated $\hat h^{\p}_u(\vw_{\hat
A}|\vw_{\hat A^c})$, then uniformly in $u$:

\begin{equation}
       \label{eq:}
\hat \mu^{\p}_{\abs;\hat A,u}(\Ii_{|S^{\p}_A(\vw)|\ge
\ga^{1/8}|A| })\le e^{-c\ga^{3/8}|A|}
       \end{equation}
where:
\begin{eqnarray*}
S^{\p}_A(\vw):=\{i\in A\sqcup\delta_{\out}^{\ga^{-1}}[A]:
\|\vw^{(\ell_0)}(i)-\vec \rho^{\p}\|_{\star}\ge
\ga^{1/8}\}
\end{eqnarray*}

\end{thm}
Now collecting all these results, we get an estimate for the ratio of partition functions 
of two abstract models, 
We  then  prove the following theorem:
\begin{thm}
   \label{thm:Peierls-2}
There is $\bar \ga$, and a constant $\kappa_1$ such that
for any $\ga<\bar \ga$, there is a value of $\beta$, $\bcg$:
\begin{equation}
       \label{eq:Peierls-2}
\left|\ln\bigl(
       \frac{Z^{(+)}_{\abs, \bcg,\La}(\vec\rho^{(+)})
       e^{R^{\mf,\q}_{\ga,\La}}}{
        Z^{(-)}_{\abs,\bcg,\La}(\vec\rho^{(-)})e^{R^{\mf,\p}_{\ga,\La}}}
        \bigr)\right|\le \kappa_1\ga^{1/8}|\Ga|
       \end{equation}

       \end{thm}

where $R^{\mf,\p}_{\ga,\La}$ are the mean-field finite volume corrections to the pressures,
\begin{eqnarray}
       \label{eq:4correction}
R^{\mf,\p}_{\ga,\La}:=\frac{\beta}{2}\sumtwo{x\in \La}{y\in\La^c}J_\ga(x,y)
(\vec\rho^{\p}\cdot\vec\rho^{\p})
       \end{eqnarray}
Deriving a factorization theorem similar to Theorem \ref{thm:Peierls-factoriz-b} for the abstract 
contours models lead then to a bound for the abstract weights as in Theorem \ref{thm:Peierls-0}.
Hence, using Theorem \ref{thm:ab-true}, we can identify abstract and true weights at temperature $\beta=\bcg$
and in turn prove Theorem \ref{thm:Peierls-0} whiich lead to our result.

 In the next sections we will proceed by proving all results presented here,
 but in a different order. Precisely we will postpone the proof of the Theorem
 \ref{thm:Peierls-factoriz-b} and start the analysis
 of the abstract models, their pressures and the uniqueness of the associated measures.
 Finally we will prove the large deviation estimate needed in
 \ref{thm:Peierls-0}.
As a preliminary step we introduce the mean field functional whose
minimizers define the unperturbed states above which the
Pirogov-Sinai analysis is developed, and discuss its properties.

\section{Coarse graining and mean
field functional} \label{sec:freeenergy} \vskip .5cm
\noindent

Let $\ell$ a large positive integer and
 $\cD^{\ell}$ a partition of $\Rr^{d}$ in cubes of size
 $\ell$. For all $x$ in $\Rr^d$,
we denote by $C_x^{\ell}$ the cube of $\cD^{\ell}$ containing $x$.
We define a coarse-grained configuration on $\cD^{\ell}$ as follows:
for each configuration $\vw\in{\Omega}$ and any $x\in \Rr^{d}$, the
coarse-grained configuration (at scale $\ell$) is the Q-dimensional
vector,
\begin{eqnarray}
\label{def:p-ell0-b}
\vw^{\,\ell}(x) = \ell^{-d} \sum_{i\in
C_x^{\ell}} {\vw}(i)
\end{eqnarray}
where we make use of our notational conventions.
The $k^{th}$ component $\w_k^{\ell}(x)$ is the
empirical density of color $a_k$ in $C_x^{\ell}$.
 Due to the underlying discretization, $\vw^{\ell}(x)$
takes values in the finite set $M_{\ell^d}^Q$
\begin{eqnarray*}
M_{\ell^d}^Q=\bigl\{(r_1,\cdots,r_Q); (\ell^d r_k)
 \in
\Nn, \sum_{k=1}^Q r_k=1 \bigr\}
\end{eqnarray*}
Let $\La$ a $\cD^{\ell}$-measurable subset of $\Rr^d$.
The set of coarse-grained configurations in $\La$ is denoted
by ${\cXx}_\La^{\ell}$ and corresponds to the set of
$\cD^{\ell}$-measurable functions on $\La$ with values in
$M_{\ell^d}^Q$.
We extend the discrete set $M_{\ell^d}^Q$ to the simplex $S_Q$ in $\Rr^Q$ defined in \eqref{def:Kac-Ham-c-2}.
Thus all coarse-grained configurations in ${\cXx}_\La^{\ell}$ are also elements of
$L^\infty(\La,S_Q)$. Conversely we will approximate any
function in $L^\infty(\La,S_Q)$ by a coarse grained
configuration. For any $\vec\rho\in L^\infty(\La,S_Q)$, we
will denote by $\vec\rho^{\,\ell}$ its
$\cD^{\ell}$-measurable approximation
\begin{eqnarray}
\label{def:rhol0}
\vec\rho^{\,\ell}(x) = \ell^{-d}
\int_{C_x^{\ell}} {\vec\rho(y)} dy
\end{eqnarray}
for all $x$ in $\La$, and by $[\vec\rho\,]^{\ell}$ the
only function  in ${\cXx}_\La^{\ell}$ such that
\begin{eqnarray}
\label{def:parteintera}
-\frac{1}{2\ell^d}< [\vec\rho\,]_k^{\ell}(x)-\rho_k^{\ell}(x)\le
\frac{1}{2\ell^d}
\end{eqnarray}
for all $k$ and all $x$ in $\La$.

For $\La$ a finite
$\cD^{\ell}$-measurable region in $\Rr^d$, we define the
mean field free energy functional on $L^\infty(\La,S_Q)$
by,
\begin{eqnarray}
\label{def:freeenergy-b}
F_{{\ga,\beta,\La}}(\vec\rho_\La|\vec\rho_{\La^c}) =
V_{\ga,\La}(\vec\rho_\La|\vec\rho_{\La^c}) -\frac{1}{\beta}
I(\vec\rho_\La)
\end{eqnarray}
where $\vec\rho_{\La^c}\in L^\infty(\La^c,S_Q)$ defines the boundary conditions.
The two functionals $V_{\ga,\La}(\vec\rho_\La|\vec\rho_{\La^c})$
and $I(\vec\rho_\La)$ are respectively the energy and the
entropy of configuration $\vec\rho_\La$,
\begin{eqnarray*}
V_{{\ga,\La}}(\vec\rho_\La|\vec\rho_{\La^c}) =
-\frac{1}{2}\int_\La\int_\La {J_\ga(x,y)} \bigl(
\vec\rho_\La(x)\cdot\vec\rho_\La(y) \bigr) dx dy
-\int_\La\int_{\La^c} {J_\ga(x,y)} \bigl(
\vec\rho_\La(x)\cdot\vec\rho_{\La^c}(y) \bigr) dx dy
\end{eqnarray*}
\begin{eqnarray*}
I(\vec\rho_\La) = - \int_\La \sum_{q=1}^Q \rho_{\La,q}(x)
\ln (\rho_{\La,q}(x))   dx{=:- \int_\La\vec\rho_{\La}(x)
\cdot\ln \vec\rho_{\La}(x)}  dx
\end{eqnarray*}
For all $\cA\sqsubset{\cXx}_\La$ we define the constrained
partition function $Z_{\ga,\beta,\La}(\cA|\vw_{\La^c})$
\begin{eqnarray*}
Z_{\ga,\beta,\La}(\cA|\vw_{\La^c})= \sum_{\vw_\La\in \cA}
e^{-\beta {H_{\ga,\La}}(\vw_\La|\vw_{\La^c})}
\end{eqnarray*}
We now state a theorem relating constrained partition functions
and mean field free energy:

\begin{thm}
\label{thm:app} There exists a constant $c>0$ such that for
all $\ga>0$, $\ell \in (1,\ga^{-1})$ and all bounded
$\cD^{\ell}$-measurable region $\La$ of $\Rr^d$, the
following inequalities hold:\par For all subsets $\cA$ of
${\cXx}_\La^{\ell}$,
\begin{eqnarray*}
\ln Z_{\ga,\beta,\La}(\{\vw^{\,\ell}_\La\in
\cA\}|\vw_{\La^c}) +\beta \inf_{\vec\rho_\La\in \cA}
F_{{\ga,\beta,\La}}(\vec\rho_\La|\vw^{\,\ell}_{\La^c}) \le
\beta c \eps(\ga,\ell) |\La|
\end{eqnarray*}
and for all $\vec\rho_\La\in L^\infty(\La,S_Q)$,
\begin{eqnarray*}
\ln Z_{\ga,\beta,\La}(\{\vw_\La : \vw^{\,\ell}_\La=
[\vec\rho_\La]^{\ell}\}|\vw_{\La^c}) +\beta
F_{{\ga,\beta,\La}}(\vec\rho_\La|\vw^{\,\ell}_{\La^c})
\ge -\beta c \eps(\ga,\ell) |\La|
\end{eqnarray*}
where
\begin{eqnarray*}
\eps(\ga,\ell) = \ga \ell + \frac{\ln
\ell}{\ell^d}
\end{eqnarray*}
\end{thm}
\begin{proof}
We first estimate the difference between the energy
${H_{\ga,\La}}(\vw_\La|\vw_{\La^c})$ and its coarse-grained
approximation
$V_{{\ga,\La}}(\vw^{\,\ell}_\La|\vw^{\,\ell}_{\La^c})$.
Given two cubes $C_1$ and $C_2$ of the partition
$\cD^{\ell}$, for any two points $i\in C_1$ and  $j\in
C_2$, we have
\begin{eqnarray*}
|{J_\ga(i,j)} - \frac{1}{\ell^{2 d}}\int_{C_1\times
C_2} {J_\ga(x,y)} ~dx ~dy| \le 2\sqrt{d} \|\nabla \cJ\|_\infty
\ga^{d+1} \ell \Ii_{d(C1,C2)\le\ga^{-1}}
\end{eqnarray*}
where $d(C1,C2)=\inf_{x\in C_1,y\in C_2} |x-y|$.

Hence
\begin{eqnarray*}
&&\big|{H_{\ga,\La}}(\vw_\La|\vw_{\La^c}) -
V_{{\ga,\La}}(\vw^{\,\ell}_\La|\vw^{\,\ell}_{\La^c})\big|
\\
&&\hskip1cm\le
\sumtwo{C_1\in\cD^{\ell}_\La}{C_2\in\cD^{\ell}}
\big|\sumtwo{i\in C_1\cap\Zz^d}{j\in C_2\cap\Zz^d}
{J_\ga(i,j)}\vw_\La(i)\cdot \vw_\La(j)
 -\int_{C_1\times C_2} {J_\ga(x,y)}
\vw^{\,\ell}_\La(x)\cdot \vw^{\,\ell}_\La(y)\big|\\
&&\hskip1cm\le
\sumtwo{C_1\in\cD^{\ell}_\La}{C_2\in\cD^{\ell}}
\big|\sumtwo{i\in C_1\cap\Zz^d}{j\in C_2\cap\Zz^d}
\Big({J_\ga(i,j)}
 -\frac{1}{\ell^{2d}}\int_{C_1\times C_2} {J_\ga(x,y)}~ dx
~dy\Big)\vw_\La(i)\cdot \vw_\La(j)
 \big|\\
&&\hskip1cm\le 2\sqrt{d} \|\nabla \cJ\|_\infty \ga^{d+1} \ell
\sum_{C_1, C_2} |C_1| |C_2| \Ii_{d(C_1,C_2)\le\ga^{-1}}\\
&&\hskip1cm\le c_d \ga \ell |\La|
\end{eqnarray*}
where $c_d$ is a constant independent on $\ga$,
$\ell\le\ga^{-1}$ and $\La$
\begin{eqnarray}
\label{def:cd} c_d=2 \, 3^d\sqrt{d}\|\nabla \cJ\|_\infty
\end{eqnarray}
Thus for any profile $\vec\rho_\La$ in
${\cXx}_\La^{\ell}$, we have
\begin{eqnarray*}
\big| \ln
\frac{Z_{\ga,\beta,\La}(\{\vw_\La:\vw^{\ell}_\La=\vec\rho_\La\}
|\vw_{\La^c})} {|\{\vw_\La:
{\vw^{~\ell}}_\La=\vec\rho_\La \}| e^{-\beta
V_{{\ga,\La}}(\vec\rho_\La|\vw^{\ell}_{\La^c})}}\big|\le
c_d \ga \ell |\La|
\end{eqnarray*}
The cardinality of $\{\vw_\La: \vw^{\ell}_\La=\vec\rho_\La
\}$ can be related to the entropy of $\vec\rho_\La$. The
error bounds for the Stirling formula
\begin{eqnarray}
\frac{1}{12N+1}\le
\frac{\ln(N!)}{N(\ln(N)-1)+\ln(\sqrt{2\pi N})}\le
\frac{1}{12N}
\end{eqnarray}
lead to the following estimate
\begin{eqnarray*}
\big|\ln |\{\vw_\La: {\vw^{~\ell}}_\La=\vec\rho_\La \}| -
I(\vec\rho_\La)\big| \le d Q |\La| \frac{\log
\ell}{\ell^d}
\end{eqnarray*}
which implies
\begin{eqnarray*}
\big| \ln Z_{\ga,\beta,\La}(\{\vw_\La:
\vw^{\,\ell}_\La=\vec\rho_\La\}|\vw_{\La^c}) +\beta
F_{{\ga,\beta,\La}}(\vec\rho_\La|\vw^{\,\ell}_{\La^c})\big|
\le \beta c_d \ga \ell |\La| + Q d |\La| \frac{\log
\ell}{\ell^d}
\end{eqnarray*}
Now an easy upper bound for the cardinality of
${\cXx}_\La^{\ell}$ gives for all
$\cA\sqsubset{\cXx}_\La^{\ell}$
\begin{eqnarray*}
\ln|\cA|\le \ln |{\cXx}_\La^{\ell}| \le {Q
d\frac{|\La|}{\ell^d}}\log \ell
\end{eqnarray*}
where $|\cA|$ denotes the cardinality of the set $\cA$.
Combining these two last inequalities gives the first part
of Theorem  \ref{thm:app} with $c=\max(c_d,\frac{2Qd}{\beta})$. On the other side, for all
$\vec\rho\in L^\infty(\La,S_Q)$, one has
\begin{eqnarray*}
\log Z_{\ga,\beta,\La}
(\vw^{\,\ell}_\La=[\vec\rho_\La\,]^{\ell}|\vw_{\La^c}) \ge
-\beta
F_{{\ga,\beta,\La}}([\vec\rho]^{\ell}|\vw^{\,\ell}_{\La^c})
-c \eps(\ga,\ell) |\La|
\end{eqnarray*}
Now using
\begin{eqnarray*}
|V_{{\ga,\La}}(\vec\rho_\La|\vw_{\La^c}^{\,\ell})
-V_{{\ga,\La}}(\vec\rho_\La^{\,\ell}|
\vw_{\La^c}^{\,\ell})|\le c_d \ga \ell |\La|
\end{eqnarray*}
and the concavity of the entropy, one gets
\begin{eqnarray*}
F_{{\ga,\beta,\La}}(\vec\rho_\La|\vw_{\La^c}^{\,\ell})
\ge F_{{\ga,\beta,\La}}(\vec\rho_\La^{\,\ell}|
\vw_{\La^c}^{\,\ell}) - c_d \ga \ell |\La|
\end{eqnarray*}
Furthermore, approximating $\vec\rho_\La^{\,\ell}$ by
$[\vec\rho_\La\,]^{\ell}$ gives
\begin{eqnarray*}
F_{{\ga,\beta,\La}}(\vec\rho_\La^{\,\ell}|
\vw_{\La^c}^{\,\ell})\ge
F_{{\ga,\beta,\La}}([\vec\rho_\La\,]^{\ell}|
\vw_{\La^c}^{\,\ell}) - d Q \frac{|\La|}{\ell^d}
\end{eqnarray*}
we thus get:
\begin{eqnarray*}
\log Z_{\ga,\beta,\La}
(\vw^{\,\ell}_\La=[\vec\rho\,]^{\ell}|\vw_{\La^c}) \ge
-\beta
F_{{\ga,\beta,\La}}(\vec\rho_\La|\vw^{\,\ell}_{\La^c}) - \beta c
\eps(\ga,\ell) |\La|
\end{eqnarray*}
which gives the second part of the theorem with the same constant as before.
\end{proof}

We will use theorem \ref{thm:app} mostly in the following weaker form:
\begin{coro}
There exists a constant $c>0$ such that for all $\ga>0$,
$\ell$ an integer in $(1,\ga^{-1})$ and all bounded
$\cD^{\ell}$-measurable regions $\La$ in $\Rr^d$,
\begin{eqnarray*}
\big|\log Z_{\ga,\beta,\La} (\vw_{\La^c}) +\beta
\inf_{\vec\rho_\La}
 F_{{\ga,\beta,\La}}(\vec\rho_\La|\vw^{\,\ell}_{\La^c}) \big| \le \beta c \eps(\ga,\ell)
|\La|
\end{eqnarray*}
\end{coro}
Theorem \ref{thm:app} leads also to the Lebowitz-Penrose
limit for the Potts model
\begin{thm}{[Lebowitz-Penrose]}
\label{thm:L-P}
There exists the limit
\begin{eqnarray*}
\lim_{\ga\to 0} \lim_{\La\to\Rr^d} \frac{\log
Z_{\ga,\beta,\La}}{|\La|}={P^{\mf}_\beta}
\end{eqnarray*}
where $P^{\mf}_\beta$ is the mean field pressure.
\end{thm}
\vskip .5cm \noindent

\begin{proof}
The free energy functional on $L^\infty(\La,S_Q)$ with
boundary conditions $\vec\rho_{\La^c}$ can be rewritten as
\begin{eqnarray*}
F_{{\ga,\beta,\La}}(\vec\rho_\La|\vec\rho_{\La^c})=
\cF_{{\ga,\beta,\La}}(\vec\rho_\La|\vec\rho_{\La^c})
-\frac{1}{2}\int_\La\int_{\La^c} {J_\ga(x,y)}
{|\vec\rho_{\La^c}(y)|^2} dx dy
\end{eqnarray*}
where
\begin{eqnarray*}
\label{proof:lp:2}
\cF_{{\ga,\beta,\La}}(\vec\rho_\La|\vec\rho_{\La^c}) =&&
\int_\La \phi^{{\mf}}(\vec\rho_\La(x)) dx
+\frac{1}{4}\int_\La\int_\La {J_\ga(x,y)}
|\vec\rho_{\La}(x)-\vec\rho_{\La}(y)|^2 dx dy
\\
&&+\frac{1}{2}\int_\La\int_{{\La^c}} {J_\ga(x,y)}
|\vec\rho_{\La}(x)-\vec\rho_{\La^c}(y)|^2 dx dy
\end{eqnarray*}
with $\phi^{\mf}_\beta(\vec v)$ the mean field free
energy density on $S_Q:=\bigl\{\vec v\in\Rr^Q_+,\sum_q \rho_q =1\bigr\}$
\begin{eqnarray*}
\phi_\beta^{\mf}(\vec
v)&:=&\frac{1}{2}(\vec v\cdot \vec v)-\frac{1}{\beta}\vec v\cdot\ln(\vec v)
\end{eqnarray*}
(see \eqref{def:mean-field-free-energy}).

We have clearly
\begin{eqnarray*}
\inf_{\vec\rho_\La\in L^\infty(\La,S_Q)}
\cF_{{\ga,\beta,\La}}(\vec\rho_\La|\vec\rho_{\La^c}) \ge
\inf_{\vec\rho_\La\in L^\infty(\La,S_Q)}  \int_\La
\phi^{\mf}_\beta(\vec\rho_\La(x)) dx
\end{eqnarray*}
which gives a lower bound for the free energy as
\begin{eqnarray*}
\inf_{\vec\rho_\La\in L^\infty(\La,S_Q)}
F_{{\ga,\beta,\La}}(\vec\rho_\La|\vec\rho_{\La^c}) \ge
-{P^{\mf}_\beta} |\La| -c\ga^{-1} |\partial\La|
\end{eqnarray*}
with
\begin{eqnarray}
P^{\mf}_\beta=-\inf_{\vec v\in S_Q}\phi^{\mf}_\beta(\vec v)
\end{eqnarray}
{}From \ref{thm:app} with $\ell=\ga^{-\frac{1}{2}}$, one gets
\begin{eqnarray*}
\frac{\log Z_{\ga,\beta,\La}}{\beta |\La|}\le
{P^{\mf}_\beta} + c \eps(\ga,\ga^{-1/2}) + \ga^{-1}
\frac{|\partial\La|}{|\La|}
\end{eqnarray*}
and hence
\begin{eqnarray*}
\limsup_{\ga\to 0}\limsup_{\La\nearrow\Rr^d} \frac{\log
Z_{\ga,\beta,\La}}{\beta |\La|} \le {P^{\mf}_\beta}
\end{eqnarray*}
where the limit $\limsup_{\La\nearrow\Rr^d}$ is taken on a sequence
of Van Hove subsets of $\Rr^d$. On the other side, writing
\eqref{proof:lp:2} for $\vec\rho_\La =\vec\rho^{\p} \Ii_\La$, where
$\vec\rho^{\p}$ is an absolute minimizer of $\phi^{\mf}_\beta$, we
get
\begin{eqnarray*}
F_{{\ga,\beta,\La}}(\vec\rho^{\,*}
\Ii_\La|\vec\rho_{\La^c}) \le
\cF_{{\ga,\beta,\La}}(\vec\rho^{\,*}
\Ii_\La|\vec\rho_{\La^c}) \le -{P^{\mf}_\beta} |\La|
+c\ga^{-1} |\partial\La|
\end{eqnarray*}
From the second inequality in \ref{thm:app} one gets
\begin{eqnarray*}
\frac{\log Z_{\ga,\beta,\La}}{\beta |\La|}\ge
{P^{\mf}_\beta} - c \eps(\ga,\ga^{-1/2}) + \ga^{-1}
\frac{|\partial\La|}{|\La|}
\end{eqnarray*}
and finally
\begin{eqnarray*}
\liminf_{\ga\to 0}\liminf_{\La\nearrow \Rr^d} \frac{\log
Z_{\ga,\beta,\La}}{\beta |\La|} \ge {P^{\mf}_\beta}
\end{eqnarray*}
\end{proof}

\vskip 1.5cm \noindent
\setcounter{equation}{0}
\section{Analysis of the abstract contour models: equality of the pressures}
\label{sec:equalitypresure}
In this section and in the following, we will analyse the abstract contours models.  
A preliminary technical step is a partial cluster expansion of the contours contribution
to the partition function against fixed configurations.

\vskip 1.5cm \noindent
\centerline{\em A partial cluster expansion}
\vskip .5cm \noindent

From a technical point of view the we will take advantage of dealing with truncated
weights \eqref{cutoff-weight-b}
making a partial cluster expansion of the contours, against fixed configurations
getting  a partition function with
an extra interaction,
an ``effettive hamiltonian " $\cH^{\p}_{\ga,\La}$, with infinite range but exponentially decaying.

     \begin{equation}
       \label{8.1a.4.1}
\sum_{\und \Ga \in \mathcal B^{\p}_\La}\hat
W^{\p}_\ga(\und\Ga;\vw)= e^{-\beta
\cH^{\p}_{\ga,\La}(\vw_\La)}
    \end{equation}

A precise statement is given in the following theorem:
  \begin{thm}
  \label{thmz4.1}
If the weights $W^{\p}(\Ga,\vw)$  satisfy the Peierls
bounds with a constant $\fK_\ga$ large enough
then, for any
bounded, $\cD^{\ell_{+,\ga}}$-measurable region $\La$ and any
$\vw \in \mathcal X^{\p}$,
\begin{equation}
      \label{z4.7b}
Z_{\abs,\beta,\La}^{\p}(\und\vw_{\La^{c}})=  \sum_{\und \Ga
\in \mathcal B^{\p}_\La} \sum_{{\vw'\in \cX_\La^{\p}}}
\hat W^{\p}_\ga(\und\Ga,;\vw') e^{-\blue{\beta} H_{\ga,\La}(  \vw'_\La
| \vw_{\La^c})}= \sum_{{\vw'\in \cX_\La^{\p}}}
 e^{-\blue{\beta}\tilde H_{\ga,\La}^{\p}(\vw'_\La |
\vw_{\La^c})}
    \end{equation}
 \begin{eqnarray}
 \label{def:tildeH}
\tilde H_{\ga,\La}^{\p}(\vw'_\La |
\vw_{\La^c}):=H_{\ga,\La}(\vw'_\La | \vw_{\La^c}) +
\cH^{\p}_{\ga,\La}(\vw'_\La)
\end{eqnarray}

\begin{eqnarray}
\label{def:cH}
\cH^{\p}_{\ga,\La}(\vw)=\sum_{\Delta\sqsubseteq
\La}U^{\p}_\Delta(\vw)
\end{eqnarray}
where the potentials $U^{\p}_\Delta( \vw_{\Delta})$ are defined by \eqref{z4.15} 
and satisfy:
      \begin{eqnarray}
          \label{z4.11a}
& &U^{\p}_\Delta=0 ~~~~\mbox{if  $\Delta$ is not connected}
\\
& &\beta\sum_{\Delta\ni x}|U^{\p}_\Delta( \vw_{\Delta})|\le e^{-\frac{\fK_\ga}{2}}
\label{new-z4.11b}
\\
\label{new-z4.11c}
& &\beta\sum_{\Delta\ni x, \Delta\sqcap A\ne \emptyset}
\le 3^d e^{-\frac{\fK_\ga}{2}N_{x,A}} \hskip1cm \forall A\sqsubset \Rr^d
    \end{eqnarray}

where $N_{x,A}$ is the minimal number of $\mathcal D^{\ell_{+,\ga}}$-cubes needed
to cover the distance between $x$ and the set $A$:
let $\Delta$ a $\mathcal D^{\ell_{+,\ga}}$-measurable region and
$N_\Delta$ the   number of $\mathcal D^{\ell_{+,\ga}}$-cubes in
$\Delta$

\begin{eqnarray}
\label{def:NxA} N_{x,A}:=\min_{\Delta} \{N_\Delta:
\Delta\sqcap(C_x\sqcup\delta_{\out}^{\ell_{+,\ga}}[C_x])\ne
\emptyset, \Delta\sqcap(A\sqcup\delta_{\out}^{\ell_{+,\ga}}[A])\ne
\emptyset; \exists \Ga: \ssp_+(\Ga)=\Delta\}\hskip.7cm
\end{eqnarray}

$C_x$ a $\mathcal D^{\ell_{+,\ga}}$-cube containing the point $x$.
\end{thm}

\vskip .5cm \noindent

  The proof of the Theorem \ref{thmz4.1} is standard and  is given in appendix \ref{app:cluster expansion}.
  We notice
   that $N_{x,A}$ in \eqref{new-z4.11c}-\eqref{def:NxA} is s.t. there is a constant $k$, s.t.:
\begin{eqnarray}
\label{bound:NxA}
N_{x,A}\ge \max
\{ 3^d, k\frac{\dist(x,A)}{\ell_{_+,\ga}}\}
\end{eqnarray}

\vskip .5cm \noindent
The  Gibbs measures relative to $\tilde
H_{\ga,\La}^{\p}(\vw'_\La |\vw_{\La^c})$: 

\begin{eqnarray}
\label{absspec}
\tilde\mu_{\abs,\La}^{\p}(\vw_\La|\und\vw_{\La^{c}}):=
\frac{e^{-\beta\tilde H_{\ga,\La}^{\p}(\vw_\La
|\vw_{\La^c})}}{Z_{\abs,\beta,\La}^{\p}(\und\vw_{\La^{c}})}
\end{eqnarray}
 are  the marginals
on $\cX^{\p}$ of the measures
$\mu_{\abs,\La}^{\p}(\vw,\und \Ga|\und\vw_{\La^{c}})$.

In the remainig part of this section and in the next one, we will only consider the two
abstract models for $\p=\pm 1$ since all the others can be deduced
from $\p=1$ by symmetry, and write a superscript $\pm$ instead of
$\p$ to distinguish them. We denote by
\begin{equation*}
P^\pm_{\abs,\La,\ga,\beta}(\und\vw_{\La^c}):=\frac{1}{\beta
|\La|}\ln Z_{\abs,\beta,\La}^{\pm}(\und\vw_{\La^c})\end{equation*}
the ``finite volume pressures" of the two abstract models.
\vskip .5cm
The proof of Theorem \ref{thm:Peierls-2} requires the
proof of the
following theorem to control the bulk contribution to the ratio
in \eqref{eq:Peierls-2}:
\begin{thm}
  \label{thm:equalitypresure}
  {Let $\{\La_n\}$ a sequence of sets in $\Rr^d$ of side $2^n\ell_{+,\ga}$.
 There exist the two limits:
\begin{eqnarray}
\label{def:P-abs-b} P^\pm_{\abs,\ga,\beta} :=
\lim_{n\to\infty}\frac{1}{\beta |\La_n|}\ln
P^\pm_{\abs,\La,\ga,\beta}(\vec\rho^{\pm}_{\beta})
\end{eqnarray}
that are continuous in $\beta$, moreover there are
constants} $c_b$, $\bar \ga$ s.t. for any $\ga<\bar \ga$
there is a value of $\beta$, $\bcg$, s.t.
      \begin{equation}
      \label{eq:abs1-b}
P^+_{\abs,\ga,\bcg}=P^-_{\abs,\ga,\bcg}\hskip1cm
|\bcmf-\bcg|<c_b\ga^{1/2}
   \end{equation}

\end{thm}

\vskip .5cm \noindent

The proof is obtained by a continuity argument, and it is based on
the following mean field result: there exists an inverse temperature
$\bcmf$ such that the mean field free energy density satisfies:

\begin{eqnarray}\
\label{eq:equalitymeanfiel}
\phi_{\bcmf}(\vec\rho^{\, +}_{\bcmf})=\phi_{\bcmf}(\vec\rho^{\, -}_{\bcmf})=
\inf_{\vec\rho\in S_Q}\phi_{\bcmf}(\vec\rho)
\end{eqnarray}
and
 \begin{eqnarray}
 \label{eq:lemmaa.2}
\frac{d}{d\beta}\left[\phi_\beta^{\mf}(\vec\rho^{~+}_\beta)-
\phi_\beta^{\mf}(\vec\rho^{~-}_\beta)\right]
\bigg|_{\beta=\bcmf}\neq 0
 \end{eqnarray}
This result is well known  (see  \cite{Wu})  and  for completeness
it is also shown in appendix \ref{app:meanfield}.

\vskip 1.5cm \noindent

\begin{proof}[Proof of Theorem \ref{thm:equalitypresure-b}]

The proof of existence of the two pressures
$P^\pm_{\abs,\ga,\bcg}$ and their continuity in $\beta$ is
given in Appendix \ref{app:exist-pressure}, while the proof
of
\eqref{eq:abs1} is an immediate consequence the following
lemma :
\begin{lemma}
\label{lemma:a} There are constants $\kappa$ and  $\bar \ga$ such
that for any $\ga<\bar \ga$ and for any $\beta$ such that
$|\beta-\bcmf|\le \ga^{2a}$:

\vskip .5cm \noindent

\begin{eqnarray}
 \label{eq:lemmaa.1}
|P^{\pm}_{\abs,\ga,\beta}+\phi_\beta^{\mf}(\vec\rho^{~\pm}_\beta)|
\leq\frac{\kappa}{\beta}\ga^{1/2}
 \end{eqnarray}
\end{lemma}
that implies:
\begin{eqnarray}
\label{eq:starstar}
|P^{+}_{\abs,\ga,\beta}-P^{-}_{\abs,\ga,\beta}-(\phi_\beta^{\mf}(\vec\rho^{~-})-\phi_\beta^{\mf}(\vec\rho^{~+}))|<
2\frac{\kappa}{\beta}\ga^{1/2}
\end{eqnarray}

\eqref{eq:starstar}, \eqref{eq:lemmaa.2},
\eqref{eq:equalitymeanfiel}, and the continuity in $\beta$
of the pressures prove \eqref{eq:abs1} and complete the
proof of the Theorem  \ref{thm:equalitypresure-b}
\end{proof}
\vskip .5cm \noindent

\begin{proof}[proof of Lemma \ref{lemma:a}]

The proof of \eqref{eq:lemmaa.1}  could be  obtained as a
byproduct of a more detailed analysis contained in the next
section but since a direct proof  is quite shorter we sketch it here.
We first prove an upper bound for $P^{\pm}_{\abs,\ga,\beta}$.

\begin{eqnarray*}
P^{\pm}_{\abs,\ga,\beta}&=&\lim_{n\to\infty}\frac{1}{\beta |\La_n|}\ln
Z_{\abs,\beta,\La}^{\pm}(\vec\rho^{\pm}_{\beta})
\end{eqnarray*}
and denoting by $\hat Z_{\abs,\beta,\La}^{\pm}(\vec\rho^{\pm}_{\beta})$ the abstract partition function
with interactions $H_{\ga,\La}(\vw_\La|\vw_{\La^c})$, we get:
\begin{eqnarray*}
P^{\pm}_{\abs,\ga,\beta}& \le&\lim_{n\to\infty}\frac{1}{\beta |\La_n|}\ln
\hat Z_{\abs,\beta,\La}^{\pm}(\vec\rho^{\pm}_{\beta})+\sup_{\vw\in \cX^{\pm}}
\sum_{\Delta:\Delta\ni 0} |U^\pm_\Delta(\vw)|
\end{eqnarray*}
By \eqref{new-z4.11b} the last term is bounded as
$e^{-\frac{\fK_\ga}{2}}$   and by
 Theorem \ref{thm:app} we have:

\begin{eqnarray*}
P^{\pm}_{\abs,\ga,\beta}\le -\lim_{n\to \infty}\inf_{\vec\rho\in \cX_{\La_n}}
\frac{F_{\beta,\ga,\La_n}(\vec\rho|\vec\rho^{\pm}_\beta)}{\beta|\La_{n}|}+c_d \ga^{1/2}
\end{eqnarray*}
we postpone at the end of this section the proof of the
following bound that follows by the concavity of the
entropy:
\begin{eqnarray}
\label{eq:star1}
P^{\pm}_{\abs,\ga,\beta}\le c_d\ga^{1/2} -\lim_{n\to \infty}\inf_{\vec\rho\in \cX_{\La_n}}
\frac{1}{|\La_{n}|}\int_{\La_{n}}\phi_{\beta}
\bigg(J_{\ga}*(\vec\rho\Ii_{\La_n}+\vec\rho^{\pm}_\beta \Ii_{\La_n^c})\bigg)dr
\end{eqnarray}

\noindent where $\Ii_A:= \Ii_{\{x\in A\}}$.
Let
    \begin{eqnarray*}
J_\ga^{(\ell_{-,\ga})}(x,y) := \frac{1}{(\ell_{-,\ga})^d}\int_{z\in C^{\ell_{-,\ga}}_y} J_\ga(x,z)
     \end{eqnarray*}
     where $C^{\ell_{-,\ga}}_y$ is the cube of the partition  $\cD^{\ell_{-,\ga}}$ containing
     the point $y$. Then for any $\vec s\in \cX^{\pm}$:

    \begin{eqnarray*}
& &\|J_\ga * \vec s(r)-\vec\rho^{\pm}\|_\star\le \|\int
J_\ga^{(\ell_{-,\ga})}(r,r')(\vec s(r')-\vec \rho^{\pm})\|_\star
\\
& &\hskip5cm+\|\int
[J_\ga^{(\ell_{-,\ga})}(r,r')-J_\ga(r,r)](\vec s(r')-\vec \rho^{\pm})\|_\star
     \end{eqnarray*}

By the assumptions on $J_\ga$  (see \eqref{def:J}) 
 the second term is bounded as :
\begin{eqnarray*}
\|\int
[J_\ga^{(\ell_{-,\ga})}(r,r')-J_\ga(r,r')]
(\vec s(r')-\vec \rho^{\pm})\|_\star\le \kappa_3 \ga\ell_{-,\ga}=\kappa_3\ga^{\al}
\end{eqnarray*}

while the first term, since $\vec s\in \cX\pm$ and $J^{(\ell_{-,\ga})}(r, r')$ is
constant w.r.t. the second variable in  each cube of $ \cD^{\ell_{-,\ga}}$:
\begin{eqnarray*}
 \|\int
J_\ga^{(\ell_{-,\ga})}(r,r')(\vec s(r')-\vec \rho^{\pm})\| _{\star}\le
\ga^a
\end{eqnarray*}
going back to \eqref{eq:star1} we have:

\begin{eqnarray}
\label{eq:star2}
P^{\pm}_{\abs,\ga,\beta}\le c\ga^{1/2} -
\inf_{\|\vec s-\vec\rho^\pm\|_\star <\ga^{a}+\kappa_3\ga^{\al} }
\phi_{\beta}(\vec s)
\end{eqnarray}

 In appendix \ref{app:meanfield}, is shown that
$\inf_{\vec s}\phi_{\bcmf}(\vec s)= \phi_{\bcmf}(\vec
\rho^{\q}_{\bcmf})$, for any $\q\in\{-1,1,\dots Q\}$. By continuity
for $|\beta-\bcmf|<\ga^{2a}$, $\ga$ small enough $\phi_{\beta}(\vec
s)$
has   $Q+1$ local minima $\vec \rho^{\q}_{\beta}$ s.t. $|\vec
\rho^{\q}_{\beta}-\vec \rho^{\q}_{\bcmf}|<c|\beta-\bcmf|<c
\ga^{2a}$. The are respectively then absolute minimizers  in the
sets $\|\vec s-\vec\rho^\pm\|_\star <\ga^{a}+\kappa_3\ga^{\al}$. We
then get:

\begin{eqnarray}
\label{eq:star3}
P^{\pm}_{\abs,\ga,\beta}\le c\ga^{1/2} -
\phi_{\beta}(\vec \rho^{\pm}_{\beta})
\end{eqnarray}

We now prove a lower bound for the pressures.
By Theorem \ref{thm:app} with $\ell=\ell_0$, we have for any $\vec\rho^{(n)}\in \cX^{\pm}_{\La_n}$:

\begin{eqnarray*}
P^{\pm}_{\abs,\ga,\beta}\ge -\lim_{n\to \infty}
\frac{F_{\beta,\ga,\La_n}(\vec\rho^{(n)}|\vec\rho^{\pm}_\beta)}{\beta|\La_{n}|}-c\ga^{1/2}
\end{eqnarray*}

and for $\ga$ small enough the
same argument as before shows that $\rho^{\pm}_{\beta}\Ii_{\La_n}\in\cX^{\pm}_{\La_n} $. We then get:
\begin{eqnarray}
\label{eq:star4}
P^{\pm}_{\abs,\ga,\beta}\ge -c\ga^{1/2} -
\phi_{\beta}(\vec \rho^{\pm}_{\beta})
\end{eqnarray}

\end{proof}

\vskip .5cm \noindent
Proof of \eqref{eq:star1}:

\begin{eqnarray*}
F_{\beta,\ga,\La_n}(\vec\rho|\vec\rho^{\pm}_\beta)&=&
F_{\beta,\ga}(\vec\rho\Ii_{\La_n}+\vec\rho^{\pm}_\beta \Ii_{\La_n^c})-F_{\beta,\ga}(\vec\rho^{\pm}_\beta \Ii_{\La_n^c})
\\& =&\int_{\La_{n}\sqcup \delta_{\out}^{\ga^{-1}}[\La_{n}]}\phi_{\beta}
\big({\st{ J_\ga*(\vec\rho\Ii_{\La_n}+\vec\rho^{\pm}_\beta \Ii_{\La_n^c})}}\big)dr\\
&& +\frac{1}{\beta}\int_{\La_{n}\sqcup \delta_{\out}^{\ga^{-1}}[\La_{n}]}
\bigg[I\big({\st{J_\ga*(\vec\rho\Ii_{\La_n}+\vec\rho^{\pm}_\beta \Ii_{\La_n^c})}}\big)-
J_\ga*I({\st{\vec\rho\Ii_{\La_n}+\vec\rho^{\pm}_\beta \Ii_{\La_n^c}}})\bigg]dr
\\&&
-\int_{\delta_{\out}^{\ga^{-1}}[\La_{n}]}\phi_{\beta}
\big({\st J_\ga*(\vec\rho^{\pm}_\beta \Ii_{\La_n^c})}\big)dr-\frac{1}{\beta}
\int_{\delta_{\out}^{\ga^{-1}}[\La_{n}]}
\bigg[I({\st J_\ga*\vec\rho^{\pm}_\beta \Ii_{\La_n^c}})-
J_\ga*I({\st\vec\rho^{\pm}_\beta \Ii_{\La_n^c}}) \bigg]dr
\end{eqnarray*}

where we have used the fact that $\phi_{\beta}(0)=I(0)=0$. By concavity of
 $I(\cdot)$ the second term is non negative and since $|\phi_{\beta}(\vec s)|$ is bounded in
 $\vec s\in S_Q$ we have:

 \begin{eqnarray*}
F_{\beta,\ga,\La_n}(\vec\rho|\vec\rho^{\pm}_\beta)\ge
\int_{\La_{n}}\phi_{\beta}
\big({\st{ J_\ga*(\vec\rho\Ii_{\La_n}+\vec\rho^{\pm}_\beta \Ii_{\La_n^c})}}\big)dr-
c|\delta_{\out}^{\ga^{-1}}[\La_n]|
 \end{eqnarray*}

\vskip 1.5cm \noindent
\setcounter{equation}{0}
\section{Analysis of the abstract contour models:  finite volume \\ corrections to the pressures}
\label{sect:finitevolumecorrections}

At the critical value of inverse temperature $\bcg$, the theorem
\ref{thm:equalitypresure-b} holds, and the bulk term of the ratios
\eqref{ratios} is null. In this case, in order to get  estimates on
Peierls weights 
it is needed refined a control of the finite volume corrections to
the thermodynamical pressure $P^{\pm}_{\abs,\ga}$  here  denoted by
$R^{\pm}_{\abs,\La}$:
     \begin{equation*}
R^{\pm}_{\abs,\La} :=\log Z^{\pm}_{\abs,\beta,\La}(\vec\rho^{(\pm
)}) -\beta |\La|P^{\pm}_{\abs,\ga} \quad \quad \quad  \lim_{\La
\nearrow \mathbb Z^d}
  \frac {R^{\pm}_{\abs,\La}}{|\La|}=0
    \end{equation*}

    Let $\La$ a $\mathcal
D^{(\ell_{+,\ga})}$ bounded region  and denoted by
$P^{\pm}_{\abs,\ga,\La}$ the ``finite volume pressure":

\begin{eqnarray*}
P^{\pm}_{\abs,\ga,\La}:= \ln  Z^{\pm}_{\abs,\beta,\La}(\vec\rho^{\pm})
\end{eqnarray*}

we  prove the following  theorem: \vskip.5cm

\begin{thm}
   \label{thm:surfacecorrections}


There is a constant $c>0$ so that
      \begin{eqnarray}
 &&
\Big|P^{\pm}_{\abs,\ga,\La} - \{\beta
|\La|P^{\pm}_{\abs,\ga} + \frac \beta 2
\sumtwo{i\in\La}{j\in \La^c} J_\ga(i,j)\, (\vec\rho^{\pm}\cdot
\vec\rho^{\pm})\} \Big| 
\le c \ga^{1/8}|\delta_{\rm
in}^{\ell_{+,\ga}}[\La]|\hskip1cm
        \label{8.0.0.1}
    \end{eqnarray}
\end{thm}

Notice that the leading contribution to $R^{\pm}_{\abs,\La}$ is the finite volume correction
to the mean field pressure (with $\rho^{\pm}$ b.c.) 
\begin{eqnarray}
\label{def:Rmf}
R_{\ga,\La}^{\mf,{\pm}}:=\frac{\beta}{2}\sumtwo{i\in\La}{j\in \La^c} J_\ga(i,j)\,
(\vec\rho^{\pm}\cdot
\vec\rho^{\pm})
\end{eqnarray}

\vskip .5cm \noindent

The proof of Theorem \ref{thm:surfacecorrections}
is the outcome of two main estimates:
the first one is a bound on the decay of correlations
and the second step is a small deviation estimate inside ``correct
regions" to control the contribution coming from regions near the
boundary.

\vskip 1.5cm \noindent\vskip 1.5cm \noindent
\centerline{\em Dobrushin interpolations}
\vskip .5cm \noindent


In this section we refer to the models with interpolating hamiltonians \eqref{def:h-u}
\begin{eqnarray}
\label{def:hu}
\hat h^{\pm}_u(\vw):=u \tilde H^{\pm}(\vw|\vec\rho^{\pm})+(1-u)\fH_{\ga}^{\pm}(\vw) \hskip1.6cm
\end{eqnarray}
where $\fH^{\pm}$ are the one body ``mean field"
Hamiltonians, defined in \eqref{def:H0-b}.

\vskip .5cm \noindent
For any $u\in [0,1]$ we denote by
$Z^{\pm}_{\abs,\beta,\La;u}$  the partition function
relative to the hamiltonian $\hat h^{\pm}_u(\vw)$.
In particular,
$Z^{\pm}_{\abs,\beta,\La;0}$ corresponds to the interpolating one-body
hamiltonian $\fH_{\ga,\La}^{\pm}(\vw)$ in
\eqref{def:H0-b}.

We recall also the expression of the [finite volume] pressure $P^{\pm}_{\abs,\ga,\La }$ \eqref{interpol}) in terms of correlation functions w.r.t.
the measures $\hat\mu^{\pm}_{\abs,\La;u}$:
     \begin{eqnarray}
P^{\pm}_{\abs,\ga,\La }:=&& \frac{1}{\beta|\La|}\ln Z_{\abs,\beta,\La;1}^{\pm}( \vec\rho^{\pm}) \nn\\
=&&
\frac{1}{\beta|\La|}\ln  Z_{\abs,\beta,\La;0}^{\pm}
- \frac{1}{|\La|}\int_0^1 du\; \left\langle \tilde H^{\pm}_\La( \vw'_\La|\vec\rho^{\pm})-
 \fH^{\pm}_{\La}( \vw'_\La) \right\rangle_{\hat\mu^{\pm}_{\abs,\La;u}}
          \label{interpol-b}
    \end{eqnarray}
\noindent
and prove the following proposition:
\vskip .5cm \noindent
\begin{prop}
\label{prop:1} The exist $g_{\ga,i,\La}, ~g_{\ga,i}: \Zz^{d}\to\Rr$:
\begin{eqnarray}
\label{eq:prop1-a}
& &\left[\tilde H^{\pm}_\La( \vw'_\La|\vec\rho^{\pm})-
\fH^{\pm}_{\La}( \vw'_\La)\right]= \sum_{i\in\La}g_{\ga,i,\La}
\\
\label{eq:prop1-b}
& &\lim_{\La\nearrow \Rr^{d}}g_{\ga,i,\La}=g_{\ga,i}
\end{eqnarray}
\end{prop}

\vskip .5cm \noindent
\begin{proof}[Proof] 

\begin{eqnarray*}
\tilde H^{\pm}_\La( \vw_\La|\vec\rho^{\pm})-
 \fH^{\pm}_{\La}( \vw_\La) = \frac{1}{2}
 \sum_{i\in\La}\vvw_i\sumtwo{j\in\La}{j\ne i}J_\ga(i,j)\vvw_j+ \cH^{\pm}_{\ga,\La}(\vw_\La)
\end{eqnarray*}
Recalling \eqref{def:cH}:
\begin{eqnarray*}
\cH^{\pm}_{\La}(\vw)=\sum_{\Delta\sqsubset \La}U^{\pm}_\Delta(\vw)=\sum_{i\in \La}
\sumtwo{\Delta\sqsubset\La}{\Delta\ni i} \frac{1}{|\Delta|}U^{\pm}_\Delta(\vw)
\end{eqnarray*}
Defining:
\begin{eqnarray}
\label{def:ggaLa}
g_{\ga,i,\La}(\vw):= \frac{1}{2}\vvw_i\sumtwo{j\in\La:}{j\ne i}J_\ga(i,j)\vvw_j
    +\sumtwo{\Delta\ni i}{\Delta\sqsubset \La} \frac{1}{|\Delta|}U^{\pm}_\Delta(\vw)
\end{eqnarray}
we can write:
\begin{eqnarray*}
\tilde H^{\pm}_\La( \vw_\La|\vec\rho^{\pm})-
 \fH^{\pm}_{\La}( \vw_\La) =\sum_{i\in\La}
g_{\ga,i,\La}(\vw)
\end{eqnarray*}

\vskip .5cm \noindent
Recalling \eqref{new-z4.11b}, the  limit \eqref{eq:prop1-b} exists:
\begin{eqnarray}
\label{def: gx}
\lim_{\La\nearrow \Rr^{d}}g_{\ga,i,\La}(\vw)=
-\frac{1}{2}\vvw_i\sum_{j:j\ne i}J_\ga(i,j)\vvw_j
+\sum_{\Delta\ni i} \frac{1}{|\Delta|}U^{\pm}_\Delta(\vw)=:g_{\ga,i}(\vw)
\end{eqnarray}
\end{proof}
Postponing the proof of the existence  of the limit
of the Gibbs measure $\tilde \mu^{\pm}_{\abs,\La;u}$
when $\La\nearrow \Rr^d$
and denoting it by $\tilde \mu^{\pm}_{\abs;u}$,
the limit of \eqref{interpol} gives:
\begin{eqnarray}
\label{eq:press-abst}
P^{\pm}_{\abs,\ga} =
P^{\pm}_{\abs,\ga,0}
- \int_0^1 du\;
 \left\langle g_{\ga,0} \right\rangle_{\tilde \mu^{\pm}_{\abs;u}}
    \end{eqnarray}

and  \eqref{interpol} can be rewritten as:
\begin{eqnarray*}
&&P^{\pm}_{\abs,\ga,\La }=P^{\pm}_{\abs,\ga}+
\left[\frac{1}{\beta |\La|}\ln  Z_{\abs,\beta,\La;0}^{\pm}-P^{\pm}_{\abs,\ga,0} \right]
\\
&&\hskip3cm
- \int_0^1 du\;
 \bigg[\frac{1}{|\La|}\sum_{i\in\La}\left\langle g_{\ga, i,\La}
 \right\rangle_{\tilde \mu^{\pm}_{\abs,\La;u}}
-
 \left\langle g_{\ga,0} \right\rangle_{\tilde \mu^{\pm}_{\abs;u}}\bigg]
 \end{eqnarray*}
\vskip .5cm \noindent

The finite volume corrections are then given by:

\begin{eqnarray}
\label{eq:surf-corr}
R_{\abs,\La}^{\pm}=R_{\abs,\La,0}^{\pm}-\beta \int_0^1 du\;\sum_{i\in\La}
\left[\left\langle g_{\ga, i,\La}
\right\rangle_{\tilde \mu^{\pm}_{\abs,\La;u}}-
 \left\langle g_{\ga,i} \right\rangle_{\tilde \mu^{\pm}_{\abs;u}}\right]
\end{eqnarray}
where
\begin{equation*}
R_{\abs,\La,0}^{\pm}:=\ln Z_{\abs,\beta,\La;0}^{\pm}- \beta |\La|P^{\pm}_{\abs,\ga,0}
\end{equation*}

In appendix \ref{app:meanfieldpressurelimit} it is proven that:

\begin{eqnarray}
\label{eq:Rmf}
R_{\abs,\La,0}^{\pm}=R_{\ga,\La}^{\mf,{\pm}}
\end{eqnarray}
and then the proof of the Theorem \ref{thm:surfacecorrections} follows by estimating the
remaining terms in \eqref{eq:surf-corr}. In the next subsections we will prove that
there are positive constant $c, \om$ such that:
\begin{itemize}
  \item[1)]
  \begin{eqnarray}
\label{eq:decay}
&&\left|
\left\langle g_{\ga, i,\La} \right\rangle_{\tilde \mu^{\pm}_{\abs,\La;u}}- \left\langle g_{\ga,i} \right\rangle_{\tilde \mu^{\pm}_{\abs;u}}
 \right|\le c~e^{-\om\ga\dist(i,\La)}
 \end{eqnarray}
\vskip .5cm \noindent

  \item[2)]
\begin{eqnarray}
 \label{eq:smalldev}
&&\hskip-.3cm\sum_{i\in \delta_{\ins}^{\ell_{+,\ga}}[\La]}
\{|\langle g_{\ga,i}\rangle_{\tilde \mu^{\pm}_{\abs;u}}|
+|\langle g_{\ga,i,\La}\rangle_{\tilde \mu^{\pm}_{\abs,\La;u}}|\}
\le c\ga^{1/8}|\delta_{\ins}^{\ell_{+,\ga}}[\La]|
\end{eqnarray}

\end{itemize}

\vskip .5cm \noindent

{The proof of \eqref{eq:decay} and the existence of the limit of
$\hat \mu^{\pm}_{\abs,\La;u}$
when $\La\nearrow \Rr^d$ 
follow by the proof
of the exponentially decay of the correlations given in next subsection, while the estimate
\eqref{eq:smalldev} follows from  the small deviations estimates
proved in the last subsection.}

\vskip .5cm \noindent
\begin{proof}[Proof of Theorem \ref{thm:surfacecorrections}]

Collecting \eqref{eq:surf-corr}, \eqref{eq:Rmf}, \eqref{eq:decay},
\eqref{eq:smalldev} and  using the last one for estimating the contribution to the correction
coming from the boundary
and \eqref{eq:decay} for estimating the contribution
to the corrections coming from the volume inside, we get \eqref{8.0.0.1}.
Details are omitted.

\end{proof}

\vskip 1.5cm \noindent

\subsection{Decay of the correlations}
\label{subsect:decayofthecorrelations}

In order to prove \eqref{eq:decay}  we state the following Theorem:

\begin{thm}

 \label{thmz5.1}

There are $c$ and $\om$ positive so that for $u \in (0,1)$ and for
any bounded sets $\La$ and $\Delta$, $\La$
$\cD^{\ell_{+,\ga}}$-measurable, and any $\und \vw \in
\cX^{({\pm})}$, there is a coupling $Q_u$ of
$\tilde\mu^{({\pm})}_{\abs,\La;u}(\cdot|\und \vw)$ and $\tilde
\mu^{({\pm})}_{\abs;u}(\cdot)$ such that
     \begin{equation}
          \label{3.K19.5}
Q_u(\vw_\Delta \ne \vw'_\Delta) \le c |\Delta| e^{-  \om\ga
\text{dist}(\Delta,\La^c )}
    \end{equation}

    \end{thm}

It follows that:

\begin{corol}
\label{cor:mu-limit} For any $u\in[0,1]$ there is a unique
DLR measure $\tilde \mu^{\pm}_{\abs;u}$ with
hamiltonian $h^{\pm}_u$ and for any local function $f$ with
support in $\Delta$ :
\begin{eqnarray*}
\lim_{\La\nearrow \Zz^{d}} \langle f \rangle_{\tilde
\mu^{\pm}_{\abs,\La;u}}= \langle f \rangle_{\tilde
\mu^{\pm}_{\abs;u}}
\end{eqnarray*}
and, for any  $\Delta\sqsubset \La$,  there are positive
constants $c, \om$:
\begin{eqnarray*}
\left| \left\langle f \right\rangle_{\tilde
\mu^{\pm}_{\abs,\La;u}}-
 \left\langle f \right\rangle_{\tilde \mu^{\pm}_{\abs;u}}
 \right|\le c~|\Delta|\sup_{x\in\Delta}\{f(x)\}~~e^{-\om\ga\dist(\Delta,\La^c)}
\end{eqnarray*}
\end{corol}

\vskip .5cm \noindent

Corollary \ref{cor:mu-limit} applied
to our case, proves inequality \eqref{eq:decay}.
\vskip .5cm \noindent

\begin{proof}
The proof of Theorem
\ref{thmz5.1} requires an extension of the Dobrushin high
temperature uniqueness Theorem.
In reference \cite{bkmp2}, the Dobrushin
uniqueness  criterium is extended to the case when the
``classical Dobrushin condition"
 is not satisfied uniformly  in the boundary conditions,
but only for {``most of the configurations".} This is  the case for
the abstract contour models, where, due to the constrain on the
space of the configurations $\cX^{\pm}$, the Vasenstein distance
between two Gibbs measures on a single spin, with different boundary
conditions is not small uniformly in all the boundary conditions.

In reference \cite{bkmp2} it is shown that, provided two main
assumptions are verified, the Dobrushin criterium can be extended to
cover such a case. A further assumption provides an exponential
decay for the correlations. Two other assumptions trivially hold in
our case and are not reported here. We refer to \cite{bkmp2} and the
Theorem \ref{thmz5.1} will be proved through the demonstration that
the two abstract models $\tilde \mu^{({\pm})}_{\abs;u}(d\vw')$
fulfill the requirements of the extended Dobrushin criteron.

\vskip .5cm \noindent
\subsubsection{First requirement:}
\label{subsec:first-assumption} \vskip .5cm \noindent

First, e need to prove that for any $i\in \Zz^d$ there is a
measurable set
$G^{\pm}_i\sqsubset \cX^{\pm}$ 
depending only on $\{\und\vw_j,\; j\in C^{\ell_{-,\ga}}_i\setminus i\}$, such that there exists
$b(i,j)$ with the following properties:
\begin{eqnarray}
\label{firstassumption-b}
&&b(i,j)\ge 0 \;\; ;\;\; b(i,i)=0 \nonumber\\
&&\sup_{i\in\Zz^d}\sum_{j\in C^{\ell_{-,\ga}}_i} b(i,j)<\delta< 1 \\
&&
\cR(\mu^{\pm}_{\abs,i;u}(\vw_i|\und\vw),\mu^{\pm}_{\abs,i;u}(\vw_i|\und\vw'))\le
\sum_j b(i,j)\; \dist(\und\vw_j,\und\vw_j') \hskip1.6cm \text{for
any $\und\vw,\und\vw'\in G_i$} \nonumber
\end{eqnarray}
where $\dist(\vw_j,\vw_j')$ is a distance defined on the
configuration space and\linebreak
$\cR(\mu^{\pm}_{\abs,i;u}(\vw_i|\und\vw),\mu^{\pm}_{\abs,i;u}(\vw_i|\und\vw'))$
is the associated Vaserstein distance. Here, we consider the
following distance between configurations:

\begin{eqnarray}
\label{def:dist2-b}
\dist(\vw_i,\vw'_i):=
\frac{1}{2}\sum_i|\w_i(i)-\w'_i(i)|
\end{eqnarray}
 and define  $G_i$ as :
\begin{eqnarray}
\label{Gx-b}
G^{\pm}_i:=\{\vw\in \cX^{({\pm})}:\vw^{(i,q)}\in \cX^{({\pm})} \forall q\}
\end{eqnarray}
 where we have denoted
\begin{eqnarray}
\hskip1cm \vw^{(i,q)}_j=
  \begin{cases}
    \vw_j & \text{$j\ne i$}, \\
    \vec u_q & \text{$j=i$}.
  \end{cases}
\end{eqnarray}

\vskip .5cm \noindent
{\bf Remark:}
$G^{\pm}_i$ is the set of configurations $\vw$ which belong to
$\cX^{({\pm})}$ independently of the value of
$\vw_i$ and is measurable on $\vw_{i^c}$.
When $\und\vw,\und\vw'$ are not in $G^{\pm}_i$, the probability
measures for  $\vw_i$,
have support on a strict subset of $\Om$
and the Vaserstein distance can be larger than the bound
in \eqref{firstassumption}.
\vskip .5cm \noindent

\begin{thm}

There are $\bar \ga,  \varsigma$, so that for
$\ga<\bar\ga$, and for any $\beta:~|\beta-\bcmf|<c_b \ga^{\frac{1}{2}}$, there exists $b(i,j)$ satisfying the relations
\eqref{firstassumption} with  $G_i$ as in \eqref{Gx}.
$b(i,j)$ has the expression:
\begin{eqnarray*}
b(i,j)=r\left[J_{\ga}({i,j})+ 3^d e^{-\frac{\fK_\ga}{2} N_{i,j}}\right]
\end{eqnarray*}
where $r<1$, 
and  $N_{i,j}$ as in \eqref{def:NxA} and satisfying the bound \eqref{bound:NxA} 
\end{thm}

\begin{proof}
The Vaserstein distance between the two measures $\mu_{i}(\vw_i|\vw_1),\mu_i(\vw_i|\vw_2)$
is defined as

\begin{eqnarray*}
\cR(\mu_{i}(\vw_i|\vw_1),\mu_i(\vw_i|\vw_2)):= \inf_{\mathcal q}E_{\mathcal q}(\dist(\vw_i,\vw'_i))
\end{eqnarray*}
where the infimum is taken over all couplings between
$\mu_{i}(\vw_i|\vw_1),\mu_i(\vw_i|\vw_2)$.
Recalling \eqref{def:dist2},
the infimum is realized on the couplings which have the maximal mass on the diagonal
\begin{eqnarray*}
\mathcal q(\vw_i=\vw'_i=\vec u_q)=\min\{\mu_i(\vec u_q|\vw_1),\mu_i(\vec u_q|\vw_2)\}
\end{eqnarray*}
Let { $\mathcal q$} be such a coupling for
$\mu_{\abs,i;u}(\vw_i|\vw_1),\mu_{\abs,i;u}(\vw_i|\vw_2)$. We have:
\begin{eqnarray*}
\cR(\mu_{\abs,i;u}(\vw_i|\vw_1),\mu_{\abs,i;u}(\vw_i|\vw_2))&=&
E_{\mathcal q}\bigg(\dist(\vw_i,\vw'_i)\bigg)
=\mathcal q(\vw_i\ne\vw'_i)\\
&=&1-\sum_{q}\min\{\mu_{\abs,i;u}(\vec u_q|\vw_1),\mu_{\abs,i;u}(\vec u_q|\vw_2)\}
\\
&=&\frac{1}{2}\sum_{q}\bigl(\mu_{\abs,i;u}(\vec u_q|\vw_1)+\mu_{\abs,i;u}(\vec u_q|\vw_2)\\
& &\hskip3cm-2\min\{\mu_{\abs,i;u}(\vec u_q|\vw_1),\mu_{\abs,i;u}(\vec u_q|\vw_2)\}\bigr)\\
&=& \frac{1}{2}\sum_{q}|\mu_{\abs,i;u}(\vec u_q|\vw_1)-\mu_{\abs,i;u}(\vec u_q|\vw_2)|
\end{eqnarray*}

Since $\und\vw$ and $\und\vw'$ are in $G_i$, the constraint of being
in $\cX^{\pm}$ does not affect the values of $\vw_i$, and the
conditional measures $\mu^{\pm}_{\abs,i;u}(\cdot|\vw)$ are given by
the Gibbs conditional measures. Using \eqref{absspec}, we get,

\begin{eqnarray*}
\cR(\mu^{\pm}_{\abs,i;u}(\vw_i|\und\vw),\mu^{\pm}_{\abs,i;u}(\vw_i|\und\vw'))&=&
\frac{1}{2}\sum_{q}\left|\frac{e^{\beta
k^{\pm;u}_q(\und\vw)}}{\sum_{p=1}^{Q} e^{\beta
k^{\pm;u}_{p}(\und\vw)}}- \frac{e^{\beta
k^{\pm;u}_{q}(\und\vw')}}{\sum_{p=1}^{Q}e^{\beta
k^{\pm;u}_{p}(\und\vw')}}\right|
\\
&=&\frac{1}{2}\sum_{q}|g_q(\vec k^{\pm;u}(\und\vw))-g_q(\vec
k^{\pm;u}(\und\vw'))|
\end{eqnarray*}
Where $\vec g(\cdot)$ is defined in \eqref{mf:g} and $\vec
k^{\pm;u}_i(\und\vw)\equiv \vec k^{\pm;u}_i(\und\vw_{i^c})$ is the
vector
\begin{eqnarray*}
\vec k^{\pm}_i(\und\vw)&:=&
u \vL_{\ga}(\vw;i) + (1-u) \vL_{\ga}(\vec \rho^\pm;i) - u
\sum_q\bigl(\cH^{\pm}(\und\vw^{(i,q)}) -\cH^{\pm}(\und\vw_{i^c})\bigr) \vec u_q\nn\\
&=& u \sum_{j\ne i}J_{\ga}({i,j})\vw_j +(1-u) \sum_{j\ne i}J_{\ga}({i,j})\vec\rho^\pm_j
- u \sum_q \sum_{\Delta\ni i}U^{\pm}_\Delta(\vw^{(i,q)}) \vec u_q
\end{eqnarray*}

Using \eqref{new-z4.11b}, we have:
\begin{eqnarray}
\label{eq:vec-k}
\lim_{\ga\to 0}\sup_{\und\vw\in G_i}
\|\vec k^{\pm}(\und\vw)-\vec \rho^{\pm}\|\le {\lim_{\ga\to 0}}\left(c\ga^a+\sum_{\Delta\ni
i}Q\|U^{\pm}_\Delta(\cdot)\|_{\infty}\right)=0
\end{eqnarray}
Hence , for $\ga$ small enough, we can use theorem
\ref{thm:mflocalstability} and get

\begin{eqnarray*}
\cR\bigg(\mu^{\pm}_{\abs,i;u}(\vw_i|\und\vw),\mu^{\pm}_{\abs,i;u}(\vw_i|\und\vw')\bigg)
&\le&\frac{1}{2} (1-\frac{1}{2Q})
\sum_q|k^{\pm;u}_q(\und\vw)-k^{\pm;u}_q(\und\vw')|
\end{eqnarray*}
On the other hand, we have the following bound for all $\Delta\ni x$
    \begin{equation*}
 |U^{\pm}_\Delta(\vw^{(i,q)})-
 U^{\pm}_\Delta(\vw'^{(i,q)})| \le
2\|U^{\pm}_\Delta(\cdot)\|_{\infty} \sum_{j\in \Delta\setminus
i}\dist(\und\vw_j,\und\vw'_j)
      \end{equation*}
and thus

\begin{eqnarray}
\label{k-dist} \sum_q|k^{\pm;u}_q(\und\vw)-k^{\pm;u}_q(\und\vw')|\le
2 \sum_{j\ne i} \left(u J_{\ga}(i,j)+\sum_{\Delta \ni i,j}
Q\|U^{\pm}_\Delta(\cdot)\|_{\infty}\right)\;\;\dist(\und\vw_j,\und\vw'_j)
\end{eqnarray}

Recalling \eqref{new-z4.11c} we
bound the last term in \eqref{k-dist}
as:
    \begin{equation}
       \label{6.8b}
  \sum_{\Delta\ni i,j}\|U^{\pm}_\Delta(\cdot)\|_{\infty} \le   \sum_{\Delta\ni
i,j} e^{-(\frac{\fK_\ga}{2} -b)N_{\Delta}} \le 3^d
e^{-(\frac{\fK_\ga}{2} ) N_{i,j}}
\le 3^d e^{-\frac{\fK_\ga}{2}  N_{i,j}}
     \end{equation}
Since
$\fK_\ga = c \ga^{-(1-\alpha)d + 2a}$,
\begin{eqnarray*}
\lim_{\ga\to 0}  \sum_{\Delta\ni
i,j}Q\|U^{\pm}_\Delta(\cdot)\|_{\infty}=0
\end{eqnarray*}

hence, we take for all $u$,

\begin{eqnarray}
\label{eq:b-xy} b(i,j):=(1-\frac{1}{2Q}) \left(u J_{\ga}(i,j)+3^d
e^{-\frac{\fK_\ga}{2}  N_{i,j}}\right)
\end{eqnarray}
We have
\begin{eqnarray}
\label{eq:blim}
\lim_{\ga\to 0}\sup_{i\in \mathbb Z^d} \sum_{j\in
C_i^{\ell_{-,\ga}}\setminus i} b(i,j)=0
\end{eqnarray}
so that \eqref{firstassumption} holds for all $\ga$ small enough.
\end{proof}

\vskip .5cm \noindent
\subsubsection{Second Requirement:}
\vskip .5cm \noindent

We need now a bound on the probability of ${G_i^{\pm}}^c$, where
$G^{\pm}_i$ is defined in \eqref{Gx}. Here we deal with a bounded
state space and the required bound may be written as:
\begin{eqnarray}
\label{assumpt2} \sup_{ \vw^*\in \cX^{\pm}}\mu^{({\pm})}_{\abs,D,u}
({G_i^{\pm}}^{c}|\vw^{~*})\le
 e^{-c\ga^{2a}\ell^{d}_{-,\ga}}
\end{eqnarray}
$c$ a positive constant and $D=C^{\ell_{-,\ga}}_i$.
The proof of \eqref{assumpt2} requires a Peierls estimate inside the
restricted set $\cX^{\pm}$.
Let $\zeta'>0$, we define the  ``bad" set
\begin{eqnarray*}
\cAA_i :=\{\vw\in \cX^{({\pm})}: \|\vw^{(\ell_{-,\ga})}(i)- \vec\rho^{\pm}\|_{\star}>(1-\zeta')\ga^a\}
\end{eqnarray*}
Notice that for any $\zeta'>0$ and  $\ga$ small enough, we have for all $i$
\begin{eqnarray*}
{G^\pm}^c_i\sqsubset \cAA_i
\end{eqnarray*}
so that $\cAA_i$ does not depend on $\vw_i$, and we have
\begin{eqnarray*}
\sup_{\vw^{~*}\in \cX^{\pm}}\mu^{({\pm})}_{\abs,D,u}
({G_i^{\pm}}^{c}|\vw^{~*})\le \sup_{ \vw^{~*}\in
\cX^{\pm}}\mu^{({\pm})}_{\abs,D,u} (\cAA_i|\vw^{~*})
\end{eqnarray*}

\vskip 1.5cm \noindent

Let
\begin{eqnarray*}
{F_{\ga,D,u}}&:=&-\frac{u}{2}\int_{D\times D}dr~dr'~~
J_\ga(r,r')\vec\rho_{D}(r)\cdot \vec\rho_{D}(r')-(1-u)\int_{D}dr~~
\vec\rho_{D}(r)
\cdot\vec \rho^{\pm}
\\
& &-\int_{D\times D^c}dr~~
J_\ga(r,r')\vec\rho_{D}(r)
\cdot \vec\rho_{D^{c}}(r')+\frac{1}{\beta}\int_{D} dr
\vec\rho_D(r)\cdot\ln\vec\rho_D(r)
\end{eqnarray*}
The proof of the bound \eqref{assumpt2} is thus based on the
following proposition whose proof is given at the end of subsection
\ref{subsection:smalldeviation}:
\begin{prop}
\label{prop:smalldeviation}
There is a constant $c>0$ so that for all  $\ga$ small enough, any $x$:
\begin{eqnarray}
\label{eq:smalldev01} && \hskip-1cm
\sup_{\vw^{*}\in\cX^{({\pm})}}\ln \; \mu_{\abs,D,u} (\cAA|\vw^{~*})
\le -\hskip-.3cm\inf_{\vec \rho_{D^{c}}\in \cX_{D^{c}}^{\pm}}
\Bigg(\inftwo{\vec\rho_{D} \in \cX_{D}^{\pm},}{
\left|\vec\rho^{(\ell_{-,\ga})}(r) - \vec\rho^{\pm}\right|
>(1-\zeta')\zeta}
\hskip-.7cm F_{\ga,D,u}(\vec\rho_{D}|\vec \rho_{D^{c}})
 \nonumber\\&&\hskip3cm
-\inf_{\vec\rho_D \in \cX_{D}^{\pm}} F_{\ga,D,u}(\vec\rho_{D}|\vec
\rho_{D^{c}})\Bigg)
 + c \ga^{1/2} \ell_{-,\ga}^d
\end{eqnarray}
\end{prop}

where, by an abuse of notation, we have denoted by the same symbol $\cX^{\pm}$,
the restricted ensemble:
         \begin{eqnarray}
 \cX^{\pm}:= \Big\{ \vec\rho\in L^\infty(\Rr^d,S_Q):\|\vec \rho^{(\ell_{-,\ga})}(r)
 - \vec\rho^{\pm}\|_{\star} \le \ga^a, ~~\forall r\in \Rr^d\Big\}
       \label{7.3}
     \end{eqnarray}
and denote by $\cX^{\pm}_\La$ the above expression \eqref{7.3}
when the constrain is imposed on
$\La$, with   $\La \sqsubset \Rr^d$, a $\cD^{(\ell_{-,\ga})}$-measurable set.

\vskip 1.5cm \noindent
\begin{proof}
Using a result similar to Theorem \ref{thm:app} but with a slightly different functional,
one gets for all $\vw^{*}\in\cX^{({\pm})}$
\begin{eqnarray*}
&&\ln \; \mu_{\abs,D,u}(\cAA|\vw^{~*})
= \ln Z_{\abs,D,u}(\cAA|\vw^{~*})- \ln Z_{\abs,D,u}(\vw^{~*})\\
& &\le
\inftwo{\vec\rho_{D} \in \cX_{D}^{\pm},}{
\left|\vec\rho^{(\ell_{-,\ga})}(r) - \vec\rho^{\pm}\right|
>(1-\zeta')\zeta}
\hskip-.7cm F_{\ga,D,u}(\vec\rho_{D}|(\vw^{~*})^{(\ell_0)})
-\inf_{\vec\rho_D \in \cX_{D}^{\pm}}
F_{\ga,D,u}(\vec\rho_{D}|(\vw^{~*})^{(\ell_0)})
 + c \ga^{-1} \ell_0 |D|\\
 & &\le
-\inf_{\vec \rho_{D^{c}}\in
\cX_{D^{c}}^{\pm}}
\Bigg(\inftwo{\vec\rho_{D} \in \cX_{D}^{\pm},}{
\left|\vec\rho^{(\ell_{-,\ga})}(r) - \vec\rho^{\pm}\right|
>(1-\zeta')\zeta}
\hskip-.7cm F_{\ga,D,u}(\vec\rho_{D}|\vec \rho_{D^{c}})
-\inf_{\vec\rho_D \in \cX_{D}^{\pm}} F_{\ga,D,u}(\vec\rho_{D}|\vec
\rho_{D^{c}})\Bigg)
 + c \ga^{1/2} \ell_{-,\ga}^d
\end{eqnarray*}

\end{proof}

\vskip 1.5cm \noindent

We need now an estimate for the right hand side of
the previous equation.
We define:

\begin{eqnarray*}
F^0_{\ga,D,u}(\vec\rho_{D}| \vec\rho_{D^{c}}) := -\int_{D}dr~~
\left( \vec\cL^{u}(r,\vec\rho_{D^{c}}) \cdot
\vec\rho_{D}-\frac{1}{\beta}\vec\rho_D\cdot\ln\vec\rho_D\right)
\end{eqnarray*}
where $\vec\cL^u(r,\vec\rho_{D^{c}})$ is the external field:
\begin{eqnarray}
\label{eq:hi}
\vec\cL^{u}(r,\vec\rho_{D^{c}}):= (1-u)\vec \rho^{\pm}+ u\int_{D^{c}}
J_\ga(r,r')\vec\rho_{D^{c}}(r')dr'
\end{eqnarray}

Since $F^0_{\ga,\beta,D}(\vec\rho_{D}|
\vec\rho_{ D^{c}   })$ differs from
 $F_{\ga,\beta,D}(\vec\rho_{D}|
\vec\rho_{ D^{c}   })$ by the self interaction
energy, 
which is bounded
proportionally to $|D|^2$,
there is a constant $c'>0$ such that:
\begin{eqnarray}
\label{eq:smalldev1} |F_{\ga,D,u}(\vec\rho_{D}| \vec\rho_{
D^{c}})-F^0_{\ga,D,u}(\vec\rho_{D}| \vec\rho_{D^{c}})|\le c' u
\ga^{d}|D|^2
\end{eqnarray}

 $F^0_{\ga,D}(\vec\rho_{D}|
\vec\rho_{D^{c}})$ is a convex functional on $L^\infty(D,S_Q)$,
and has thus a unique minimizer that we denote by
$\vec\rho^{~*}(r;\vec\rho_{D^c})\equiv \vec\rho^{~*}(r)$, whose components are  given by:
\begin{eqnarray}
\label{barmD} \rho^{~*}_k(r)=\frac{e^{\beta
\cL^u_k(r,\vec\rho_{D^{c}})}} {\sum_l e^{\beta
\cL^u_l(r,\vec\rho_{D^{c}})}}
\end{eqnarray}
We need to evaluate the difference:
\begin{eqnarray*}
& &F^0_{\ga,D,u}(\vec\rho_{D}|\vec\rho_{D^c})-F^0_{\ga,D,u}(\vec\rho^{~*}_D|\vec\rho_{D^c})=\\
& &\hskip2cm\int_D dr' \left(- \vec\rho_{D}(r') \cdot
\vec\cL^{u}(r,\vec\rho_{D^c})+\frac{1}{\beta}
\vec\rho_{D}(r')\cdot\ln \vec\rho_{D}(r')\right)-
\\
&
&\hskip3cm\left(-\vec\rho^{~*}_{D}(r')\cdot\vec\cL^{u}(r,\vec\rho_{D^c})+\frac{1}{\beta}
\vec\rho^{~*}_{D}(r')\cdot\ln\vec\rho^{~*}_{D}(r')\right)
\end{eqnarray*}
Using \eqref{barmD}, we write $\vec\cL^{u}(r,\vec\rho_{D^c})$ in
terms of $\vec\rho^{~*}$ and get:
\begin{eqnarray}
\label{relativentropy} &
&F^0_{\ga,D,u}(\vec\rho_{D}|\vec\rho_{D^c})-F^0_{\ga,D,u}(\vec\rho^{~*}_D|\vec\rho_{D^c})=
 \frac{1}{\beta}\int_D dr'
\sum_{k=1}^{Q}\rho_{k}(r') \ln\frac{\rho_{k}(r')}{\rho^{~*}_{k}(r')}
\end{eqnarray}
Thus by the Kullback-Leibler inequality, one gets
\begin{eqnarray*}
\label{relativentropy2}
F^0_{\ga,D,u}(\vec\rho_{D}|\vec\rho_{D^c})-F^0_{\ga,D,u}(\vec\rho^{~*}_D|\vec\rho_{D^c})
& \ge &\frac{1}{2\beta}\int_D dr'|
 \vec\rho_{D}(r')- \vec\rho^{~*}_{D}(r')|^{2}
\end{eqnarray*}

\vskip .5cm \noindent
We claim that there is $\eps>0$ such that,  for any
$\vec\rho_{D^{c}}$ and for $\ga>0$ small enough,
\begin{eqnarray}
\label{eq:smalldev2}
\|\vec \rho^{~*}- \vec \rho^{\pm}\|_{\star}\le (1-\eps)\ga^a
\end{eqnarray}
Using Cauchy-Schwartz inequality, we thus get (taking $\zeta'< \eps$):
\begin{eqnarray}
\label{eq:smalldev3}
F^0_{\ga,D,u}(\vec\rho_{D}|\vec\rho_{D^c})-F^0_{\ga,D,u}(\vec\rho^{~*}_D|\vec\rho_{D^c})
\ge \frac{1}{2\beta}|D|(\eps-\zeta')^2\ga^{2a}
\end{eqnarray}

\vskip .5cm \noindent
Postponing the proof of \eqref{eq:smalldev2},
 we get the following bound by using proposition
\ref{prop:smalldeviation} together with \eqref{eq:smalldev1} and \eqref{eq:smalldev3}:

\begin{eqnarray*}
\sup_{\vw^{*}\in\cX^{({\pm})}}\ln \;
\mu_{\abs,D,u}(\cAA_x|\vw^{~*})&\le& \exp\left\{-\left(
\frac{1}{2\beta}|D|(\eps-\zeta')^2\ga^{2a}-c'\ga^{d}|D|^2 \right)+
c\ga^{1/2}\ell_{-,d}^{d}\right\}
\\
&=&\exp\left\{-\left(
\frac{1}{2\beta}(\eps-\zeta')^2\ga^{2a}-c'\ga^{d\al} +c\ga^{1/2}\right)\ell_{-,d}^{d}\right\}
\end{eqnarray*}
The bound \eqref{assumpt2} is then proven for our abstract models.
We now turn to the proof of \eqref{eq:smalldev2}:

\vskip .5cm \noindent
In order to prove \eqref{eq:smalldev2}, we show that there is $b\in (0,1)$:
\begin{eqnarray}
\label{eq:star-L}
|\rho^{~*}_k(r)-  \rho^{\pm}_k|\le b |\cL^u_k(r)- \rho^{\pm}_k|
\end{eqnarray}
while $\|\vec\cL^u(r)- \vec\rho^{\pm}\|_{\star}$ is small and close enough to $\ga^a$.

We first prove that $| \cL^u_k(r)-\rho_k^{\pm}|$ is small enough.
We define

     \begin{equation*}
     J^{(\ell_{-,\ga})}_{\ga}(r,\lng r' \rng):= \frac{1}{|C^{\ell_{-,\ga}}_{r'}|}\int J_{\ga}(r,r'')dr''
    \end{equation*}
For $\ga$ small enough, by hypothesis on $\mathcal J$ (see \eqref{def:J})

    \begin{equation}
      \label{p4.3.1.5}
\big| J_{\ga}(r,r')-J^{(\ell_{-,\ga})}_{\ga}(r,\lng r' \rng)\big| \le
 \sqrt{d} \|\nabla \cJ\|_\infty\ga^{d+1}\ell_{-,\ga} \text{\bf
1}_{\{|r-r'|\le 2\ga^{-1}\}}
     \end{equation}
Then, for any $\vec\rho\in L^\infty(\mathbb
R_+^d,S_Q)$, there is a constant $c_d=2^d  \sqrt{d} \|\nabla \cJ\|_\infty$
    \begin{eqnarray*}
\big\| J_\ga* \vec\rho - J_{\ga}^{(\ell_{-,\ga})}* \vec\rho \big\|
&\leq& {c_d~  \ga\ell_{-,\ga}= c_d\ga^{\al}}
    \end{eqnarray*}

Using the fact that $\vec\rho_{D^c}\in \cX^\pm_{D^c}$, we write:

\begin{eqnarray*}
\big\|\vec\cL^u(r) - \vec\rho^{\pm} \big\|
&=&u\big\|\int_{D^c}J_\ga(r,r') \vec\rho_{D^c}(r') dr' - \vec\rho^{\pm} \big\|\\
&\le&\big\|\int_{D^c}\bigl(J_\ga(r,r')-J_{\ga}^{(\ell_{-,\ga})}(r,r')\bigr) \vec\rho_{D^c}(r') dr' \big\|
\\
& & \hskip1cm+\big\| \int_{D^c}J_{\ga}^{(\ell_{-,\ga})}(r,r') \bigl(\vec\rho_{D^c}(r')-
\vec\rho^{\pm}  \bigr)dr'\big\|
+\big\|\int_{D}J_{\ga}^{(\ell_{-,\ga})}(r,r') \vec\rho^{\pm} dr' \big\|\\
&\le& c_d\ga^{\al} +\ga^a + c'_d \ga^{\al d}
\end{eqnarray*}
with $c_d'=\|\cJ\|_\infty$.
Recalling \eqref{barmD}, we have
\begin{eqnarray}
\big\|\vec\rho^{~*}_{D}(r)-  \vec\rho^{\pm}\big\|=\big\|\vec g(\vec\cL^u(r)) - \vec g(\vec\rho^{\pm}) \big\|
\end{eqnarray}
Applying Theorem \ref{thm:mflocalstability}, we get
\begin{eqnarray*}
\big\|\vec\rho^{~*}_{D}(r)-  \vec\rho^{\pm}\big\|&\le&
(1-\frac{1}{2Q})\bigg(\ga^{a}+c_d\ga^{\al}+c'_d\ga^{\al d}\bigg)
\\
&\le& (1-\frac{1}{4Q})\ga^{a}-\bigg(\frac{1}{4Q}\ga^{a}-(1-\frac{1}{2Q})(c_d\ga^{\al}+c_d'\ga^{\al d})\bigg)
\\
& \le& \ga^{a}(1-\eps)
\end{eqnarray*}
for $\ga$ small enough and $a<\al$, taking $\eps=1-\frac{1}{4Q}$.

\vskip .5cm \noindent

\subsubsection{Third requirement}
\vskip .5cm \noindent

Let
\begin{eqnarray*}
\cB(n):=\{m\in (\ell_{-,\ga}\Zz)^d:
\dist(C^{\ell_{-,\ga}}_n,C^{\ell_{-,\ga}}_m)\le\ga^{-1} \}
\end{eqnarray*}
there is $\tilde r_{nm}: (\ell_{-,\ga}\Zz)^{d}\times
(\ell_{-,\ga}\Zz)^{d}\to\Rr^{+}$ s.t. for any
$\vw^{1},\vw^{2}:\vw^{1}(j)=\vw^{2}(j)~~\forall j\in [\Zz^d\sqcap
\cB(n)]\setminus i $, it holds that:
\begin{eqnarray}
& & R_{C_n}(\mu_C(\cdot|\vw_{C_n^c}^{(1)}),
\mu_{C_n}(\cdot|\vw_{C_n^c}^{(2)}))\le
\sum_{m\in (\ell_{-,\ga}\Zz)^d\setminus \cB(n)}
\tilde r_{nm} \dist(\vw_{C_m}^{(1)},\vw_{C_m}^{(2)}) \label{eq:a3-a}
\\
&&\sup_{n}\sum_{m\notin \cB(n)} \tilde r_{nm}<1 \label{eq:a3-b}
\end{eqnarray}

This is a condition on the tail of the interaction, which in our
case is satisfied because of the exponential decay of the
interaction $h_u^{\pm}$ due to the Peierls estimates.

\vskip .5cm \noindent
\subsubsection{Fourth requirement}
\vskip .5cm \noindent

Let $r(n,m): (\ell_{-,\ga}\Zz)^{d}\times (\ell_{-,\ga}\Zz)^{d}\to\Rr^{+}$ 
defined as follows:

\begin{eqnarray*}
& &r(n,m):=
\begin{cases}
    0 & \text{if  }m=n\\
    \tilde r_{nm} & \text{if } m\notin \cB(n)\\
 \tilde r_{nm}^*    &
    \text{if } m\in \cB(n)\setminus n
  \end{cases}
\end{eqnarray*}
where:
\begin{eqnarray*}
\tilde r_{nm}^*:=\dis{\sup_{j\in C^{\ell_{-,\ga}}_m}\sum_{i\in
C^{\ell_{-,\ga}}_n}\sum_{k>0}
    b_{C^{\ell_{-,\ga}}_n}^{k}(i,j)+2e^{-c\ga^{2a}\ell^{d}_{-,\ga}} \bigg(\ell_{-,\ga}^d+
    \sum_{i,i'\in C^{\ell_{-,\ga}}_n}\sum_{k>0}b_{C^{\ell_{-,\ga}}_n}^{k}(i,i')\bigg)}
\end{eqnarray*}
with
\begin{eqnarray*}
b_{C^{\ell_{-,\ga}}_n}(i,j):=b(i,j)\Ii_{i\in C_{n}} 
\end{eqnarray*}
and $b_{C^{\ell_{-,\ga}}_n}^{k}(i,j)$ is the $k$-th convolution of
$b_{C^{\ell_{-,\ga}}_n}(i,j)$. Assumption $4$ then states:
\begin{eqnarray}
\label{eq:as-4} \sup_{n}\sum_{m}r(n,m)\le\delta     \hskip3cm
0<\delta<1
\end{eqnarray}
\vskip .5cm \noindent
\subsubsection{Fifth Requirement:}
\vskip .5cm \noindent

There is a constant $b>0$ such that for all $n \in
(\ell_{-,\ga}\Zz)^d$,
\begin{eqnarray}
\label{eq:as-5} \sum_{m\not=n}r(n,m)e^{b\ga |n-m|}<1
\end{eqnarray}

\vskip .5cm \noindent
\subsubsection{Conclusion of subsection \eqref{subsect:decayofthecorrelations}}
\vskip .5cm \noindent

We have proved that the two abstract models fulfill the requirements
of the extended Dobrusin criterion of reference \cite{bkmp2}, which
imply uniqueness of the measures. Moreover,  $5$ holds and implies
exponential decay of correlations for the measures $\tilde
\mu^{\pm}_{\abs,u}$. Hence Theorem \ref{thmz5.1} and Corollary
\ref{cor:mu-limit} are proven.
\end{proof}

\vskip 2.5cm \noindent
\subsection{Small deviation estimates}
\label{subsection:smalldeviation}
\vskip .5cm \noindent

In this subsection we prove the estimate \eqref{eq:smalldev}
\begin{eqnarray*}
\int_0^1 du\sum_{i \in A}
\{|\langle g_{\ga,i}\rangle_{\tilde \mu^{\pm}_{\abs;u}}|
+|\langle g_{\ga,i,\La}\rangle_{\tilde \mu^{\pm}_{\abs,\La;u}}|\}
\le c\ga^{1/8}|\delta_{\ins}^{\ell_{+,\ga}}[\La]|
\end{eqnarray*}
Let $A:=\delta_{\ins}^{\ell_{+,\ga}}[\La]$ and define the set $S^{\pm}(\vw)$ as:
\begin{eqnarray}
\label{def: S}
S^{\pm}(\vw):= \{i\in\check A: \|\vw^{(\ell_0)}(i)-\vec \rho^{\pm}\|_{\star}\ge \ga^{1/8}\}
\end{eqnarray}
where
\begin{eqnarray}
\label{def: hatA}
\check A:= A\sqcup\delta_{\out}^{\ga^{-1}}[A]
\end{eqnarray}
We first prove the following bound:
\begin{eqnarray}
\label{eq:sm-dev-1}
&&\int_0^1 du\sum_{i\in A}
\{|\langle g_{\ga,i}\rangle_{\tilde \mu^{\pm}_{\abs;u}}|
+|\langle g_{\ga,i,\La}\rangle_{\tilde \mu^{\pm}_{\abs,\La;u}}|\}
\\
&&\hskip1cm\le \fJ_\ga\int_0^1~ du
\left[\langle \Ii_{|S^\pm|\ge\ga^{1/8}|A|}\rangle_{\tilde \mu^{\pm}_{\abs;u}}
+\langle \Ii_{|S^\pm|\ge\ga^{1/8}|A|}\rangle_{\tilde \mu^{\pm}_{\abs,\La;u}}\right]
+ c(\ga^{1/2}|A|+\ga^{1/8}|A|)\nn
\end{eqnarray}
where

we have denoted by $\fJ_\ga$ the normalization constant of the
interaction kernel $J_\ga$ on $\Zz^d$:
\begin{eqnarray}
\label{def:Jb} \fJ_\ga:= \sum_{j\in \Zz^d
}{J_\ga(0,j)}\hskip1.5cm\lim_{\ga\to 0}\fJ_\ga=1
\end{eqnarray}

 \vskip .5cm \noindent
\begin{proof}[proof of \eqref{eq:sm-dev-1}]

\vskip .5cm \noindent
We recall the definition of $g_{\ga,i,\La}(\vw)$ \eqref{def:ggaLa} and $g_{\ga,i}(\vw)$
\eqref{def: gx}:
\begin{eqnarray*}
g_{\ga,i,\La}(\vw)&:=&-\frac{1}{2}\vvw_i\sumtwo{j\in\La}{j\ne i}J_\ga(i,j)\vvw_j
    +\sumtwo{\Delta\ni i}{\Delta\sqsubset \La} \frac{1}{|\Delta|}U^{\pm}_\Delta(\vw)
\\
g_{\ga,i}(\vw)&:=&\lim_{\La\nearrow \Rr^{d}}g_{\ga,i,\La}(\vw)=
-\frac{1}{2}\vvw_i\sum_{j\ne i}J_\ga(i,j)\vvw_j
+\sum_{\Delta\ni i} \frac{1}{|\Delta|}U^{\pm}_\Delta(\vw)
\end{eqnarray*}

\vskip .5cm \noindent

By \eqref{new-z4.11b},
\begin{eqnarray*}
\dis{\sum_{\Delta\ni i} \frac{1}{|\Delta|}U^{\pm}_\Delta(\vw)\le
e^{-\frac{\fK_\ga}{2} }}
\end{eqnarray*}
Hence, we have:
\begin{eqnarray*}
&&\sum_{i\in A}\big| g_{\ga,i}(\vw)\big|\le
\frac{1}{2}\sum_{i\in A}\big|\vvw_i\sum_{j\ne i}J^{(\ell_0)}_\ga(i,j)\vvw_j\big| + c \ga\ell_0|A|\\
&&\le \sum_{i\in A}\sum_{j\in\check A}J^{(\ell_0)}_\ga(i,j)\|\vvw_j^{(\ell_0)}\|_\star +c \ga^{1/2}|A|\\
&&\le \fJ_\ga \sum_{j\in\check A}\|\vvw_j^{(\ell_0)}\|_\star +c \ga^{1/2}|A|\\
&&\le \fJ_\ga |S^\pm(\vw)|+ c(\ga^{1/2}|A|+\ga^{1/8}|A|)
\end{eqnarray*}
where we have defined
\begin{eqnarray*}
\blue{J^{(\ell_0)}_\ga(i,\lng j \rng)}:=\frac{1}{|C^{\ell_0}|}
\sum_{j'\in C_j^{\ell_0}} J_\ga(i,j)
\end{eqnarray*}
and used that by definition \eqref{def: S},
$\|\vvw_j\|_{\star}\le \ga^{1/8}$  for all $j\notin S$.
A similar estimate for $g_{\ga,i,\La}$ holds and \eqref{eq:sm-dev-1} follows.
\end{proof}

\vskip .5cm \noindent
Since the two terms in the integral in the right hand side of \eqref{eq:sm-dev-1} are very similar, we give
the derivation of an estimate for the first term only.
We prove the following
\begin{eqnarray}
\label{eq:bound8}
\langle \Ii_{|S^{\pm}|\ge\ga^{1/8}|A|}\rangle_{\tilde \mu^{\pm}_{\abs;u}}
\le e^{-c\ga^{3/8}|A|}
\end{eqnarray}

\begin{proof}[proof of \eqref{eq:bound8}]

\vskip .5cm \noindent

Let

\begin{eqnarray*}
\hat A:=A\sqcup \delta_{\out}^{\ell_{+,\ga}}[A]
\end{eqnarray*}
\begin{eqnarray}
\label{eq:proto8}
\langle \Ii_{|S^{\pm}|\ge\ga^{1/8}|A|}\rangle_{\tilde \mu^{\pm}_{\abs;u}}
=\langle
\frac{\dis{\hat \mu^{\pm}_{\abs;\hat A,u}(\Ii_{|S^{\pm}|\ge\ga^{1/8}|A|}
e^{-\beta\sum_{\Delta:\Delta\sqcap \hat A\not=\varnothing} U^\pm_\Delta(\vw)}|\vw_{\hat A^c})}}
{\dis{\hat \mu^{\pm}_{\abs;\hat A,u}(
e^{-\beta\sum_{\Delta:\Delta\sqcap \hat A\not=\varnothing} U^\pm_\Delta(\vw)}|\vw_{\hat A^c})}}
\rangle_{\tilde \mu^{\pm}_{\abs;u}}
\end{eqnarray}
where $\hat \mu^{\pm}_{\abs;\hat A,u}$ is the measure on $\cX^{\pm}_{\hat A}$
associated to the finite range interpolating Hamiltonian $\hat h^{\pm}_u(\vw_{\hat A}|\vw_{\hat A^c})$
\begin{eqnarray*}
\hat h^{\pm}_u(\vw_{\hat A}|\vw_{\hat A^c}):=u H_{\ga,\hat A}(\vw_{\hat A}|\vw_{\hat A^c})
+(1-u)\fH_{\ga,\hat A}^{\pm}(\vw)
\end{eqnarray*}
Recalling \eqref{new-z4.11b}, we get:
\begin{eqnarray}
\langle \Ii_{|S^{\pm}|\ge\ga^{1/8}|A|}\rangle_{\tilde
\mu^{\pm}_{\abs;u}} \le\langle \hat \mu^{\pm}_{\abs;\hat
A,u}(\Ii_{|S^{\pm}|\ge\ga^{1/8}|A|}|\vw_{\hat A^c}) \rangle_{\tilde
\mu^{\pm}_{\abs;u}}\times e^{\beta|\hat A| e^{-\frac{\fK_\ga}{2} }}
\end{eqnarray}
we write
\begin{eqnarray}
\hat \mu^{\pm}_{\abs;\hat A,u}(\Ii_{|S^{\pm}|\ge\ga^{1/8}|A|}|\vw_{\hat A^c})
=\frac{\dis{\hat Z_{\abs,\beta,\hat A;u}^{\pm}(S^{\pm}|\vw_{\hat A^c})}}
{\dis{\hat Z_{\abs,\beta,\hat A;u}^{\pm}(\vw_{\hat A^c})}}
\end{eqnarray}
where
\begin{eqnarray*}
\hat Z_{\abs,\beta,\hat A;u}^{\pm}(S^{\pm}|\vw_{\hat A^c}):=
&&\sum_{\vw_{\hat A}\in \cX^{\pm}_{\hat A}} \Ii_{|S^{\pm}|\ge\ga^{1/8}|A|}
e^{-\beta \hat h^{\pm}_u(\vw_{\hat A}|\vw_{\hat A^c})}\\
\hat Z_{\abs,\beta,\hat A;u}^{\pm}(\vw_{\hat A^c}):=
&&\sum_{\vw_{\hat A}\in \cX^{\pm}_{\hat A}}
e^{-\beta \hat h^{\pm}_u(\vw_{\hat A}|\vw_{\hat A^c})}
\end{eqnarray*}
Now the partition function $\hat Z_{\abs,\beta,\hat A;u}^{\pm}(S^{\pm}|\vw_{\hat A^c})$
can be estimated using an approximation to the continuum. Defining
\begin{eqnarray*}
&&F_{\ga,\hat A,u}(\vw_{\hat A}|\vw^{(\ell_0)}_{\hat A^c}):=
-\frac{1}{2}\int_{\hat A\times \hat A}J_\ga(r,r')
\big( u \vw_{\hat A}(r)\cdot\vw_{\hat A}(r') +(1-u) \vec\rho^{\pm}\cdot \vec\rho^{\pm}\big)dr dr'\\
&&\hskip3cm-\int_{\hat A\times \hat A^c}J_\ga(r,r')
\big(u \vw_{\hat A}(r)\cdot\vw^{(\ell_0)}_{\hat A^c}(r') +(1-u) \vec\rho^{\pm}\cdot \vec\rho^{\pm}\big) dr dr'\\
&&\hskip3cm-(1-u)\int_{\hat A}\big(\vw_{\hat A}(r)\cdot\vec\rho^{\pm}\big) dr
+\frac{1}{\beta}\int_{\hat A}\vw_{\hat A}(r)\cdot\ln\vw_{\hat A}(r) dr
\end{eqnarray*}
and
\begin{eqnarray*}
\fZ^{\pm}:=\left\{\vw_{\hat A}\in \cXx_\La^{\ell_0}: \vw_{\hat A}(r)=\vw^{(\ell_0)}(r)~;~
\eta(\vw;r)=a_{\pm}~~ r\in \hat A~;~\int_{\check A}
\Ii_{\|\vw_{\hat A}-\vec\rho^{\pm}\|_{\star}\ge\ga^{1/8}}\ge\ga^{1/8}|A| \right\}
\end{eqnarray*}
A result similar to theorem \ref{thm:app} holds for the above functional and leads to
\begin{eqnarray}
\label{eq:sm-dev-2}
\ln\hat Z_{\abs,\beta,\hat A;u}^{\pm}(S^{\pm}|\vw_{\hat A^c})\le-\beta\inf_{\vw_{\hat A}\in\fZ^{\pm}}
F_{\ga,\hat A,u}(\vw_{\hat A}|\vw^{(\ell_0)}_{\hat A^c})+c\ga^{1/2}|\hat A|
\end{eqnarray}
where $c$ is a  constant independent on $u$.

We are then reduced to study the variational problem in \eqref{eq:sm-dev-2}
for $F_{\ga,\hat A,u}(\vw_{\hat A}|\vw^{(\ell_0)}_{\hat A^c})$ on  $\fZ^{\pm}$.
We define the ``excess free energy functional":
\begin{eqnarray}
\label{def:F-eff-u}
& &\cF^{\eff}_{\ga,\hat A,u}(\vw_{\hat A}|\vw^{(\ell_0)}_{\hat A^c}):=
F_{\ga,\hat A,u}(\vw_{\hat A}|\vw^{(\ell_0)}_{\hat A^c})+\fA
\\
& &\fA:=-\inf_{\vec v: \|\vec v\|=1}\hskip-.3cm\phi^{\mf}_{u}(\vec v)|\hat A|
+\frac{u}{2}\int_{\hat A\times \hat A^c}\hskip-.73cmJ_\ga(r,r')\big(\vw^{(\ell_0)}_{\hat A^c}(r')\cdot
\vw^{(\ell_0)}_{\hat A^c}(r')\big)\; dr\ dr'
\\
\label{def:fmfu}
& &\hskip-.73cm\phi^{\mf,\pm}_{u}(\vec v):=-u\frac{(\vec v\cdot\vec v )}{2}-
(1-u)(\vec v\cdot \rho^{\pm})+\frac{1}{\beta}
(\vec v\cdot \ln\vec v)
\end{eqnarray}
$\cF^{\eff}_{\ga,\hat A,u}(\vw_{\hat A}|\vw^{(\ell_0)}_{\hat A^c}) $ is positive and
 differs from $F_{\ga,\hat A,u}(\vw_{\hat A}|\vw^{(\ell_0)}_{\hat A^c})$
by a constant and hence has the same minimizers and its minimum is
finite

\vskip .5cm \noindent
Denoting by:
\begin{eqnarray}
\label{def:Fmfu}
\Phi^{\eff,\pm}_u(\vw_{\hat A}):=\phi^{\mf,\pm}_u(\vw_{\hat A})-\inf_{\vec v: \|\vec v\|=1}\phi^{\mf,\pm}_{u}(\vec v)
\end{eqnarray}
$\cF^{\eff}_{\ga,u}(\vw_{\hat A}|\vw^{(\ell_0)}_{\hat A^c})$ can be rewritten then as:
\begin{eqnarray*}
& &\cF^{\eff}_{\ga,\hat A,u}(\vw_{\hat A}|\vw^{(\ell_0)}_{\hat A^c})=
\int_{\hat A} \Phi^{\eff,\pm}_u(\vw_{\hat A})
+\frac{u}{4}\int_{\hat A\times \mathbb{B}}J_\ga(r,r')[\vw_{\hat A}(r)-\vw_{\hat A}(r')]^{2}
\\
& &\hskip7cm+\frac{u}{2}\int_{\hat A\times \hat A^c}
J_\ga(r,r')[\vw_{\hat A}(r)-\vw^{(\ell_0)}_{\hat A^c}(r')]^{2}
\end{eqnarray*}

\vskip .5cm \noindent The analysis in  appendix \ref{app:dinamica},
see corollary \ref{corol:dimamica},  proves that there are positive
constants $\om, c_\om$, so that for any $\vw\in \fZ^{\pm}$ there is
$\vec\psi_{\hat A}$:
\begin{eqnarray}
\label{eq:dinamica}
\cF^{\eff}_{\ga,\hat A,u}(\vw_{\hat A}|\vw^{(\ell_0)}_{\hat A^c})\ge \cF^{\eff}_{\ga,\hat A,u}
(\vec \psi_{\hat A}|\vw^{(\ell_0)}_{\hat A^c})-
c_\om e^{-\om\ga\ell_{+,\ga}/4}|\hat A|
\end{eqnarray}
where $\vec\Psi_{\hat A}$ has the following properties:
\begin{eqnarray}
& &\eta(\vec \psi_{\hat A};r)=a_{\pm} ~~~\forall r\in \hat A
\nn \\
& & \vec \psi_{\hat A}(r) =\vec \rho^{\pm} ~~~\forall r\in \blue{\Si:=}
\blue{\big(\hat A\setminus A\big)\setminus \delta_{\ins}^{\ell_{+,\ga}/4}\big(\hat A\setminus A\big)}
\label{def: Si} \\
& & \vec \psi_{\hat A}(r)=\vw(r) ~~~\forall r\in \check A 
 \nn\\
& &
\sup_q|\psi_{\hat A,q}(r)-\rho_q^{\pm}|\le(1-\kappa_0)\ga^{a} ~~~
\forall r\in\delta_{\ins}^{\ell_{+,\ga}/4}[\hat A]
~~~~~~~~~~~~~~~~\kappa_0>0 \nn
\end{eqnarray}

We can write:

\begin{eqnarray}
\label{eq:u1-0}
\ln\hat Z_{\beta,\hat A;u}^{\pm}(S^{\pm}|\vw_{\hat A^c})\le -\beta \inf_{\vec\psi_{\hat A}\in\cB^0}
\cF^{\eff}_{\ga,\hat A,u}(\vec\psi_{\hat A}|\vw_{\hat A^c}^{(\ell_0)})+c\ga\ell_0|\hat A|
+\beta c_\om e^{-\om\ga\ell_{+,\ga}/4}|\hat A|+\fA\hskip.73cm
\end{eqnarray}
where

\begin{eqnarray}
\label{def: B0} &&\hskip-1.6cm\cB^{0}:=\{\vec\psi_{\hat A}:
\sup_q|\psi_{\hat A,q}-\rho_q^{\pm}|\le(1-\kappa_0)\ga^{a},
~r\in\delta_{\ins}^{\ell_{+,\ga/4}}[\hat A]; ~~ \eta(\vec\psi_{\hat
A};r)=a_{\pm}, ~r\in \hat A;\\ \nn &&\hskip-.5cm \vec\psi_{\hat
A}=\vec\rho^{\pm} ~ r\in \Si; ~~
 \vec\psi_{\hat A}(r)=\vec\psi_{\hat A}^{(\ell_0)}(r), ~r\in \check A; ~~
 \int_{\check A}\Ii_{\| \vec\psi_{\hat A,q}-\vec\rho^{\pm}\|_{\star}}\ge \ga^{1/8}>\ga^{1/8}|A|\}
\end{eqnarray}

In appendix \ref{app:u1} it is proved that for any $\psi_{\hat A}\in\cB^0$ there is
$\vec\psi^{*}_{\hat A}:$
\begin{eqnarray}
\label{eq:u1-a}
\vec \psi^*_{\hat A}=
  \begin{cases}
    \vec \psi_{\hat A} & \text{on}~~\delta_{\ins}^{\ell_{+,\ga}/4}[\hat A], \\
    \vec\rho^{\pm} & \text{elsewhere}.
  \end{cases}
\end{eqnarray}
so that:
\begin{eqnarray}
\label{eq:u1-b}
\cF^{\eff}_{\ga,\hat A,u}(\vec\psi_{\hat A}|\vw^{(\ell_0)}_{\hat A^c})\ge
\cF^{\eff}_{\ga,\hat A,u}(\vec\psi^*_{\hat A}|\vw^{(\ell_0)}_{\hat A^c})+c\ga^{1/4}(\ga^{1/8}|A|)
\end{eqnarray}
Let $[\psi^{*}]^{\ell_0}$ defined as in \eqref{def:parteintera}.
By theorem \ref{thm:app} we then have:

\begin{eqnarray*}
-\beta F_{\ga,\hat A,u}(\vec\psi^{~*}_{\hat A}|\vw^{(\ell_0)}_{\hat A^c})\le
\ln \hat Z_{\beta,\hat A;u}\big(\{\vw_{\hat A}^{\ell_0}=[\vec\psi^{~*}_{\hat A}]^{\ell_0}\}\big|\vw_{\hat A^c}\big)+\ga\ell_0|\hat A|
\end{eqnarray*}
and since by definition :
\begin{eqnarray*}
\|[\vec\psi^{~*}_{\hat A}]^{\ell_0}-\vec\psi^{~*}_{\hat A}\|_{\star}<\frac{1}{2} \ell_0^{-d}
\end{eqnarray*}
we have:

\begin{eqnarray*}
\|[\vec\psi^{~*}_{\hat
A}]_q^{\ell_0}-\vec\rho^{~\pm}\|_{\star}<\frac{1}{2}\ell_0^{-d}+
(1-\kappa_0)\ga^{a}
\end{eqnarray*}
\blue{so that the set $\{\vw_{\hat A}^{\ell_0}=[\vec\psi^{~*}_{\hat A}]^{\ell_0}\}\in \cX^{\pm}$ and
}
\begin{eqnarray*}
-\beta F_{\ga,\hat A,u}(\vec\psi^{~*}_{\hat A}|\vw^{(\ell_0)}_{\hat A^c})\le
\ln \hat Z^\pm_{\abs,\beta,\hat A;u}\big(\vw_{\hat A^c}\big)+\ga\ell_0|\hat A|
\end{eqnarray*}

By \eqref{def:F-eff-u} and \eqref{eq:u1-0}
\begin{eqnarray*}
\ln\hat Z_{\abs,\beta,\hat A;u}^{\pm}(S^{\pm}|\vw_{\hat A^c})\le
\ln \hat Z_{\abs,\beta,\hat A;u}^\pm\big(\vw_{\hat A^c}\big)
+\beta c_\om e^{-\om\ga\ell_{+,\ga}/4}|\hat A|-c\ga^{1/4}(\ga^{1/8}|A|)+c'\ga\ell_0|\hat A|
\end{eqnarray*}
Inserting this inequality in \eqref{eq:proto8}, we get \eqref{eq:bound8} for $\ga$ small enough.

\end{proof}

\vskip .5cm \noindent
\setcounter{equation}{0}
\section{Factorization theorem and large deviation estimate}
\label{section:largedeviation}
In this section we prove the factorization theorem \ref{thm:Peierls-factoriz-b} in a slightly different form,
proving at once factorization and control through the mean field functional, for which we prove  
the large deviation estimate needed to get the Peierls bound of Theorem \ref{thm:Peierls-0}.

Let $\Ga$ a $\p$-contour and define
\begin{eqnarray}
\label{def:gp}
\hG^{\p}:=\ssp(\Ga)\bigsqcup_{\q\ne \p} A^{\q}
\end{eqnarray}

\vskip .5cm \noindent
\begin{thm}
   \label{thm:Peierls-factoriz}
  \red{ There are $\bar \ga,
   \bar b$ and a constant $c$ such that for
all $\ga<\bar \ga, |\beta-\bcmf|<\bar b $}
:
       \begin{eqnarray}
       \label{eq:Peierls-factoriz}
       w^{\p}_{\ga,\beta}(\Ga;\vw_{A^{\p}})&\le &
       \frac{\prod_{\q\ne \p} Z^{\q}_{\ga,\beta,\Int^{q}(\Ga)\setminus A^{\q}}
       ({\vec\rho^{\q}})}{
       \prod_{\q\ne \p} Z^{\p}_{\ga,\beta,\Int^{\q}(\Ga)\setminus A^{\q}}(\vec\rho^{\p})}
       ~~\exp\{c\ga^{\frac{1}{2}}|\Ga|\}
       \\ \nn
       &&\hskip-1cm
       \cdot\exp\bigg\{-\beta\bigg[\inftwo{\vec\rho_{\hG^{\p}}:
       \eta(\vec\rho_{\hG^{\p}})=\eta_\Ga}{\vec\rho_{A^{\q}}=\vec\rho^{~\q}}
       F_{\ga,\beta,\hG^{\p}}
       (\vec\rho_{\hG^{\p}}|\vw^{(\ell_0)}_{A^{\p}})
-\inftwo{\vec\rho_{\hG^{\p}}:
       \eta(\vec\rho_{\hG^{\p}})=a^{p}}{\vec\rho_{A^{\q}}=\vec\rho^{~\p}}
       F_{\ga,\beta,\hG^{\p}}
       (\vec\rho_{\hG^{\p}}|\vw^{(\ell_0)}_{A^p})
       \bigg]\bigg\}
       \end{eqnarray}
\end{thm}
\vskip .5cm \noindent

In order to short  notations
we define $\cR(\Ga)$ and  $\cR(\not\Ga)$
as the two subsets of $L^\infty(\hG^{\p},S_Q)$ such that
\begin{eqnarray*}
\cR(\Ga):= \{\vec\rho\in L^\infty(\hG^{\p},S_Q):
\eta_x(\vec\rho)=\eta_\Ga(x)\; \forall x\in \ssp(\Ga);\quad \vec
\rho(x)= \vec\rho^{\,\q} \;\forall x\in A^{\q}\;\forall \q\ne\p\}
\end{eqnarray*}
\begin{eqnarray*}
\cR(\not\Ga):= \{\vec\rho\in L^\infty(\hG^{\p},S_Q):
\eta(\vec\rho,x)=a_{\p}\; \forall x\in \Ga;\quad
\rho (x)= \vec\rho^{\,\p} \;\forall x\in \bigsqcup_{\q\ne\p} A^{\q}\}
\end{eqnarray*}

We will prove also the following Theorem:

\begin{thm}
\label{thm:Peierls-3}
There exists $\bar\ga>0$ and a constant $c_f>0$ such that for all $\ga\le\bar\ga$
and all $\beta$ such that $|\beta-\bcmf|\le c_b \ga^{\frac{1}{2}}$,
the following large deviation estimate holds:
\begin{eqnarray*}
\inf_{\vec\rho\in \cR(\Ga)}
F_{\ga,\beta,\hG^{\p}}(\vec\rho|\vw^{(\ell_0)}_{A^{\p}})
- \inf_{\vec\rho\in \cR(\not\Ga)}
F_{\ga,\beta,\hG^{\p}}(\vec\rho|\vw^{(\ell_0)}_{A^{\p}})
\ge
 c_f |\Ga| \ga^{2a}\bigl(\frac{\ell_{-,\ga}}{\ell_{+,\ga}}\bigr)^d
+\sum_{\q\ne \p} \cI_{A^{\q}}(\vec \rho_{\q}^2 -\vec \rho_{\p}^2)
\end{eqnarray*}
where
\begin{eqnarray*}
\cI_{A^{\q}} = \frac{1}{2} \int_{A^{\q}}\int_{(\Int^{\q}(\Ga)\setminus A^{\q})} J_\ga(x,y) dx dy
\end{eqnarray*}
\end{thm}
\vskip .5cm \noindent

\vskip 1cm \noindent
\centerline{\em Proof of Theorem \ref{thm:Peierls-factoriz}}

\vskip .5cm \noindent
\label{proofofPeierls-factoriz}

\vskip .5cm \noindent
We first recall the definition of the weight of a $\p$-contour given in \eqref{weight}:
\begin{eqnarray*}
w^{\p}_{\ga,\beta}(\Ga;\vw_{A^{\p}}):=
\frac{\mu_{\ga,\beta,c(\Ga)\setminus
\Int^{\p}(\Ga)}(\cE(\Ga,\p)|{\vw_{A^{\p}}})}{\mu_{\ga,\beta,c(\Ga)\setminus
\Int^{\p}(\Ga)}(\cE(\not\Ga,\p)|{\vw_{A^{\p}}})}
\end{eqnarray*}
For each set $A^{\q}$, $\q\ne\p$, we denote by:

\begin{eqnarray}
\label{def: hatAq}
\hat A^{\q}&:=& A^{\q}\sqcup\delta_{\out}^{{\ell_{+,\ga}}}[A^{\q}]\\
\label{def: checkAq}
\check A^{\q}&:=& A^{\q}\sqcup\delta_{\out}^{{\ell_{+,\ga}/2}}[A^{\q}]\\
\label{def: Siq}
\Si^{\q,e}&:=&\bigl(\hat A^{\q}\setminus \check A^{\q}\bigr)\sqcap \ssp(\Ga)^c\\
\bar\Si^{\q,e}&:=&\bigl(\check A^{\q}\setminus A^{\q}\bigr)\sqcap \ssp(\Ga)^c\\
\Si^{e}&:=&\sqcup_{\q\ne\p}\Si^{\q,e}
\label{def: Siqe}
\end{eqnarray}

and also:
\begin{eqnarray*}
& &\Delta^{\p}:=\ssp(\Ga)\sqcup_{\q\ne\p}\check A^{\q}
\\
& &V^{\q}:=\Int^{\q}(\Ga)\setminus \hat A^{\q}
\end{eqnarray*}

We write:
\begin{eqnarray*}
w^{\p}_{\ga,\beta}(\Ga;\vw_{A^{\p}})&:=&
\frac{\dis{
\sum_{\vw_{\Si^{e}}:\vw_{\Si^{\q,e}}\in\cX^{\q}}e^{-\beta \sum_{\q\ne \p} H_\ga(\vw_{\Si^{\q,e}})}
Z_{\beta,\Delta^{\p}}(\cE(\Ga)|\vw_{\Si^{e}};\vw_{A^{\p}})
\prod_{\q\ne\p}
Z^{\q}_{\beta,V^{\q}}(\vw_{\Si^{\q,e}})
}}{\dis{\sum_{\vw_{\Si^{e}}\in\cX^{\p}}e^{-\beta H_\ga(\vw_{\Si^{e}})}
Z_{\beta,\Delta^{\p}}(\cE(\not\Ga)|\vw_{\Si^{e}};\vw_{A^{\p}})
\prod_{\q\ne\p}
Z^{\p}_{\beta,V^{\q}}(\vw_{\Si^{\q,e}})}}
\end{eqnarray*}

\vskip .5cm \noindent
By Theorem \ref{thm:app} with $\ell=\ell_0$, we have:
\begin{eqnarray}
\label{prf:app1}
& &w^{\p}_{\ga,\beta}(\Ga;\vw_{A^{\p}})\le e^{c\ga\ell_0|\Ga|}\cdot\\
&&\hskip.3cm
\frac{\dis{
\sum_{\vw_{\Si^{e}}:\vw_{\Si^{\q,e}}\in\cX^{\q}}e^{-\beta \sum_{\q\ne \p} H_\ga(\vw_{\Si^{\q,e}})}
\exp\{-\beta\inftwo{\rho: \eta(\rho)=\eta_\Ga ~r\in\ssp(\Ga)}{~\eta(\rho)=a_{\q} ~r\in\check A^{\q}}
F_{\ga,\beta,\Delta^{\p}}(\vec\rho|\vw^{\ell_0}_{\Si^{e}};
\vw^{\ell_0}_{A^{\p}})\}
\prod_{\q\ne\p}
Z^{\q}_{\beta,V^{\q}}(\vw_{\Si^{\q,e}})
}}{\dis{\sum_{\vw_{\Si^{e}}\in\cX^{\p}}e^{-\beta H_\ga(\vw_{\Si^{e}})}
\exp\{-\beta\inf_{\rho: \eta(\rho)=a_{\p}}
F_{\ga,\beta,\Delta^{\p}}(\vec\rho|\vw^{\ell_0}_{\Si^{e}};
\vw^{\ell_0}_{A^{\p}})\}
\prod_{\q\ne\p}
Z^{\p}_{\beta,V^{\q}}(\vw_{\Si^{\q,e}})}}
\nn\end{eqnarray}
Using Corollary \ref{corol:dimamica} with $\La=\check A^{\q}$, we can find a lower bound
for the free energy term in the numerator by considering density profiles $\vec\rho$ identically
equal to $\vec\rho^{\q}$ on $A^{\q}$,at the expense of a small error term; we have:
\begin{eqnarray*}
& &\inftwo{\rho: \eta(\rho)=\eta_\Ga ~r\in\ssp(\Ga)}{~\eta(\rho)=a_{\q} ~r\in\check A^{\q}}
F_{\ga,\beta,\Delta^{\p}}(\vec\rho|\vw^{\ell_0}_{\Si^{e}};\vw^{\ell_0}_{A^{\p}})\ge
\infthree{\rho: \eta(\rho)=\eta_\Ga \; r\in\ssp(\Ga)}{~\eta(\rho)=a_{\q} \; r\in\check A^{\q}}
{\vec\rho\equiv\vec\rho^{\q} \;  r\in  A^{\q}}
F_{\ga,\beta,\Delta^{\p}}(\vec\rho|\vw^{\ell_0}_{\Si^{e}};\vw^{\ell_0}_{A^{\p}})-c_\om e^{-\om \ga\ell_{+,\ga}/2}
|\check A|\\
& &=\inftwo{\rho: \eta(\rho)=\eta_\Ga \; r\in\ssp(\Ga)}{\vec\rho\equiv\vec\rho^{\q} \;  r\in  A^{\q}}
F_{\ga,\beta,G^{\p}}(\vec\rho|\vw^{\ell_0}_{\Si^{e}};\vw_{A^{\p}})
 +\sum_{\q}\inf_{\rho: \eta(\rho)=a_{\q}}
F_{\ga,\beta,\bar\Si^{\q,e}}(\vec\rho|\vw^{\ell_0}_{\Si^{e}};\vec\rho^{q}_{A^{\q}})
-c_\om e^{-\om \ga\ell_{+,\ga}/2}|\Ga|\\
& &\ge\inftwo{\rho: \eta(\rho)=\eta_\Ga \; r\in\ssp(\Ga)}{\vec\rho\equiv\vec\rho^{\q} \;  r\in  A^{\q}}
F_{\ga,\beta,G^{\p}}(\vec\rho|\vw^{\ell_0}_{\Si^{e}};\vw_{A^{\p}})
 -\frac{1}{\beta}\sum_{\q}\log\bigl(Z_{\beta,\bar\Si^{\q,e}}(\vw_{\Si^{e}};\vec\rho^{q}_{A^{\q}})\bigr)
- \bigl(c_\om e^{-\om \ga\ell_{+,\ga}/2} +c \ga\ell_0\bigr)|\Ga|
\end{eqnarray*}
The free energy term in the denominator of \eqref{prf:app1} can be directly bounded from above, as
\begin{eqnarray*}
& &\inf_{\rho: \eta(\rho)=a_{\p}}
F_{\ga,\beta,\Delta^{\p}}(\vec\rho|\vw^{\ell_0}_{\Si^{e}};\vw^{\ell_0}_{A^{\p}})
\le
\inftwo{\rho: \eta(\rho)=a_{\p}}{\vec\rho\equiv\vec\rho^{\q} \;  r\in  A^{\q}}
F_{\ga,\beta,\Delta^{\p}}(\vec\rho|\vw^{\ell_0}_{\Si^{e}};\vw^{\ell_0}_{A^{\p}})
\\
& &=\inf_{\rho: \eta(\rho)=a_{\p}}
F_{\ga,\beta,G^{\p}}(\vec\rho|\vw^{\ell_0}_{\Si^{e}};\vw_{A^{\p}})
 +\sum_{\q}\inf_{\rho: \eta(\rho)=a_{\p}}
F_{\ga,\beta,\bar\Si^{\q,e}}(\vec\rho|\vw^{\ell_0}_{\Si^{e}};\vec\rho^{\p}_{A^{\q}})
\\
& &\le
\inf_{\rho: \eta(\rho)=a_{\p}}
F_{\ga,\beta,G^{\p}}(\vec\rho|\vw^{\ell_0}_{\Si^{e}};\vw_{A^{\p}})
 -\frac{1}{\beta}\sum_{\q}\log\bigl(Z_{\beta,\bar\Si^{\q,e}}(\vw_{\Si^{e}};\vec\rho^{\p}_{A^{\q}})\bigr)
 +c \ga\ell_0|\Ga|
\end{eqnarray*}
Inserting both estimates in \eqref{prf:app1} we get the following factorization:
\begin{eqnarray*}
& &w^{\p}_{\ga,\beta}(\Ga;\vw_{A^{\p}})\\
& &\le e^{\beta(3 c\ga\ell_0+ c_\om e^{-\om \ga\ell_{+,\ga}/2})|\Ga|}\quad
\frac{\dis{
\sum_{\vw_{\Si^{e}}:\vw_{\Si^{\q,e}}\in\cX^{\q}} \prod_{\q\ne\p} e^{-\beta H_\ga(\vw_{\Si^{\q,e}})}
Z_{\beta,\bar\Si^{\q,e}}(\vw_{\Si^{e}};\vec\rho^{q}_{A^{\q}})
Z^{\q}_{\beta,V^{\q}}(\vw_{\Si^{\q,e}})}}
{\dis{
\sum_{\vw_{\Si^{e}}\in\cX^{\p}} \prod_{\q\ne\p} e^{-\beta H_\ga(\vw_{\Si^{\q,e}})}
Z_{\beta,\bar\Si^{\q,e}}(\vw_{\Si^{e}};\vec\rho^{\p}_{A^{\q}})
Z^{\p}_{\beta,V^{\q}}(\vw_{\Si^{\q,e}})
}}
\\& &\qquad\times
\exp\bigl\{-\beta \bigl(\inftwo{\rho: \eta(\rho)=\eta_\Ga \; r\in\ssp(\Ga)}{\vec\rho\equiv\vec\rho^{\q} \;  r\in  A^{\q}}
F_{\ga,\beta,G^{\p}}(\vec\rho|\vw^{\ell_0}_{\Si^{e}};\vw_{A^{\p}})- \inf_{\rho: \eta(\rho)=a^{p}}
F_{\ga,\beta,G^{\p}}(\vec\rho|\vw^{\ell_0}_{\Si^{e}};\vw_{A^{\p}})\bigr)\bigr\}
\\
& & = e^{3 \beta c\ga\ell_0|\Ga|+ \beta c_\om e^{-\om \ga\ell_{+,\ga}/2}}\quad
       \frac{\prod_{\q\ne \p} Z^{\q}_{\ga,\beta,\Int^{q}(\Ga)\setminus A^{\q}}
       ({\vec\rho^{\q}})}{
       \prod_{\q\ne \p} Z^{\p}_{\ga,\beta,\Int^{\q}(\Ga)\setminus A^{\q}}(\vec\rho^{\p})}
       \\ \nn
       &&\hskip-1cm
       \cdot\exp\bigg\{-\beta\bigg[\inftwo{\vec\rho_{\hG^{\p}}:
       \eta(\vec\rho_{\hG^{\p}})=\eta_\Ga}{\vec\rho_{A^{\q}}=\vec\rho^{~\q}}
       F_{\ga,\beta,\hG^{\p}}
       (\vec\rho_{\hG^{\p}}|\vw^{(\ell_0)}_{A^{\p}})
-\inftwo{\vec\rho_{\hG^{\p}}:
       \eta(\vec\rho_{\hG^{\p}})=a^{p}}{\vec\rho_{A^{\q}}=\vec\rho^{~\p}}
       F_{\ga,\beta,\hG^{\p}}
       (\vec\rho_{\hG^{\p}}|\vw^{(\ell_0)}_{A^p})
       \bigg]\bigg\}
\end{eqnarray*}

\begin{proof}[Proof of Theorem \ref{thm:Peierls-3}]

For each $\cD^{\ell_{+,\ga}}$-measurable cube $C$
in $\ssp(\Ga)$,
at least one of the two following events occurs (recall definition \eqref{def:Theta}):

$(a)$ there is $x$ in $\Ga$ such that $C_x^{\ell_{+,\ga}}\sim C$ and
$\eta_x^{\ell_{-,\ga}}=0$.\par

$(b)$ there are $x_1$, $x_2$ in  $\Ga$ such that  $C_{x_1}^{\ell_{+,\ga}}\sim C$,
$C_{x_2}^{\ell_{+,\ga}}\sim C$, $C_{x_1}^{\ell_{-,\ga}}\sim C_{x_2}^{\ell_{-,\ga}}$ and
$\eta_{x_1}^{\ell_{-,\ga}}\not=0$,$\eta_{x_2}^{\ell_{-,\ga}}\not=0$,$\eta_{x_1}^{\ell_{-,\ga}}\not=\eta_{x_2}^{\ell_{-,\ga}}$.\par

Let $N_\Ga$ the number of cubes of
$\cD^{\ell_{+,\ga}}$ contained in  $\ssp(\Ga)$. Since a single event
can be associated to at most $3^d$ cubes, there are
at least $3^{-d} N_\Ga$ distinct and nonintersecting
events.
Let consider a maximal family of nonintersecting events and
denote by $(x_0^i)_{i\in I_a}$, respectively
$(x_1^j,x_2^j)_{j\in I_b}$  a set of points (respectively a
set of pairs of points) characterizing the events of type
$a$ (respectively of type $b$). We also define
\begin{eqnarray}
&&B_a=\bigsqcup_{i\in I_a} C_{x_0^i}^{\ell_{-,\ga}}\\
&&B_b=\bigsqcup_{j\in I_b} (C_{x_1^j}^{\ell_{-,\ga}}\sqcup C_{x_2^j}^{\ell_{-,\ga}})
\end{eqnarray}

We have
\begin{eqnarray}
|I_a| +|I_b|\ge \frac{1}{3^d} N_\Ga
\end{eqnarray}
\begin{eqnarray}
|B_a| +|B_b|\ge \frac{1}{3^d} \bigl(\frac{\ell_{-,\ga}}{\ell_{+,\ga}}\bigr)^d|\Ga|
\end{eqnarray}

We separate $\hG^{\p}$ in two components $\hG^{\p}=\hG_1 \sqcup \hG_2$,
\begin{eqnarray}
\hG_1 &=& \delta_{\out}^{\ell_{+,\ga}}(A^{\p})\sqcap \ssp(\Ga)\\
\hG_2  &=& \hG^{\p} \setminus \hG_1
\end{eqnarray}
and write for all $\vec \rho$ in $\cR(\Ga)$
\begin{eqnarray}
F_{\ga,\beta,\hG^{\p}}(\vec\rho|\vw^{(\ell_0)}_{A^{\p}})=
F_{\ga,\beta,\hG_1}(\vec\rho_{\hG_1}|\vw^{(\ell_0)}_{A^{\p}})
+F_{\ga,\beta,\hG_2}(\vec\rho_{\hG_2}|\vec\rho_{\hG_1})
\end{eqnarray}
We first apply corollary \ref{corol:dimamica} with $\La =
\hG_1\setminus \delta_{\ins}^{\ga^{-1}}(\hG_1)$,
$\rho^*=\rho$ and $u=0$. Thus there are positive constants
$\omega$ and $c_\omega$ and there exists $\psi\in\cR(\Ga)$
such that
\begin{eqnarray}
F_{\ga,\beta,\hG_1}(\vec\rho_{\hG_1}|\vw^{(\ell_0)}_{A^{\p}})\ge
F_{\ga,\beta,\hG_1}(\vec\psi_{\hG_1}|\vw^{(\ell_0)}_{A^{\p}})
- c_\omega |\hG_1|e^{-\omega \ell_{+,\ga}/4}
\end{eqnarray}
with
\begin{eqnarray}
\label{ld-psiG1}
&&\vec\psi_{\hG_1} =
  \begin{cases}
    \vec\rho_{\hG_1} & \text{ on } \delta_{\ins}^{\ga^{-1}}(\hG_1), \\
    \vec\rho^{\p} & \text{ on } \hG_1\setminus \delta_{\ins}^{\ell_{+,\ga}/4}(\hG_1)
  \end{cases}
\end{eqnarray}
and
\begin{eqnarray}
F_{\ga,\beta,\hG_2}(\vec\rho_{\hG_2}|\vec\rho_{\hG_1}) =
F_{\ga,\beta,\hG_2}(\vec\rho_{\hG_2}|\vec\psi_{\hG_1})
\end{eqnarray}
Thus we have
\begin{eqnarray}
\label{ld-st1}
F_{\ga,\beta,\hG^{\p}}(\vec\rho|\vw^{(\ell_0)}_{A^{\p}})\ge
F_{\ga,\beta,\hG^{\p}}(\vec\psi|\vw^{(\ell_0)}_{A^{\p}})
-c_\omega |\hG_1|e^{-\omega \ell_{+,\ga}/4}
\end{eqnarray}
For $i\in I_a$, we write $C_i = C_{x_i}^{\ell_{-,\ga}}$ as a shorthand notation,
and define the function $\vec\psi^0$ on $\hG^{\p}$ as
\begin{eqnarray}
\vec\psi^0(r)=
  \begin{cases}
    \dis{\frac{1}{|C_i|}\int_{C_r^{\ell_{-,\ga}}} \vec\psi(r') dr'} & \text{ if } r\in B_a, \\
    \psi(r)  & \text{otherwise}.
  \end{cases}
\end{eqnarray}
From the definition of a contour, it follows that $\dist(B_a,\ssp(\Ga)^c)\ge\ga^{-1}$.
Thus the energy terms in the free energies of $\vec\psi$ and $\vec\psi^0$ differ by
the quantity
\begin{eqnarray}
&&|U_{\ga,\hG^{\p}}(\vec\psi|\vw^{(\ell_0)}_{A^{\p}})
-U_{\ga,\hG^{\p}}(\vec\psi^0|\vw^{(\ell_0)}_{A^{\p}})|\nn\\
&&=\frac{1}{2}\bigl|\int_{\ssp(\Ga)\times\ssp(\Ga)} J_\ga(x,y)
\vec\psi(x)\cdot\vec\psi(y) dx dy - \int_{\ssp(\Ga)\times\ssp(\Ga)}
J_\ga(x,y) \vec\psi^0(x)\cdot\vec\psi^0(y) dx dy\bigr|\nn\\
&&=\frac{1}{2}\bigl|\int_{\ssp(\Ga)\times\ssp(\Ga)} J_\ga(x,y)
(\vec\psi(x)+\vec\psi^0(x))\cdot(\vec\psi(y)-\vec\psi^0(y))
dx dy\bigr|\nn\\
&&=\frac{1}{2}\sum_{i\in I_a} \bigl|\frac{1}{|C_i|}
\int_{\ssp(\Ga)\times C_i} dx dy J_\ga(x,y)
(\vec\psi(x)+\vec\psi^0(x)) \cdot \int_{C_i} dy'
(\vec\psi(y)-\vec\psi(y'))\bigr|\nn\\
&&=\frac{1}{2}\sum_{i\in I_a} \bigl|\frac{1}{|C_i|}
\int_{\ssp(\Ga)\times C_i} dx dy (\vec\psi(x)+\vec\psi^0(x))
\cdot \vec\psi(y) \int_{C_i} dy'
(J_\ga(x,y)- J_\ga(x,y'))\bigr|\nn\\
&&\le\frac{1}{2}\sum_{i\in I_a} \frac{1}{|C_i|}
\int_{\ssp(\Ga)\times C_i} dx dy |(\vec\psi(x)+\vec\psi^0(x))
\cdot \vec\psi(y)| \int_{C_i} dy'
|J_\ga(x,y)- J_\ga(x,y')|\nn\\
&&\le 2^d\sqrt{d} \ell_{-,\ga} \ga \|\nabla \cJ\|_\infty
|B_a|\nn
\end{eqnarray}
where in the last inequality we used
\begin{eqnarray}
|(\vec\psi(x)+\vec\psi^0(x)) \cdot \vec\psi(y)|\le 2
\end{eqnarray}
and
\begin{eqnarray}
|J_\ga(x,y)- J_\ga(x,y')|\le \sqrt{d} \ell_{-,\ga}
\ga^{d+1} \|\nabla \cJ\|_\infty {\bf 1}_{\{|x-y|\le
2\ga^{-1}\}}
\end{eqnarray}
that for all $y$, $y'$ in $C_i$.
By concavity of the entropy, $I(\vec\psi)\le I(\vec\psi^0)$
and we get
\begin{eqnarray}
\label{ld-st2}
F_{\ga,\beta,\hG^{\p}}(\vec\psi|\vw^{(\ell_0)}_{A^{\p}})\ge
F_{\ga,\beta,\hG^{\p}}(\vec\psi^0|\vw^{(\ell_0)}_{A^{\p}})
-2^d\sqrt{d} \ell_{-,\ga} \ga \|\nabla \cJ\|_\infty |B_a|
\end{eqnarray}
We look for a lower bound of the free energy of
$\vec\psi^0$.
We again divide $\hG^{\p}$ in two parts
$\hG^{\p}=\hG_1'\sqcup\hG_2'$ where
\begin{eqnarray}
&&\hG_1'=\delta_{\out}^{\ell_{+,\ga}/2}(A^{\p})\sqcap\ssp(\Ga)\\
&&\hG_2'= \hG\setminus \hG_1'
\end{eqnarray}
We have
\begin{eqnarray}
F_{\ga,\beta,\hG^{\p}}(\vec\psi^0|\vw^{(\ell_0)}_{A^{\p}})=
F_{\ga,\beta,\hG_1'}(\vec\psi^0_{\hG_1'}|\vw^{(\ell_0)}_{A^{\p}})
+F_{\ga,\beta,\hG_2'}(\vec\psi^0_{\hG_2'}|\vec\psi^0_{\hG_1'})
\end{eqnarray}
We write the second term as:
\begin{eqnarray}
&&F_{\ga,\beta,\hG_2'}(\vec\psi^0_{\hG_2'}|\vec\psi^0_{\hG_1'})=
\int_{\hG_2'} dx \phi^{mf}_{\beta}(\vec\psi^0(x))
+\frac{1}{4}\int_{{\hG_2'}\times{\hG_2'}} dx dy J_\ga(x,y)
|\vec\psi^0(x)-\vec\psi^0(y)|^2 \nn\\
&&-\int_{{\hG_2'}\times{\hG_1'}} dx dy J_\ga(x,y) (\vec\psi^0(x)\cdot\vec\psi^0(y))
+\frac{1}{2}\int_{{\hG_2'}\times{\hG_2'^c}} dx dy J_\ga(x,y) |\vec\psi^0(x)|^2
\end{eqnarray}
The last two terms can be calculated since,
by \eqref{ld-psiG1}, $\vec\psi^0$ is constant and equal to $\vec\rho^{\p}$ on both
$\delta_{\out}^{\ga^{-1}}(\hG_1')\sqcap \hG_2'$ and $\hG_1'\sqcap\delta_{\out}^{\ga^{-1}}(\hG_2')$,
and equal to $\vec\rho^{\q}$ on $A^{\q}$, $\q\ne \p$. We get:
\begin{eqnarray}
F_{\ga,\beta,\hG_2'}(\vec\psi^0_{\hG_2'}|\vec\psi^0_{\hG_1'})=
\int_{\hG_2'} dx \phi^{\mf}_{\beta}(\vec\psi^0(x))
+\frac{1}{4}\int_{{\hG_2'}\times{\hG_2'}} dx dy J_\ga(x,y)
|\vec\psi^0(x)-\vec\psi^0(y)|^2 \nn\\
+\frac{1}{2}\sum_{\q\ne\p}\int_{{\hG_2'}\times (\Int^{\q}(\Ga)\setminus A^{\q})} dx dy J_\ga(x,y) |\vec\rho^{\q}|^2\nn
-\frac{1}{2}\int_{{\hG_2'}\times {\hG_1'}} dx dy J_\ga(x,y) |\vec\rho^{\p}|^2
\end{eqnarray}
Since $B_b\sqsubset \hG_2'$, the second term can be bounded by the contributions of the $b$-events.
We have the following estimate:
\begin{eqnarray}
&&\frac{1}{4}\int_{\hG_2'\times\hG_2'} dx dy J_\ga(x,y)
|\vec\psi^0(x)-\vec\psi^0(y)|^2 \nn\\
&&\ge \frac{1}{4}\sum_{j\in I_b}  \bigl(\int_{\hG_2'\times C_j}
dx dy J_\ga(x,y) |\vec\psi^0(x)-\vec\psi^0(y)|^2+
\int_{\hG_2'\times C_j'} dx dy' J_\ga(x,y')
|\vec\psi^0(x)-\vec\psi^0(y')|^2\bigr)\nn\\
&&=\frac{1}{4}\sum_{j\in I_b} \frac{1}{|C_j|}\int_{\hG_2'}
dx \int_{C_j\times C_j'} dy dy' \bigl(J_\ga(x,y)
|\vec\psi^0(x)-\vec\psi^0(y)|^2+  J_\ga(x,y')
|\vec\psi^0(x)-\vec\psi^0(y')|^2\bigr)\nn
\end{eqnarray}
where we used the notations $C_j=C_{x_1^j}^{\ell_{-,\ga}}$,
$C_j'=C_{x_2^j}^{\ell_{-,\ga}}$ for all $j \in I_b$.
Using now the fact that
for all $(y,y')$ in $C_j\times C_j'$,
\begin{eqnarray}
\bigl|J_\ga(x,y)- J_\ga(x,y')\bigr|
\le 2\sqrt{d} \ell_{-,\ga} \ga^{d+1} \|\nabla \cJ\|_\infty
{\bf 1}_{\{|x-y|\le2\ga^{-1}\}}
\end{eqnarray}
we get
\begin{eqnarray}
&&\frac{1}{4}\int_{\hG_2'\times\hG_2'} dx dy J_\ga(x,y)
|\vec\psi^0(x)-\vec\psi^0(y)|^2 \\
&&\ge\frac{1}{4}\sum_{j\in I_b} \int_{\hG_2'} dx
\frac{1}{2|C_j|}\cdot \\
& &\hskip1cm
\int_{C_j\times C_j'} dy dy' (J_\ga(x,y)+J_\ga(x,y'))
\bigl(|\vec\psi^0(x)-\vec\psi^0(y)|^2+|\vec\psi^0(x)-\vec\psi^0(y')|^2\bigr)\nn\\
&&\hskip1cm-2^{d-1} \sqrt{d} \ell_{-,\ga} \ga \|\nabla \cJ\|_\infty |B_b|\nn
\end{eqnarray}

Now using successively the inequality
\begin{eqnarray}
|a-b|^2+|a-c|^2\ge\frac{1}{2}|b-c|^2
\end{eqnarray}
and $\dist(B_b,\hG_2'^c)\ge\ga^{-1}$, we can sum over the $x$ variable to obtain

\begin{eqnarray}
&&\frac{1}{4}\int_{\hG_2'\times\hG_2'} dx dy J_\ga(x,y)
|\vec\psi^0(x)-\vec\psi^0(y)|^2 \\
&&\ge\frac{1}{8}\sum_{j\in I_b} \int_{\hG_2'} dx
\frac{1}{2|C_j|}
\int_{C_j\times C_j'} dy dy' (J_\ga(x,y)+J_\ga(x,y'))|\vec\psi^0(y)-\vec\psi^0(y')|^2\nn\\
&&-2^{d-1} \sqrt{d} \ell_{-,\ga} \ga \|\nabla \cJ\|_\infty |B_b|\nn\\
&&\ge\frac{1}{8}\sum_{j\in I_b} \frac{1}{|C_j|}
\int_{C_j\times C_j'} dy dy'
|\vec\psi^0(y)-\vec\psi^0(y')|^2
 -2^{d-1} \sqrt{d} \ell_{-,\ga} \ga \|\nabla \cJ\|_\infty
|B_b|\nn
\end{eqnarray}
We finally get
\begin{eqnarray}
\label{ld-st3}
&&\frac{1}{4}\int_{\hG_2'\times\hG_2'} dx dy J_\ga(x,y)
|\vec\psi^0(x)-\vec\psi^0(y)|^2\nn\\
&&\ge\frac{1}{8}\sum_{j\in I_b} \frac{1}{|C_j|}
\bigl|\int_{C_j} dy \vec\psi^0(y)- \int_{C_j'}
dy'\vec\psi^0(y')\bigr|^2
 -2^{d-1} \sqrt{d} \ell_{-,\ga} \ga \|\nabla \cJ\|_\infty
|B_b|\nn
\end{eqnarray}
Now collecting all estimates \eqref{ld-st1},\eqref{ld-st2} and \eqref{ld-st3},
we get for all $\vec\rho$ in
$\cR(\Ga)$:
\begin{eqnarray}
&&F_{\ga,\beta,\hG^{\p}}(\vec\rho|\vw^{(\ell_0)}_{A^{\p}})\ge
F_{\ga,\beta,\hG_1'}(\vec\psi^0_{\hG_1'}|\vw^{(\ell_0)}_{A^{\p}})
+ \sum_{\q\ne\p}\cI_{A^{\q}}|\vec\rho^q|^2\nn
-\frac{1}{2}\int_{{\hG_2'}\times {\hG_1'}} dx dy J_\ga(x,y) |\vec\rho^{\p}|^2\\
&&+ \int_{\hG_2'} dx \phi_\beta^{\mf}(\vec\psi^0(x)) +
\frac{1}{8}\sum_{j\in I_b} \frac{1}{|C_j|} \bigl|\int_{C_j}
dy' \vec\psi^0(y')- \int_{C_j'} dy''\vec\psi^0(y'')\bigr|^2\nn\\
&&-c_\omega|\hG_1|e^{-\omega \ell_{+,\ga}/4}
 -2^d\sqrt{d}\ell_{-,\ga} \ga \|\nabla \cJ\|_\infty |B_a|
-2^{d-1} \sqrt{d} \ell_{-,\ga} \ga \|\nabla \cJ\|_\infty
|B_b|\nn
\end{eqnarray}
Using the definitions of $B_a$ and $B_b$ and the results of appendix A, we have
for all $\beta$ such that $\beta-\bcmf\le c_b \ga^{\frac{1}{2}}$,
\begin{eqnarray}
\phi^{mf}_{\beta}(\vec\psi^0(x))\ge \phi^{mf}_{\beta}(\vec\rho^{\p})
+ (\frac{Q}{\beta}-1) \ga^{2a}- c \ga^{\frac{1}{2}}
\end{eqnarray}
for all $x$ in $B_a$ and
\begin{eqnarray}
\frac{1}{|C_j|^2} \bigl|\int_{C_j} dy \vec\psi^0(y)-
\int_{C_j'} dy'\vec\psi^0(y')\bigr|^2 \ge (\frac{Q-1}{Q}-
2\ga^a)^2
\end{eqnarray}
on all $j$ in $I_b$
Hence
\begin{eqnarray}
&&F_{\ga,\beta,\hG^{\p}}(\vec\rho|\vw^{(\ell_0)}_{A^{\p}})\ge
F_{\ga,\beta,\hG_1'}(\vec\psi|\vw^{(\ell_0)}_{A^{\p}})
+ |\hG_2'|  \phi^{\mf}_{\beta}(\vec\rho^{\p})
+ \sum_{\q\ne\p}\cI_{A^{\q}} |\vec\rho^q|^2\nn
-\frac{1}{2}\int_{{\hG_2'}\times {\hG_1'}} dx dy J_\ga(x,y) |\vec\rho^{\p}|^2\\
&&+(\frac{Q}{\beta}-1) \ga^{2a} |B_a|
 +\frac{1}{4}(\frac{Q-1}{2Q}-\ga^a)^2 |B_b|\nn\\
  &&-c_\omega|\hG_1|e^{-\omega \ell_{+,\ga}/4}
 -2^d\sqrt{d}\ell_{-,\ga} \ga \|\nabla \cJ\|_\infty |B_a|- c\ga^{\frac{1}{2}}
-2^{d-1} \sqrt{d} \ell_{-,\ga} \ga \|\nabla \cJ\|_\infty
|B_b|\nn
\end{eqnarray}
Now for $a<\min(\frac{1}{4},\frac{\alpha}{2})$, we may choose $\ga$ so that the last three error terms are a fraction of
the respective gain terms. In addition, since we have both $|B_a|+|B_b| \ge
\frac{1}{3^d}\big(\frac{\ell_{-,\ga}}{\ell_{+,\ga}}\bigr)^d|\Ga|$
and $\hG_1\sqsubset\Ga$, the remaining error term can be also compensated for $\ga$ small enough.
We define $\bar\ga$ as the largest
value of  $\ga$ such that the following inequalities hold
simultaneously
\begin{eqnarray}
&&2^{d-1} \sqrt{d} \ell_{-,\ga} \ga \|\nabla \cJ\|_\infty\le
\frac{1}{3} (\frac{Q-1}{2Q}-\ga^a)^2\nn\\
&& 2^d\sqrt{d}\ell_{-,\ga} \ga \|\nabla \cJ\|_\infty \le
\frac{1}{3} (\frac{Q}{\beta}-1) \ga^{2a}\nn\\
&&c_\omega e^{-\omega \ell_{+,\ga}/4} \le
\frac{1}{3} (\frac{Q}{\beta}-1) \ga^{2a}\nn\\
&&\frac{1}{4}(\frac{Q-1}{2Q}-\ga^a)^2 \ge \frac{1}{3}
(\frac{Q}{\beta}-1) \ga^{2a}\nn
\end{eqnarray}
For all $\ga\le\bar\ga$, we have
\begin{eqnarray}
&&F_{\ga,\beta,\hG^{\p}}(\vec\rho|\vw^{(\ell_0)}_{A^{\p}})\ge
F_{\ga,\beta,\hG_1'}(\vec\psi|\vw^{(\ell_0)}_{A^{\p}})
 + |\hG_2'|  \phi^{mf}_{\beta}(\vec\rho^{\p})
 -\frac{1}{2}\int_{{\hG_2'}\times {\hG_1'}} dx dy J_\ga(x,y) |\vec\rho^{\p}|^2\nn\\
&&+ \sum_{\q\ne\p}\cI_{A^{\q}}|\vec\rho^q|^2+
\frac{1}{3^{d+1}}(\frac{Q}{\beta}-1) \ga^{2a}
\bigl(\frac{\ell_{-,\ga}}{\ell_{+,\ga}}\bigr)^d|\Ga|
\end{eqnarray}
Now consider the function $\vec\varphi$ in
$L^\infty(\hG^{\p},S_Q)$ defined as
\begin{eqnarray}
\vec\varphi(r) =
  \begin{cases}
    \vec\psi(r) & r\in\hG_1', \\
    \vec\rho^{\p} & r\in\hG_2'.
  \end{cases}
\end{eqnarray}
$\vec\varphi(r)$ belongs clearly to $\cR(\not\Ga)$, and
since $\vec\varphi(r)=\vec\rho^{\p}$ on
$\hG_1'\sqcap\delta_{\out}^{\ga^{-1}}(\hG_2')$, its
free energy reads
\begin{eqnarray}
&&F_{\ga,\beta,\hG^{\p}}(\vec\varphi|\vw^{(\ell_0)}_{A^{\p}})=
F_{\ga,\beta,\hG_1'}(\vec\varphi|\vw^{(\ell_0)}_{A^{\p}})\nn\\
&&+ |\hG_2'|\phi^{mf}_{\beta}(\vec\rho^{\p})
-\frac{1}{2}\int_{{\hG_2'}\times {\hG_1'}} dx dy J_\ga(x,y) |\vec\rho^{\p}|^2
+\sum_{\q\ne\p}\cI_{A^{\q}}|\vec\rho^p|^2\nn
\end{eqnarray}
Hence we
have for all $\vec\rho$ in $\cR(\Ga)$,
\begin{eqnarray}
\label{ld-st4}
&&F_{\ga,\beta,\hG^{\p}}(\vec\rho|\vw^{(\ell_0)}_{A^{\p}})\ge
F_{\ga,\beta,\hG^{\p}}(\vec\varphi|\vw^{(\ell_0)}_{A^{\p}})\nn\\
&&+\sum_{\q\ne\p}\cI_{A^{\q}}( |\vec\rho^{\q}|^2 -
|\vec\rho^{\p}|^2) + \frac{1}{3^{d+1}}(\frac{Q}{\beta}-1) \ga^{2a}
\bigl(\frac{\ell_{-,\ga}}{\ell_{+,\ga}}\bigr)^d|\Ga|
\end{eqnarray}
and since $\vec\varphi$ is in $\cR(\not\Ga)$,
\begin{eqnarray}
F_{\ga,\beta,\hG^{\p}}(\vec\varphi|\vw^{(\ell_0)}_{A^{\p}})\ge
\inf_{\rho\in \cR(\not\Ga)} F_{\ga,\beta,\hG^{\p}}(\vec\rho|\vw^{(\ell_0)}_{A^{\p}})
\end{eqnarray}
Taking now the infimum over $\cR(\Ga)$ in \eqref{ld-st4}, we get for all $\ga\le\bar\ga$,
\begin{eqnarray}
&&\inf_{\vec\rho\in\cR(\Ga)}F_{\ga,\beta,\hG^{\p}}(\vec\rho|\vw^{(\ell_0)}_{A^{\p}})\ge
\inf_{\rho\in \cR(\not\Ga)} F_{\ga,\beta,\hG^{\p}}(\vec\rho|\vw^{(\ell_0)}_{A^{\p}})\nn\\
&&+\sum_{\q\ne\p}\cI_{A^{\q}}( |\vec\rho^{\q}|^2 -
|\vec\rho^{\p}|^2) + c_f
\ga^{2a}
\bigl(\frac{\ell_{-,\ga}}{\ell_{+,\ga}}\bigr)^d|\Ga|
\end{eqnarray}
with
\begin{eqnarray}
c_f=\frac{1}{3^{d+1}}(\frac{Q}{\beta}-1)
\end{eqnarray}
\end{proof}

\vskip 2.5cm \noindent
\centerline{\bf Acknowledgments }

We thank Marzio Cassandro, Roberto Fernandez, Joel Lebowitz and
Errico Presutti for many helpful discussions. T.G. thanks the warm hospitality
of the Mathematics departments of Universities of Rome 2 and l'Aquila where most of
the work has been done.

\vskip 2.5cm \noindent
\appendix

\vskip 2.5cm \noindent
\section{Mean Field Model}
\setcounter{equation}{0}
\label{app:meanfield} In this appendix, we review briefly
the mean field theory of the Potts model
\cite{Wu}, and derive the various quantities
needed in the rest of the paper.

We consider a $Q$-state Potts model defined on a complete
graph with $N$ sites and derive its behavior in the large
$N$ limit. A variable $\sigma_i$, $\sigma_i\in
[1,\cdots,Q]$, is attached to each site $i$ of the graph so
that the space of configurations is
$\Omega_N=[1,\cdots,Q]^N$. The mean field
Hamiltonian on $\Omega_N$ is:

\begin{eqnarray*}
H^{\mf}(\si):=-\frac{1}{2 N}\sum_{i\ne j}
\Ii_{\{\si_i=\si_j\}}
\end{eqnarray*}
$H^{\mf}$ is invariant under any permutation of sites so
that its value on a given configuration depends only on the
number of sites with color $q$, says $N_q$,
$q\in[1,\cdots,Q]$. The partition function of the model is thus:

\begin{eqnarray} Z_{N,\beta}=\sum_{\{N_q\}:\sum N_q
=N}\frac{N!}{\prod_q N_q!} e^{\frac{\beta}{2N}\sum_q N_q
(N_q -1)}
\end{eqnarray} For $N$ large, $Z_{N,\beta}$ is
dominated by the configurations which realize the minimum
of the free energy density $\phi^{\mf}_\beta$ defined on
$S_Q=\{\vec\rho\in\Rr_+^Q,\sum_q\rho_q=1\}$ as
\begin{eqnarray}
\label{def:mean-field-free-energy}
\phi^{\mf}_\beta(\vec\rho) = -\frac{1}{2}\sum_q \rho_q^2
+\frac{1}{\beta} \sum_q\rho_q \ln (\rho_q)
\end{eqnarray}
with the correspondence $\rho_q=\frac{N_q}{N}$.
In fact, for our present purpose, we are also interested in all local
minimizers of $\phi^{\mf}_\beta$, which appear to be of two kinds:
one ``disordered'' (or uniform) state in which all colors have the same density
and $Q$ degenerated ``ordered'' (or colored) states in which one color dominates.
As can be expected, the first one exists for small values of $\beta$,
while the other $Q$ exist only for $\beta$ large enough. In addition,
there is a critical value of $\beta$ which determines which kind of local minimizer
is the actual absolute minimizer for $\phi^{\mf}_\beta$. We make these statements
precise in the two following theorems. We first characterize all local minimizers:
\begin{thm}
   \label{thm:mflocalminimizers}
For all $Q>2$, there exists $\beta_0<Q$ such that the mean field free energy $\phi^{\mf}_\beta$
\eqref{def:mean-field-free-energy} on $S_Q$ has exactly:
{\obeylines
$\bullet$ 1 local minimizer for $\beta\le\beta_0$,
$\bullet$ $Q+1$ local minimizers for $\beta_0<\beta<Q$, and
$\bullet$ $Q$ local minimizers for $\beta\ge Q$.
}
\noindent These minimizers are of two kinds and characterized as follows:
\begin{itemize}
  \item
  For all $\beta<Q$, the uniform state $\vec\rho^{(-1)}$, with components
  \begin{eqnarray}
  \label{mf:rhod}
\rho^{(-1)}_q=\frac{1}{Q} \quad\hbox{ for all } q\in\{1,\cdots,Q\}
  \end{eqnarray}
  and free energy
  \begin{eqnarray}
  \label{mfphi(rhod)}
\phi^{\mf}_\beta(\vec\rho^{(-1)})=\frac{-1}{2Q}-\frac{1}{\beta}\log(Q)
  \end{eqnarray}
  \item
  For all $\beta>\beta_0$, $Q$ colored states $\vec\rho^{(p)}\equiv\vec\rho^{(p)}_\beta$, $p\in\{1,\cdots,Q\}$,
  with components
  \begin{eqnarray}
  \label{mf:rhoo}
\rho^{(p)}_q=
  \begin{cases}
    \rho_A & \text{ if } q=p, \\
    \rho_B & \text{otherwise}.
  \end{cases}
  \end{eqnarray}
where $(\rho_A,\rho_B)$, $\rho_A>\rho_B$, is the solution of the set of equations
\begin{eqnarray}
\label{mf:defrhoo1}
\frac{\log(\rho_A)-\log(\rho_B)}{\rho_A-\rho_B}&=& \beta\\
\label{mf:defrhoo2}
\rho_A+ (Q-1)\rho_B &=& 1
\end{eqnarray}
which verify
\begin{eqnarray}
\label{mf:defrhoo3}
Q\beta \rho_A \rho_B <1
\end{eqnarray}
These $Q$ states are degenerate and have free energy
\begin{eqnarray}
\label{mf:phi(rhoo)}
\phi^{\mf}_\beta(\vec\rho^{(p)})=\frac{-1}{2}Q\rho_A\rho_B+\frac{1}{2\beta}\log(\rho_A\rho_B)
\end{eqnarray}
\end{itemize}
\end{thm}
The mean field first order transition is described in the following
\begin{thm}
   \label{thm:mfabsoluteminimizer}
For all $Q>2$, there exists a critical value of $\beta$, in $(\beta_0,Q)$,
\begin{equation}
\label{mf:bcmf}
\bcmf\equiv \frac{2(Q-1)}{Q-2}\log(Q-1)
\end{equation}
such that $\phi^{\mf}_\beta$ has:
{\obeylines
$\bullet$ 1 minimizer $\vec\rho^{(-1)}$ for all $\beta<\bcmf$;
$\bullet$ $Q+1$ minimizers $\vec\rho^{(\p)}$, $\p\in\{-1,1,\cdots,Q\}$ for $\beta=\bcmf$;
$\bullet$ $Q$ minimizers $\vec\rho^{(p)}$, $p\in\{1,\cdots,Q\}$ for $\beta>\bcmf$.
}
\end{thm}

Finally, the relevance of local minimizers in our problem arises from their local stability, which is stated
in the following theorem:
\begin{thm}
   \label{thm:mflocalstability}
For all $Q>2$ and all $\beta$ in $(\beta_0,Q)$, the map
       \begin{equation}
\label{contract}
\vec\rho \rightarrow \vec g(\vec\rho)
       \end{equation}
       where
       \begin{eqnarray}\dis
       \label{mf:g}
g_q(\vec\rho)=\frac{\exp(\beta\rho_q)}{\sum_{p=1}^Q\exp(\beta\rho_p)}
       \end{eqnarray}
is a contraction around the $Q+1$ local minimizers $\vec\rho^{(\p)}$ of the mean field free energy
$\phi^{\mf}_\beta$. In particular,
\begin{eqnarray}
\sup_{q,q'} \bigl|\frac{\partial g_q(\vec\rho)}{\partial\rho_{q'}}\bigr|\le 1-\frac{1}{2Q}
\end{eqnarray}
for all $\vec\rho$ such that $\sup_q|\rho_q -\rho_q^{\p}|\le \frac{1}{4\beta^2Q^2}$ for some $\p$.

\end{thm}
\vskip2.0truecm

\begin{proof}{\bf of Theorem \ref{thm:mflocalminimizers}}:

We consider the variational
problem for $\phi^{\mf}_\beta(\vec\rho)$.
\begin{eqnarray*}
\inf_{\vec \rho: \sum_{q=1}^Q\rho_q=1}\phi^{\mf}_\beta(\vec
\rho)= {\min_{\vec\rho,\la}}^{\loc}\left[\phi^{\mf}_\beta(\vec
\rho)+\la\left(\sum_{j=q}^Q\rho_q-1\right)\right]
\end{eqnarray*}
where $\lambda$ is a Lagrange parameter associated to the
constraint $\sum_q \rho_q=1$. Since the gradient of the
free energy points inward the simplex $S_Q$, the (local) minima
cannot stay on the boundary of $S_Q$ and are thus solutions
of the set of equations:
\begin{eqnarray}
\label{mf:criticalpoints}
\frac{\partial \phi^{\mf}_\beta(\vec \rho)}{\partial \rho_q}
+\la = 0,\qquad
q=1,\cdots, Q
\end{eqnarray}
together with the condition
\begin{eqnarray}
\label{mf:minima}
\sum_{q1,q2 =1}^Q\frac{\partial^2 \phi^{\mf}_\beta(\vec \rho)}{\partial \rho_{q_1}\rho_{q_2}} x_{q_1}x_{q_2} \ge 0
\end{eqnarray}
for all $\vec x\in\Rr^Q$ such that $\sum_{q=1}^Q x_q =0$.

Explicitly the first derivatives of the free energy read
\begin{eqnarray}
\frac{\partial \phi^{\mf}_\beta(\vec \rho)}{\partial \rho_q}
\equiv 1-\rho_q+\frac{1}{\beta}(\ln\rho_q+1)
\end{eqnarray}
while the Hessian matrix of $\phi^{\mf}_\beta(\vec \rho)$ is diagonal and
\begin{eqnarray}
\label{mf:hessian}
\frac{\partial^2 \phi^{\mf}_\beta(\vec \rho)}{\partial \rho_{q_1}\rho_{q_2}}
=\delta_{q_1,q_2} \bigl(-1 +\frac{1}{\beta \rho_{q_1}}\bigr)
\end{eqnarray}
As a function of $\rho_q$ alone, $\frac{\partial
\phi^{\mf}_\beta(\vec \rho)}{\partial \rho_q}$ is a strictly
concave $C^\infty$ function and hence cannot take the same value more than twice.
Thus there are two kind of solutions for \eqref{mf:criticalpoints},
depending on whether $\rho_q$ takes one or two values.

The first case correspond to a ``disordered'' solution
$\vec\rho^{(-1)}$ in which each color has the same density:
\begin{eqnarray}
\rho^{(-1)}_q &=& \frac{1}{Q}\qquad \hbox{for all } q=1,\cdots,Q\\
\phi^{\mf}_\beta(\vec\rho^{(-1)})&=&\frac{-1}{2Q}-\frac{1}{\beta}\ln(Q)
\end{eqnarray}
Using \eqref{mf:hessian}, $\vec\rho^{(-1)}$ is a local minimum of $\phi^{\mf}_\beta(\vec
\rho)$ if and only if $\beta< Q$\vskip .5cm

In the second case, let $\vec\rho$ a vector in $S_Q$ which components takes two values,
 says $\rho_A$ and $\rho_B$ with $\rho_A>\rho_B$, and let $n$, $0<n<Q$, the number of components
 equal to $\rho_A$ (the remaining $Q-n$ are thus equal to $\rho_B$).
According to \eqref{mf:criticalpoints}, $\vec\rho$ is a critical point
for $\phi^{\mf}_\beta(\vec \rho)$ in $S_Q$ whenever the constraint is
satisfied and both $\rho_A$ and $\rho_B$ are associated to
the same value of the Lagrange parameter:
\begin{eqnarray}
\label{os1}
&& n\rho_A+(Q-n)\rho_B=1\\
\label{os2}
 &&\lambda=\rho_A-1-\frac{1}{\beta}(\ln\rho_A+1)=\rho_B-1
-\frac{1}{\beta}(\ln\rho_B+1)
\end{eqnarray}
However, all such points are not local minima: equation \eqref{os2} implies the relation \eqref{mf:defrhoo1}
between $\beta$,
$\rho_A$ and $\rho_B$,
\begin{eqnarray}
\label{mf:betaAB}
\beta=\frac{\ln \rho_A -\ln \rho_B}{\rho_A - \rho_B}
\end{eqnarray}
and by concavity of the logarithmic function, we have necessarily
\begin{eqnarray}
\frac{1}{\rho_A}<\beta<\frac{1}{\rho_B}
\end{eqnarray}
Thus the second derivative of $\phi^{\mf}_\beta(\vec \rho)$ is negative in each direction in which
$\rho_q=\rho_A$. This is obviously not compatible with the condition \eqref{mf:minima} for $n\ge 2$.
In the case $n=1$, the constraint \eqref{os2} reduces to \eqref{mf:defrhoo2}:
\begin{eqnarray}
\label{mf:QAB}
\rho_A+(Q-1)\rho_B=1
\end{eqnarray}
while the condition \eqref{mf:minima} can be made explicit using \eqref{mf:hessian}:
\begin{eqnarray*}
\bigl(-1 +\frac{1}{\beta \rho_A}\bigr)+\frac{1}{Q-1}\bigl(-1 +\frac{1}{\beta \rho_B}\bigr)\ge 0
\end{eqnarray*}
or equivalently using the constraint \eqref{mf:QAB}, one gets \eqref{mf:defrhoo3}:
\begin{eqnarray}
\label{mf:stabilite}
\beta Q\rho_A\rho_B\le 1
\end{eqnarray}
We need to find out all triples $(\beta,\rho_A,\rho_B)$ which solve simultaneously
\eqref{mf:betaAB}, \eqref{mf:QAB} and \eqref{mf:stabilite}. For $\rho_A$ in $(\frac{1}{Q},1)$,
we consider the function $\tilde\beta(\rho_A)$ as
\begin{eqnarray}
\label{mf:beta(rhoA)}
\tilde\beta(\rho_A)=\frac{\ln \rho_A -\ln \rho_B}{\rho_A - \rho_B}
\end{eqnarray}
where $\rho_B$ is taken implicitly as a function of $\rho_A$, through \eqref{mf:QAB}
Its first derivative reads
\begin{eqnarray}
\frac{\partial \tilde\beta(\rho_A)}{\partial \rho_A}=\frac{1}{(Q-1)(\rho_A-\rho_B)}
\bigl[\frac{1}{\rho_A \rho_B} -\beta Q\bigr]
\end{eqnarray}
while the second derivative can be cast in the form
\begin{eqnarray}
\frac{\partial^2 \tilde\beta(\rho_A)}{\partial \rho_A^2}=
\frac{\varphi(\rho_A)-\varphi(\rho_B)}{(Q-1)^2(\rho_A-\rho_B)^3}
\end{eqnarray}
where
\begin{eqnarray}
\varphi(\rho)=2Q^2\log(\rho)+\frac{4Q}{\rho}-\frac{1}{\rho^2}
\end{eqnarray}
One can check that the function $\varphi(\rho)$ is strictly increasing and thus the second derivative
of $\tilde\beta(\rho_A)$ is always positive. It follows that
there exists a unique value $\rho_0$ in $(\frac{1}{Q},1)$ so that the first derivative is zero
in $\rho_0$, strictly negative for $\rho<\rho_0$ and strictly positive for $\rho> \rho_0$.
We define $\beta_0\equiv\tilde\beta(\rho_0)$.
Since equation \eqref{mf:betaAB} is equivalent to $\beta=\tilde\beta(\rho_A)$, solutions to \eqref{mf:betaAB}
will exist only for $\beta$ in the image of $\tilde\beta(\cdot)$, and thus for $\beta\ge\beta_0$.
Now the condition for a local minimum \eqref{mf:stabilite} is equivalent
to $\frac{\partial \tilde\beta(\rho_A)}{\partial \rho_A}\ge 0$ and thus to $\rho_A\ge\rho_0$.
Furthermore the function $\tilde\beta(\cdot)$ is invertible from  $(\rho_0,1)$ onto $(\beta_0,+\infty)$ and
therefore, for all $\beta>\beta_0$ there is a unique couple $(\rho_A,\rho_B)$ for which the vectors \eqref{mf:rhoo}
are $Q$ local minima, and there is no ``colored'' solutions for $\beta<\beta_0$.
Whenever they exist, those minima are degenerate and their mean field free energy is given by \eqref{mf:phi(rhoo)}.
We postpone the proof that $\beta_0<Q$ at the end of the proof of the next theorem.
\end{proof}
\begin{proof}{\bf of Theorem \ref{thm:mfabsoluteminimizer}}
For $\beta$ in $(\beta_0,Q)$, we consider the difference of free energy between ordered
and disordered local minima
\begin{eqnarray}
\label{mf:diff-freeenergy}
\Delta(\phi^{\mf}_\beta)\equiv
\phi^{\mf}_\beta(\vec\rho^{(p)})-\phi^{\mf}_\beta(\vec\rho^{(-1)})=
\frac{1-Q^2\rho_A\rho_B}{2Q}+\frac{1}{2\beta}\log(Q^2\rho_A\rho_B)
\end{eqnarray}
We first note that $\beta\Delta(\phi^{\mf}_\beta)$ is a strictly decreasing function
of $\beta$: using \eqref{mf:beta(rhoA)}, we have for $\beta$ in $(\beta_0,Q)$,
\begin{eqnarray}
\label{mf:Delta}
\frac{\partial}{\partial\beta}\bigl(\beta\Delta(\phi^{\mf}_\beta)\bigr)=
\bigr(\frac{\partial\tilde\beta}{\partial\rho_A}\bigr)^{-1}
\frac{\partial\tilde\beta\Delta(\phi^{\mf}_{\tilde\beta})}{\partial\rho_A}=
-\frac{(1-Q\rho_A)^2}{2Q(Q-1)}<0
\end{eqnarray}
Furthermore, $\Delta(\phi^{\mf}_\beta)$ has one zero for $Q^2\rho_A\rho_B=1$
or equivalently
\begin{eqnarray}
\rho_A=\frac{Q-1}{Q}
\end{eqnarray}
This zero is thus necessarily unique and defines a critical value of $\beta$
\begin{eqnarray}
\bcmf=\tilde\beta(\frac{Q-1}{Q})=\frac{2(Q-1)}{Q-2}\log(Q-1)
\end{eqnarray}
We complete the proof by showing that \eqref{mf:stabilite} holds at
$\beta=\bcmf$.
We have for all $Q>2$:
\begin{eqnarray}
\label{mf:bcmfstab}
Q\bcmf\rho_A\rho_B = \frac{2(Q-1)}{Q(Q-2)}\log(Q-1)=
\frac{\log(Q-1)}{\sinh(\log(Q-1))}<1
\end{eqnarray}
Thus $\bcmf>\beta_0$. On the other hand, since $\rho_A\rho_B=Q^{-2}$
at $\beta=\bcmf$, \eqref{mf:bcmfstab}  proves also that
$\bcmf<Q$, and thus $\beta_0<\bcmf<Q$. This also complete
 the proof of theorem \ref{thm:mflocalminimizers}.

\end{proof}
\begin{proof}{\bf of Theorem \ref{thm:mflocalstability}}

Let $\beta$ be in the interval $(\beta_0,Q)$ and consider the map $\vec\rho\rightarrow\vec g(\vec\rho)$
defined in \eqref{mf:g}. We have
\begin{eqnarray}
\label{mf:dg}
\frac{\partial g_q(\vec\rho)}{\partial\rho_{q'}}=\beta g_q(\vec\rho)\bigl(\delta_{q,q'}-g_{q'}(\vec\rho)\bigr)
\end{eqnarray}
and since $0\le g_q(\vec\rho)\le 1$, we have a first bound uniform in $\vec\rho$:
\begin{eqnarray}
\label{mf:uniformbounddg}
\sup_{q,q'} \bigl|\frac{\partial g_q}{\partial\rho_{q'}}(\vec\rho)\bigr|\le \beta
\end{eqnarray}
On the other hand, from \eqref{mf:dg}, one can also write
\begin{eqnarray}
\label{mf:bounddg1}
\sup_{q,q'}|\frac{\partial g_q}{\partial\rho_{q'}}(\vec\rho)|\le
\sup_{q,q'}|\frac{\partial g_q}{\partial\rho_{q'}}(\vec\rho^{\p})|+
\sup_{q,q'}|\frac{\partial g_q}{\partial\rho_{q'}}(\vec\rho)-\frac{\partial g_q}{\partial\rho_{q'}}(\vec\rho^{\p})|
\end{eqnarray}
The first term can be bounded by
\begin{eqnarray}
\sup_{q,q'} |\frac{\partial g_q}{\partial\rho_{q'}}(\vec\rho^{\p})|=
\beta \sup_{q,q'} |\rho^{\p}_q (\delta_{q,q'}-\rho^{\p}_{q'})|\le 1-\frac{1}{Q}
\end{eqnarray}
where the inequality follows from \eqref{mf:rhod} and $\beta\le Q$ for $\p=-1$,
and from \eqref{mf:defrhoo3} and \eqref{mf:defrhoo2} for $\p>0$.
The second term in \eqref{mf:bounddg1} can be bounded using \eqref{mf:uniformbounddg}
as
\begin{eqnarray}
\sup_{q,q'}
|\frac{\partial g_q}{\partial\rho_{q'}}(\vec\rho)-\frac{\partial g_q}{\partial\rho_{q'}}(\vec\rho^{\p})|\le
2\beta\sup_{q}| g_q(\vec\rho)- g_q(\vec\rho^{\p})|\le 2Q\beta^2 \sup_{q}|\rho_q- \rho^{\p}_q|
\end{eqnarray}
Thus for all $\rho$ such that $\sup_{q}|\rho_q- \rho^{\p}_q|\le\frac{1}{4\beta^2Q^2}$ for some $\p$
one gets
\begin{eqnarray}
\sup_{q,q'}|\frac{\partial g_q}{\partial\rho_{q'}}(\vec\rho)|\le 1-\frac{1}{2Q}
\end{eqnarray}
\end{proof}

We conclude this appendix by a proof of \eqref{eq:lemmaa.2}:

From equation \eqref{mf:Delta} and the definition of $\bcmf$
\eqref{mf:bcmf}, one gets explicitly:

\begin{eqnarray}
\frac{d}{d\beta}\left[P^{-}_{\mf,\beta}-P^{+}_{\mf,\beta}\right]\Big|_{\beta=\bcmf}
&=&\frac{1}{\bcmf}\frac{\partial}{\partial \beta} \bigl(\beta\Delta(\phi^\mf_\beta)\bigr)|_{\beta=\bcmf}\nn\\
&=&-\frac{(Q-2)^2}{2Q(Q-1)\bcmf}<0
\end{eqnarray}

\vskip 2.5cm \noindent
\section{Local equilibrium}
\setcounter{equation}{0}
\label{app:dinamica}

The main result of this appendix is the proof that, for suitable values of the temperature,
if a density profile is in a neighborhood of an equilibrium value in a region
$\La\sqcup\delta_{\out}^{\ga^{-1}}[\La]$, then it can be made closer to equilibrium inside $\La$
at an exponential rate from its boundary, decreasing the free energy.
This result is essentially due to the stability properties of the free energy functionals
originating from the contraction property  of the map \eqref{contract} around 
its fixed point.
The precise result is stated in  the Theorem \ref{thm:dimamica}
 below. The proof follows the lines developed in \cite{errico-leip},(see also \cite{bkmp1})
 for Ising model and continuum particle models, and we will stress here only the points
 specific for our model while we will only  sketch    the points that are
 quite analogous to the other cases.

Without lost of generality we study the local equilibrium around the phase
$``\ \p\ "$, $\p\in \{-1,1,\dots Q\}$.
Let $\La$ a bounded $\cD^{\ell_{+,\ga}}$-measurable region,
and $\eta^{\zeta}_x(\vec\rho)\in L^\infty(\Rr^d,S_Q)$ defined analogously
as $\eta_x(\vec\rho)$  in \eqref{def:eta},
but with an accuracy parameter, denoted by $\zeta$, that here we leave free
\begin{eqnarray*}
\eta^{\zeta}_x(\vec\rho)=
\begin{cases}
    a_{{\p}} & \text{if }~
    \dis{\|{\vec \rho^{\ell_{-,\ga}}(x)}-
    \red{\vec \rho_{\beta}^{~(\p)}}\|_{\star}< \zeta}
    \\
    0 & \text{otherwise}.
  \end{cases}
\end{eqnarray*}

\begin{equation*}
\cM_{\La,\zeta}^{\p}:=\{ \vec\rho\in
L^{\infty}(R^{d},S_Q): \red{\eta^{\zeta}_x(\vec\rho)=a_{\p}} ~~
\forall x\in\La\sqcup\delta^{\ga^{-1}}_{\out}[\La ]\}
\end{equation*}
and for any $\vec\rho^{~*}\in \cM_{\La}^{\p}$, we define
\begin{equation*}
\cX^{\p}_{\La,\vec\rho^{~*}}:=
    \{\vec\rho\in M_{\La,\zeta}^{\p}: \vec\rho_{\La^{c}}(x)
    \equiv \vec\rho^{~*}_{\La^{c}}(x) \}
\end{equation*}

\begin{thm}
   \label{thm:dimamica}
There are positive constant $\zeta_0$, $\om$, $c_\om$
so that for any \red{$\zeta<\zeta_0$,
$ \ga^\al<\kappa_0\zeta $} and any $\vec\rho^{~*}\in M_{\La,\zeta}^{\p}$, s.t.
\begin{itemize}
  \item there is a unique $\vec\psi\in \cX_{\La,\vec\rho^{~*}}$ s.t.:
   \begin{equation}
       \inf_{\vec\rho\in \cX^{\p}_{\La,\vec\rho^{~*}}}\cF_{\ga,u,\La}(\vec\rho|\vec\rho^{~*}_{\La^c})
       =\cF_{\ga,u,\La}(\vec\psi|\vec\rho^{~*}_{\La^c})
       \end{equation}
  \item $\vec\psi$ is the unique solution of the mean field equation and has
  the following properties:
\begin{itemize}
  \item[*] \red{$\vec\psi_\La\in C^{\infty}(\La,M_{\La,(1-\kappa_0)\zeta}^{\p})$, \;\;\;
  $\sup_{r\in\La}\|\nabla \psi_{\La} (r)\|_{\star}\le \beta\|\nabla J_\ga\|_1$}

\vskip .5cm \noindent
  \item[*] \begin{equation*}\|\vec\psi_{\La}(r)-\vec \rho^{\p}\|
  \le c_\om e^{-\om\; \dist(r,\La^c_{\ne})}\end{equation*}
 where
$\La^c_{\ne}:= \{r\in \La^c: \dist(r,\La)\le\ga^{-1}; \vec\rho^{~*}(r)\ne\vec\rho^{\p}\}$

\vskip .5cm \noindent

\end{itemize}

   \item If $\vec\psi, \vec\phi$ are minimizers resp. in
  $\cX^{\p}_{\La,\vec\rho_{1}}, \cX^{\p}_{\La,\vec\rho_{2}}$ then:
  \begin{equation*}\|\vec\psi_{\La}(r)-\vec \phi_{\La}(r)\|
  \le c_\om e^{-\om\dist(r,\La^c_{1,2,\ne})}\end{equation*} where
$\La^c_{1,2,\ne}:= \{r\in \La^c: \dist(r,\La)\le\ga^{-1}; \vec\rho_{1}(r)\ne\vec\rho_{2}(r)\}$

\end{itemize}

\end{thm}

\vskip .5cm \noindent

The proof of Theorem \ref{thm:dimamica} is obtained by defining a dynamic
$\vec T^{u,\La,\p}_t$ on
$L^{\infty}(\Rr^{d},S_Q)$, and
proving that this dynamic maps $M_{\La,\zeta}^{\p}$
into itself and that it is dissipative for the free energy $\cF_{\ga,u,\La}$.
The minimizer
$\vec \psi$ is then obtained as the limit point of the orbit
$\vec T^{u,\La,\p}_t$ as $t\to\infty$.

\red{Following \cite{errico-leip}} we define
an opportune dynamic $\vec T^{u,\La,\p}_t$ (suitable for our model)
that has the properties
that allow to conclude as in reference \cite{errico-leip}.

The essential point in the proof of the Theorem
\ref{thm:dimamica} is the contraction property of the map
$\vec \fM^{(u,\p)}(\cdot)$
parameterized by $u$, $u\in[0,1]$, defined as follows:

\begin{eqnarray*}
& & \fM^{(u,\p)}_q(\vec \rho)(r):=
\frac{e^{\beta \cL^{u,\p}_q(\vec\rho;r)}}{\sum _{q'} e^{\beta \cL_{q'}(\vec \rho;r)}}\\
& &\cL^{u,\p}_q(\vec\rho;r):=u\int dr'~J_\ga(r,r')\rho_q(r')+(1-u)\; \rho^{\p}_q\\
\end{eqnarray*}
 We state here a lemma  which proof is postponed at the end of this appendix :
\begin{lemma}
   \label{lem:dinam}
There are  $\zeta'_0$ and $\kappa_0$ positive, so that for any $\zeta<\zeta'_0$
and $\ga^{\al}<\kappa_0\zeta$, any bounded $\cD^{\ell_{-,\ga}}$-measurable region
$\La$, $r\in \La$, $\vec\rho\in M_{\La,\zeta}^{\p}$

       \begin{eqnarray}
       \label{eq:lem-dinam-1}
       &&\sup_{r\in\La}\|\vL^{u,\p}(\vec\rho)(r)-\vec\rho^{~\p}(r)\|_{\red{\star}}\le {u(1+c_d\kappa_0)}\zeta<2\zeta\\
       \label{eq:lem-dinam-2}
&&\sup_{r\in\La}\|\vec\fM^{(u,\p)}(\vec\rho)(r)-\vec\rho^{~\p}(r)\|_{\red{\star}}\le u(1-\kappa_0)\zeta
       \end{eqnarray}

\end{lemma}

\vskip .5cm \noindent

We then define a  dynamic given by the semigroup $\vec T^{u,\La,\p}_{t}(\cdot)$, $t\geq 0$ on
$L^{\infty}(\La,S_Q)$,
\begin{eqnarray}
\label{def:dynamic}
\vec T^{u,\La,\p}_t(\vec \rho):=\big(T^{u,\La,\p}_{1,t}(\rho_1),\dots,
T^{u,\La,\p}_{Q,t}(\rho_Q))\hskip3cm
\mbox{for any $\vec \rho \in L^{\infty}(\Rr^{d},S_Q)$}
\end{eqnarray}
where $T^{u,\La,\p}_{q,t}(\rho_i^*)$ are solutions of the Cauchy problem:
\begin{eqnarray}
\label{def:dynamics}
\left\{\ba{lr}
\dis{\frac{d\rho^{\La}_q(r,t)}{dt}=-\rho^{\La}_q(r,t)+\fM_q^{(u,\p)}(\vec \rho^{\La})}
&  \mbox{for any $q=1,\dots, Q$}
\\
\\
\dis{\vec\rho^{\La}(r,t)=\vec\rho^{~*}(r,t)}& (r,t)\in\left[\La^c\times \{t\ge 0\}\right]
\sqcup[\Rr^{d}\times\{t=0\}]
\ea\right.
\end{eqnarray}

 Existence, uniqueness and continuity w.r.t. the initial datum of the solution follows
by the continuity and the Lipschitz property of the r.h.s. of equation \eqref{def:dynamics}

Notice that $\vec T^{u,\La}_t(\cdot)$ maps $L^{\infty}(\Rr^{d},S_Q)$ in itself,
and has as a  fixed point
$\vec\rho^{\p}$. \red{We next  prove} the following properties:

\vskip .5cm \noindent
\begin{enumerate}
  \item 
$\vec T^{u,\La,\p}_t(M_{\La,\zeta}^{\p})\sqsubset M_{\La,\zeta}^{\p}$ for any $t\ge 0$

\vskip .5cm \noindent
  \item For any $\vec\rho_0\in L^{\infty}(\Rr^d,S_Q)$  and  $\La$ a Borel set,
   $T^{u,\La,\p}_{q,t}(\vec\rho_0)$, $q=1,\dots, Q$,
  converges by subsequences  as $t\to \infty$ to  functions
  $v_q$ that are bounded in $\La$ and with $\nabla v_q$ bounded in $\La$.
  The limit points are solutions
  of \eqref{eq:staz} below.

\vskip .5cm \noindent
  \item $\cF_{\ga,u,\La}(\vec T^{u,\La,\p}_t(\vec\rho_0))$ decrease with $t$,
  strictly unless $\vec\rho_0$
  is stationary, in which case satisfies:
\begin{eqnarray}
\label{eq:staz}
\rho_{0,q}=\frac{\dis{e^{\beta\cL^u_q(\vec\rho_0(r,t))}}}
{\dis{\sum_{q'=1}^{Q}e^{\beta\cL^u_{q'}(\vec\rho_0(r,t))}}}
\hskip3cm \forall q=1,\dots Q ~~~~~\forall r\in \La
\end{eqnarray}

\vskip .5cm \noindent
    \item  As a consequence of the  property $(3)$, for any
    $\vec\rho^{~*}(r)\in M_{\La,\zeta}^{\p}$,
    the minimizers of $\cF_{\ga,u,\La}(\cdot)$ in $\cX_{\La,\vec\rho^{~*}}:=
    \{\vec\rho\in M_{\La,\zeta}^{\p}: \vec\rho_{\La^{c}}(r)
    \equiv \vec\rho^{~*}_{\La^{c}}(r) \}$ are solutions of \eqref{eq:staz}:
    $\vec\rho=\vec\fM^{u,\p}(\vec\rho)$. By the
    contraction property of the map $\vec\fM^{u,\p}(\cdot)$ 
    we get uniqueness of the minimizer.
\end{enumerate}

\vskip 1.5cm \noindent
{\em Proof of the properties 1,2,3}
\begin{enumerate}

\item
Clearly we have
$\vec T^{u,\La,\p}_{t}(S_Q)\sqsubset S_Q$.
To prove the first point, 
let then $\tau>0$, $\vec\rho_0\in M_{\La,\zeta}^{\p}$ and
\begin{eqnarray*}
\cX_{\tau,\vec\rho_0}:=\{\vec\rho\in L^{\infty}(\Rr^{d}\times[0,\tau];S_{Q}):
\vec\rho(r,t)=\vec\rho_0(r)
~~(r,t)\in\big[ \Rr^{d}\times 0\big]\sqcup\big[\La^c\times[0,\tau]\big] \}
\end{eqnarray*}

Let  $\hat\Omega^{\p}(\cdot)$ the map from $\cX_{\tau,\vec\rho_0}$ into itself defined
for any $t\in[0,\tau]$, $r\in \La$ as

\begin{eqnarray*}
\vec\Omega^{\p}(\vec\rho)(r,t)=e^{-t}\vec\rho_0(r)+\int_{0}^{t} e^{-(t-s)}
\vec\fM^{(u,\p)}(\vec\rho) ~ds
\end{eqnarray*}
if $\tau$ is small enough $\vec \Omega^{\p}$ is a contraction and its fixed point is the solution
of \eqref{def:dynamics},
$T^{u,\La,\p}_{t}(\vec\rho_0)$, $t\in [0,\tau]$. By \eqref{eq:lem-dinam-2} the set:
\begin{eqnarray*}
\{\vec v\in \cX_{\tau,\vec\rho_0}: \vec v(\cdot ,t)\in M_{\La,\zeta}^{\p}~ \forall t\in [0,\tau]\}
\end{eqnarray*}
is invariant under the map $\hat\Omega$, and since it is closed, it contains the fixed point of
$\hat\Omega$. By induction the statement can be extended fo any $t$: $T^{u,\La,\p}_{t}(\vec\rho_0)
\sqsubset M_{\La,\zeta}^{\p}$, $t\ge 0$.

\vskip .5cm \noindent
\item
Convergence on subsequences follows by Ascoli-Arzel\`a theorem, after having written the
integral expression of\; \blue{the evolution} \eqref{def:dynamics} and observed
$T^{u,\La,\p}_{q,t}(\vec \rho)-e^{-t}\rho_q$
is bounded with bounded gradient.

\vskip .5cm \noindent

 \item The decreasing of the free energy functional, follows by observing that:
 \begin{eqnarray*}
& &\frac{d}{dt}\cF_{\ga,u,\La}([\vec T^{u,\La,\p}_t(\vec\rho)] \; |\; \vec\rho_{\La^c})<0
\end{eqnarray*}
an explicit calculation gives in fact :
\begin{eqnarray*}
&&
\frac{d}{dt}\cF_{\ga,u,\La}([\vec T^{u,\La,\p}_t(\vec\rho)] \; |\; \vec\rho_{\La^c}=
\sum_q\int_{\La}dr~\bigg(-\cL_q^{u}(\vec T^{u,\La,\p}_t(\vec\rho))+\frac{1}{\beta}
(1+\ln T^{u,\La,\p}_{q,t}(\vec\rho))\bigg)
\\
& &\hskip6cm \cdot
 \bigg(-T^{u,\La,\p}_{q,t}(\vec\rho)+\fM^{(u,\p)}_q(\vec T^{u,\La,\p}_t(\vec\rho))\bigg)
\\
&&\hskip1cm=
\sum_q\int_{\La} dr~\frac{1}{\beta}\bigg(\ln\frac{T^{u,\La,\p}_{q,t}(\vec\rho)}{\fM^{(u,\p)}_q
(\vec T^{u,\La,\p}_t(\vec\rho))}
+1-\ln\sum_{q'} e^{\beta\cL^{u}_{q'}(\vec T^{u,\La,\p}_t(\vec\rho))})\bigg)
\\ & &\hskip6cm \cdot
\bigg(-T^{u,\La,\p}_{q,t}(\vec\rho)+\fM^{(u,\p)}_q(\vec T^{u,\La,\p}_t(\vec\rho))\bigg)
\end{eqnarray*}
\begin{eqnarray*}
&&\hskip1cm =
\frac{1}{\beta}\sum_q\int_{\La} dr~\bigg(\ln\frac{T^{u,\La,\p}_{q,t}(\vec\rho)}{\fM^{(u,\p)}_q(\vec T^{u,\La}_t(\vec\rho))}\bigg)
\bigg(-T^{u,\La,\p}_{q,t}(\vec\rho)+\fM^{(u,\p)}_q(\vec T^{u,\La,\p}_t(\vec\rho))\bigg)\\& &\hskip6cm+
\frac{K}{\beta}\sum_q\int_{\La} dr~
\bigg(- T^{u,\La,\p}_{q,t}(\vec\rho)+\fM^{(u,\p)}_q(\vec T^{u,\La,\p}_t(\vec\rho))\bigg)
 \end{eqnarray*}
 with $K=(1-\ln\sum_{q'} e^{\beta\cL^{u}_{q'}(\vec\rho)})$. By normalization condition,
 last term is null, while
 the first one is negative (in fact if the first factor  inside the integral is positive,
 the last one is negative and viceversa). We denote by:
 \begin{eqnarray*}
\cD^{u,\p}_{\La}(\vec \rho):=\frac{1}{\beta}\sum_i \bigg(\ln\frac{\rho_i}{\fM^{(u,\p)}_i(\vec\rho)}\bigg)
\bigg(-\rho_i+\fM^{(u,\p)}_i(\vec\rho)\bigg)
 \end{eqnarray*}

$\cD^{u,\p}_{\La}(\vec \rho)=0$  if and only if $\vec\rho$ satisfies the equation:
\begin{eqnarray*}
\rho_q=\fM^{(u,\p)}_q(\vec\rho) \hskip3cm \forall q=1,\dots,Q
\end{eqnarray*}
\blue{To conclude we need
a lower bound on $\vec T^{u,\La,\p}_t(\vec\rho)$ that assure that the dynamic is always well
defined for any time $t>0$.
(notice  that $\cD^{u,\p}_{\La}(\vec \rho) $ diverges if one of the coordinates
$\rho_q(t)$  becomes null.)}

\begin{lemma}
   \label{lem:bound}
Let $\vec\rho\in L^{\infty}(\Rr^{d},S_Q)$, $\La$ a Borel set in $\Rr^{d}$, $\vec\rho^{\La}(\cdot,t)
=\vec T^{u,\La,\p}_t(m)$
and $v_q^{0}:=\dis{\inf_{r'\in\La}}\rho_q^{\La}(r')$.
Then for all $r\in\La$ and $t\ge 0$:
       \begin{equation}
       \label{eq:lembound}
       \rho_q(r,t)\ge\big[ v_q^0-c_q(u,\beta)\big]e^{-t}+c_q(u,\beta)
       \end{equation}
where $c_q(u,\beta)=\frac{1}{e^{\beta u}}
\frac{e^{(1-u)\beta\rho_q^{\p}}}{\sum_{q'}e^{(1-u)\beta\rho_{q'}^{\p}}} \ge
\frac{1}{Q e^{\beta [u+(1-u)]}}\ge\frac{1}{Q e^{\beta }}$
\end{lemma}

\vskip .5cm \noindent
\begin{proof}
 For any $q=1,\dots,Q$, let  $v_q(t)$ the solutions of the Cauchy problems:
\begin{eqnarray}
\label{def:lemboundproof1}
\left\{\ba{lr}
\dis{\frac{dv_q(t)}{dt}=-v_q(t)+\frac{1}{e^{\beta u}}c_q(u,\beta)}
\\
\dis{\vec v(t)=\vec v^{0}}
\ea\right.
\end{eqnarray}
Let $w_q(r,t):=v_q(t)\Ii_{r\in\La}+\vec \rho^*(r,t) \Ii_{r\in \La^{c}}$,
Since $-w_q(r,t)+\frac{1}{e^{\beta u}}c_q(u,\beta)\le -w_q(r,t)+\fM^{(u,\p)}_q(\vec w)$ uniformly in
$r$ and
 $w_q\in [0,1]$. {By Gronwall Lemma }, $\rho_q(r,t)\ge w_q(r,t)$
 for any $r\in \Rr^d,t>0$, and it
 is strictly positive for any $t>s>0$.

\end{proof}
 By Lemma \ref{lem:bound}  for any  $s>0$, the functions
$[ T_{q,t}^{u,\La,\p}(\vec \rho)]_{\La}$, $t\ge s$ , for any $q\in [1,\dots, Q]$ are
 bounded away from $0$, so that:
 \begin{eqnarray*}
\cF_{\ga,u,\La}([\vec T_{t}^{u,\La,\p}(\vec \rho)]_{\La}| \vec\rho^{~*})-
\cF_{\ga,u,\La}([\vec T_{s}^{u,\La,\p}(\vec \rho)]_{\La}| \vec\rho^{~*})=
\int_s^{t}\cD^{u,\p}_\La(\vec T_{t'}^{u,\La,\p}(\vec \rho))dt'
 \end{eqnarray*}
 Since  $\cD^{u,\p}_\La(\vec T_{s}^{u,\La,\p}(\vec \rho))$ is monotone in $s$,
by  the Lebesgue dominated convergence theorem, the limit $s\to 0$ exists, and we get:

\begin{eqnarray*}
\cF_{\ga,u,\La}([\vec T_{t}^{u,\La,\p}(\vec \rho)]_{\La}| \vec\rho^{~*})-
\cF_{\ga,u,\La}(\vec \rho_{\La}| \vec\rho^{~*})=
\int_0^{t} \cD^{u,\p}_\La(\vec T_{t'}^{u,\La,\p}(\vec \rho))dt'
 \end{eqnarray*}

 \end{enumerate}
\vskip .5cm \noindent
We omit the proof of  the following theorem that follows by previous analysis:
\begin{thm}
Let $\vec\rho\in \red{L^{\infty}(\Rr^{d} ,S_Q)}$ and $\La$ a  bounded, Borel set. Then, any limit
point of $\vec T_t^{\La}(\vec \rho)$ satisfies \eqref{eq:staz} and for any $\vec\rho^{~*}\in
\red{L^{\infty}(\La^{c} ,S_Q)}$ there is $\vec v\in \red{C^{1}(\La ,S_Q)}$,
\red{with $\nabla v_x$ bounded for any $x$ in $\La$} s.t.
       \begin{equation*}
     \cF_{\ga,u,\La}(\vec v|\vec\rho^{~*}_{\La^{c}})
     \le\cF_{\ga,u,\La}(\vec \rho^{~*}_{\La}|\vec\rho^{~*}_{\La^{c}})
       \end{equation*}
\end{thm}
\vskip .5cm \noindent

As a corollary of the Theorem \ref{thm:dimamica} we have the following result used in Subsection
\ref{subsection:smalldeviation}, and  \ref{section:largedeviation}
\begin{corol}
   \label{corol:dimamica}
There are positive constants $\om$ and $c'_\om$ so that for any
\red{$u\in [0,1]$} and any $\vec\rho^{~*}\in
M_{\La,\zeta,\ell}^{\p}$
there is $\vec\psi^{(u)}\in \cX_{\La,\vec\rho^{~*}}$
with the following properties:
\begin{eqnarray}
\ba{lr}
\vec\psi^{(u)}(r)=\vec\rho^{~*}(r) & \hskip3cm r\in\La^{c}\sqcup\delta^{\ga^{-1}}_{\ins}[\La]
\\
\vec \psi^{(u)}(r)=\vec \rho^{\p}(r)& r\in\La\setminus \delta_{\ins}^{\ell_{+,\ga}/4}[\La]
 \\
\cF_{\ga,u,\la}(\vec \psi^{(u)})\le \cF_{\ga,u,\la}(\vec\rho^{~*})+c_\om|\La|e^{-\om\ell_{+,\ga}/4}&
\ea
\end{eqnarray}

\end{corol}

\vskip .5cm \noindent
We conclude this appendix by giving the proof of the Lemma \ref{lem:dinam}

\vskip .5cm \noindent
\begin{proof}[Proof of Lemma \ref{lem:dinam}:]
\vskip .5cm \noindent
In order to prove the first statement, we define a $\cD^{(\ell_{-,\ga})}$- measurable approximation
of the interaction kernel as:
\begin{eqnarray}
J_\ga^{(\ell_{-,\ga})} (x,\lng y\rng ) = \frac{1}{|C^{(\ell_{-,\ga})}|}\int_{y'\in C_y^{(\ell_{-,\ga})}} J_\ga (x,y') dy'
\end{eqnarray}
We have for $\ell_{-,\ga}=\ga^{-1+\al}\ll \ga^{-1}$,
\begin{eqnarray}
|J_\ga^{(\ell_{-,\ga})} (x,\lng y\rng) - J_\ga (x,y)| &&\le
\sup_{y'\in C_y^{(\ell_{-,\ga})}}  |J_\ga^{(\ell_{-,\ga})} (x,\red{\lng y\rng}) - J_\ga (x,\red{y'})|\\
 &&\le \sqrt{d}~ \ell_{-,\ga} \ga^{d+1} \|\nabla J\|_\infty {\bf 1}_{\{|x-y|\}\le 2 \ga^{-1}}
\end{eqnarray}
Using this result, we can write for all $r$ and all $q$,
\begin{eqnarray*}
&&|\cL^{u}_q(\vec\rho)(r)-\rho_q^{~\p}| =
u \left| \int dr'~J_\ga(r,r')(\rho_q(r') - \rho^{\p}_q) \right|
\\ && \le
u \left| \int dr'~J^{(\ell_{-,\ga})}_\ga(r,\lng r'\rng)(\rho_q(r') - \rho^{\p}_q) \right| +
u \left| \int dr'~(J_\ga(r,r') - J^{(\ell_{-,\ga})}_\ga(r,\lng r'\rng))(\rho_q(r') -
\rho^{\p}_q) \right|
\\
&&\le
u \bigl( \sum_{j\in (\ell_{-,\ga}\Zz)^d} J^{(\ell_{-,\ga})}_\ga(r,\lng x \rng)
\int_{r'\in C_j^{(\ell_{-,\ga})}} |\rho_q(r') - \rho^{\p}_q| dr'
+ 2^d \sqrt{d} \|\nabla J\|_\infty \ga \ell_{-,\ga} \bigr)\\
&& \le u (1 + c_d \kappa_0 ) \zeta
\end{eqnarray*}
for $\ga^{\al}\le \kappa_0 \zeta $ and $c_d = 2^d \sqrt{d} \|\nabla J\|_\infty$.
Hence
\begin{eqnarray}
\sup_{r}\|\vL^{u}(\vec\rho)-\vec\rho^{~\p}\|_\star \le u (1 + c_d \kappa_0 ) \zeta \le 2  \zeta
\end{eqnarray}
for $\kappa_0$ small enough.

In order to prove \eqref{eq:lem-dinam-2},
we take $\zeta_0'$ small enough so that theorem \ref{thm:mflocalstability} holds
\blue{( for example $\zeta_0'<\left(\frac{1}{2\beta Q}\right)^2$)}
for all $\zeta<\zeta_0'$.
We get for any $r\in \La$
\begin{eqnarray}
\|\vec\fM^{(u,\p)}(\vec\rho)(r)-\vec\rho^{~\p}(r)\|_{\star}&=&
\|\vec g(\vL^{u}(\vec\rho))(r)-\vec g(\vec\rho^{~\p})(r)\|_{\star}\nn \\
&\le& (1-\frac{1}{2Q}) \|\vL^{u}(\vec\rho)(r)-\vec\rho^{~\p}(r)\|_\star \nn \\
&\le&  (1-\frac{1}{2Q}) u (1 + c_d \kappa_0 ) \zeta \le u (1 - \kappa_0 ) \zeta
\end{eqnarray}
having chosen  $\kappa_0$ so small that
\begin{eqnarray}
\kappa_0 \le \frac{1}{2Q(1+ c_d)}
\end{eqnarray}
and \eqref{eq:lem-dinam-2}  holds.
\end{proof}

\vskip 2.5cm \noindent
\setcounter{equation}{0}
\section{Existence of the pressures $P^{\pm}_{\abs,\ga,\beta}$ of the abstract models }
\label{app:exist-pressure}

Let $\{\La_n\}$ a sequence of sets of side
$2^n\ell_{+,\ga}$ and
\begin{eqnarray*}
D_{\ga,\beta}(n):= \frac{\ln
Z^{\pm}_{\abs,\La_n\ga,\beta}(\vec\rho^{\pm}_{\beta})}{\beta|\La_n|}-\frac{\ln
Z^{\pm}_{\abs,\La_{n-1},\ga,\beta}(\vec\rho^{\pm}_{\beta})}{\beta|\La_{n-1}|}
\end{eqnarray*}

The proof of existence and continuity in $\beta$ o the
abstract pressures follows by the continuity in $\beta$ of
$D_{\ga,\beta}(n)$ and by proving that there is a constant
$\kappa_7$:
\begin{eqnarray}
\label{eq:dn} |D_{\ga,\beta}(n)|\le \kappa_7 2^{-n}
\end{eqnarray}

\vskip .5cm \noindent

\begin{proof}

\vskip .5cm \noindent
Decomposing $\La_n$ into cubes $\La_{n-1}(k)$, $k=1,\dots,2^d$,
 since the interaction energy  is bounded uniformly in $\vw$ and
 recalling \eqref{new-z4.11b}, we have:
\begin{eqnarray}
\label{eq:ab-1}
&&\ln Z^{\pm}_{\abs,\La_n\ga,\beta}(\vec\rho^{\pm}_{\beta})\ge
\\
&&\hskip1cm2^{d}\ln Z^{\pm}_{\abs,\La_{n-1}\ga,\beta}(\vec\rho^{\pm}_{\beta})-
c\ga^{-1}
(2^n\ell_{+,\ga})^{d-1}-2d(2^{n-1})^{d-1}e^{-\frac{\fK_{\ga}}{2}}\nn
\end{eqnarray}

where, denoting by
$[\delta\La_n^{0}]:=\sqcup_{k=1}^{2^d}
\delta_{in}^{\ell_{+,\ga}}[\La_{n-1}(k)] $, we have used the estimate
\begin{eqnarray*}
|\sum_{\Delta\sqsubset\La_n}U_{\Delta}(\vw)-\sum_k
\sum_{\Delta\sqsubset\La_{n-1}(k)}U_{\Delta}(\vw)|=
\sum_{\Delta: \Delta
\sqcap[\delta\La_n^{0}]\ne \emptyset}\|U_{\Delta}(\vw)\|_{\infty}
\le 2d(2^{n-1})^{d-1}e^{-\frac{\fK_{\ga}}{2}}
\end{eqnarray*}

\vskip .5cm \noindent
\eqref{eq:ab-1} gives:
\begin{eqnarray*}
D_{\ga,\beta}(n)\ge -c2^{-n}
\end{eqnarray*}
the same arguments  give also the upper bound:
\begin{eqnarray*}
D_{\ga,\beta}(n)\le -c2^{-n}
\end{eqnarray*}
\end{proof}

\vskip 2.5cm \noindent
\section{Existence of the pressure $P^{\pm}_{\abs,\ga;0}$  and surface correction.}
\label{app:meanfieldpressurelimit}
\setcounter{equation}{0}
In this appendix  we prove the following  Lemma:

\begin{lemma}
   \label{lem:0pressure}
There exists a constant $c_d$ such that for $\ga$ small enough
       \begin{eqnarray}
       \label{eq:lem0pressure}
|P^{\pm}_{\abs,\ga;0} + \phi^{\mf}_\beta(\rho^{\pm})|\le c_d \ga
\end{eqnarray}
where $P^{\pm}_{\abs,\ga;0}$ is defined in \eqref{def:H0-b}
for $u=0$, and
\begin{eqnarray}
       \label{eq:lem0correction}
R^{\pm}_{\abs,\La;0}=R^{\mf}_{\ga,\La}:=\frac{\beta}{2}\sumtwo{x\in \La}{y\in\La^c}J_\ga(x,y)
(\vec\rho^{\pm}\cdot\vec\rho^{\pm})
       \end{eqnarray}

\end{lemma}

\vskip .5cm \noindent
\begin{proof}
{\ }
\vskip .5cm \noindent
We denote by $C_0\equiv C_0^{\ell_{+,\ga}}$,
the cube of the partition $\cD^{\ell_{+,\ga}}$ that contains the point $0$,
\begin{eqnarray}
\ln  Z^{\pm}_{\abs,\beta,\La;0}&=&  \ln \{\sum_{\vw_{\La}} \prod_{x\in \La}
e^{\beta \big((\vw_x-\rho^{\pm})\cdot \vec\cL(\rho^{\pm})\big)}
\Ii_{\vw\in\cX^{\pm}_{C_0}}\}
-\beta  H_{\ga,\La}(\vec\rho^{\pm}_\La|\vec\rho^{\pm}_{\La^c})
\nn \\
& =&\ln \{\sum_{\vw_{\La}} \prod_{x\in \La}
e^{\beta \big(\vw_x\cdot \vec\cL(\rho^{\pm})\big)}\Ii_{\vw\in\cX^{\pm}_{C_0}}\}
-\beta |\La|\fJ_\ga( \rho^{\pm}\cdot \vec\rho^{\pm})
-\beta H_{\ga,\La}(\vec\rho^{\pm}_\La|\vec\rho^{\pm}_{\La^c})
\nn \\
&=&\frac{|\La|}{|C_0|} \ln \{\sum_{\vw_{C_0}} \prod_{x\in C_0}
e^{\beta \big(\vw_x\cdot \vec\cL(\rho^{\pm})\big)}\Ii_{\vw\in\cX^{\pm}_{C_0}}\}
-\frac{\beta|\La|}{2}\fJ_\ga( \rho^{\pm}\cdot \vec\rho^{\pm})\label{P-abs-0a}
\\ \nn
& &\hskip7cm+\frac{\beta|}{2}\sumtwo{x\in\La}{y\in \La^c}J_\ga(x,y)
(\vec\rho^{\pm}\cdot\vec\rho^{\pm})
\end{eqnarray}
where we recall
$H_{\ga,\La}(\vec\rho^{\pm}_\La|\vec\rho^{\pm}_{\La^c})=-\frac{1}{2}\sumtwo{x\in\La}{y\in \La}J_\ga(x,y)
(\vec\rho^{\pm}\cdot\vec\rho^{\pm})-\sumtwo{x\in\La}{y\in \La^c}J_\ga(x,y)
(\vec\rho^{\pm}\cdot\vec\rho^{\pm})=-\frac{1}{2}\sumtwo{x\in\La}{y\in \Zz^{d}}J_\ga(x,y)
(\vec\rho^{\pm}\cdot\vec\rho^{\pm})-\frac{1}{2}\sumtwo{x\in\La}{y\in \La^c}J_\ga(x,y)
(\vec\rho^{\pm}\cdot\vec\rho^{\pm})$.
Then we get:

\begin{eqnarray}
P^{\pm}_{\abs,\ga,0}&:=&\lim_{\La\nearrow \infty}\frac{1}{\beta|\La|}\ln
 Z^{\pm}_{\abs,\beta,\La;0}=\lim_{\La\nearrow \infty}\frac{1}{\beta|\La|}\ln
\{\sum_{\vw_{C_0}} \prod_{x\in C_0} e^{-\beta\fH_{\ga,\La}^{\pm}(\vw)}
\Ii_{\vw\in\cX^{\pm}_{C_0}}\}
\nn \\\label{P-abs-0b}
&\equiv& \frac{1}{\beta|C_0|}
\ln \{\sum_{\vw_{C_0}} \prod_{x\in C_0}
e^{\beta \big(\vw_x\cdot \vec\cL(\rho^{\pm})\big)}\Ii_{\vw\in\cX^{\pm}_{C_0}}\}-
\frac{1}{2}\fJ_\ga
(\vec\rho^{\pm}\cdot \vec\rho^{\pm})
\end{eqnarray}

A comparison of \eqref{P-abs-0a} and \eqref{P-abs-0b} gives directly \eqref{eq:lem0correction}.

\vskip .5cm \noindent
We now prove \eqref{eq:lem0pressure}

Let consider the first term of \eqref{P-abs-0b}:
\begin{eqnarray*}
&&\frac{1}{\beta|C_0|}
\ln \{\sum_{\vw_{C_0}} \prod_{x\in C_0}
e^{\beta \big(\vw_x\cdot \vec\cL(\rho^{\pm})\big)}\Ii_{\vw\in\cX^{\pm}_{C_0}}\}
\hskip7cm\\
&&\hskip3cm=
\frac{1}{\beta|C_0|}
\ln G_{{\pm},\ga;0}[\Ii_{\vw\in\cX^{\pm}_{C_0}}]+\frac{1}{\beta|C_0|}\ln
\sum_{\vw_{C_0}} \prod_{x\in C_0}
e^{\beta \big(\vw_x\cdot \vec\cL(\rho^{\pm})\big)}\\
&&\hskip3cm=
\frac{1}{\beta|C_0|}
\ln G_{{\pm},\ga;0}[\Ii_{\vw\in\cX^{\pm}_{C_0}}]+
  \frac{1}{\beta}\ln\sum_{i=1,Q}e^{\beta\fJ_\ga \rho^{\pm}_i}
\end{eqnarray*}
where $G_{{\pm},\ga;0}[\Ii_{\vw\in\cX^{\pm}_{C_0}}]$ is the  probability of the event
$\Ii_{\vw\in\cX^{\pm}_{C_0}}$ w.r.t. the Gibbs measure specified by:
\begin{eqnarray*}
\mu^{\pm}_{\ga;0}:=
\frac{e^{\beta\sum_x\big(\vw_x\cdot \vec\cL(\rho^{\pm})\big)}}{Z^{\pm}_{\ga;0}}
\end{eqnarray*}

Postponing the proof that the first term is negligible, as
\begin{eqnarray}
\label{app2:gpm}
G_{{\pm},\ga;0}[\Ii_{\vw\in\cX^{\pm}_{C_0}}]>1-Ce^{-c\ga^{2a}|C^{\ell_{-,\ga}}|}
\end{eqnarray}
we consider the second term. We have
\begin{eqnarray*}
|\frac{1}{\beta}\ln\sum_{i=1,Q}e^{\beta\fJ_\ga \rho^{\pm}_i} -
\frac{1}{\beta}\ln\sum_{i=1,Q}e^{\beta\rho^{\pm}_i}|\le 2^d \sqrt{d} \|\nabla\cJ\|_\infty \ga
\end{eqnarray*}
and
\begin{eqnarray*}
\frac{1}{\beta}\ln\sum_{i=1,Q}e^{\beta\rho^{\pm}_i}
&&=\frac{1}{\beta}\sum_j\rho^{\pm}_j\ln\sum_{i=1,Q}e^{\beta\rho^{\pm}_i}
\\
  &&=
  \frac{1}{\beta}\sum_j \rho^{\pm}_j\left[-\ln
  \rho^{\pm}_j+
  {\beta \rho^{\pm}_j}\right]
=-\phi^{\mf}_\beta(\vec\rho^{\pm})
\end{eqnarray*}
Going back to \eqref{P-abs-0b}, and using \eqref{app2:gpm}, we get

\begin{eqnarray*}
|P^{\pm}_{\abs,\ga,0} +\phi^{\mf}_\beta(\vec\rho^{\pm})| \le c_d \ga +Ce^{-c\ga^{2a}|C^{\ell_{-,\ga}}|}
\le c_d'\ga
\end{eqnarray*}
for $\ga$ small enough.
\vskip 1.5cm \noindent
To conclude we are then left with the proof of \eqref{app2:gpm}.
Since the derivation is similar (and simpler) to what is done in section \ref{section:largedeviation}
but with a different free energy functional, we just sketch the proof here. We consider the
one body functional $\cF_{\beta,C_0}^{(1)}$
on $C_0$,
\begin{eqnarray}
\cF_{\beta,C_0}^{(1)}(\vec\rho_{C_0})=-\frac{1}{2}\int_{C_0} (\vec\rho^\pm \cdot \vec\rho_{C_0}(x)) dx
 -\frac{1}{\beta} \int_{C_0} I(\vec\rho_{C_0}(x))dx
\end{eqnarray}
Using a result similar to \ref{thm:app} leads to an estimate valid for
$\ga$ small enough of $G_{{\pm},\ga;0}[\Ii_{\vw\notin\cX^{\pm}_{C_0}}]$ in terms of the
functional $\cF_{\beta,C_0}^{(1)}$, as
\begin{eqnarray*}
G_{{\pm},\ga;0}[\Ii_{\vw\notin\cX^{\pm}_{C_0}}]&=&
\frac{\sum_{\vw_{C_0}} \prod_{x\in C_0}
e^{\beta \big(\vw_x\cdot \vec\cL(\rho^{\pm})\big)}
\Ii_{\vw\notin\cX^{\pm}_{C_0^+}}
}{\sum_{\vw_{C_0}} \prod_{x\in C_0}
e^{\beta \big(\vw_x\cdot \vec\cL(\rho^{\pm})\big)}}
\\
&=& \frac{\sum_{\vw_{C_0}} \prod_{x\in C_0}
e^{\beta \fJ_\ga\big(\vw_x\cdot \vec\rho^{\pm}\big)}
\Ii_{\vw\notin\cX^{\pm}_{C_0}}
}{\sum_{\vw_{C_0}} \prod_{x\in C_0}
e^{\beta \fJ_\ga\big(\vw_x\cdot \vec\rho^{\pm}\big)}}
\\
&\approx&
\exp\bigl\{-\beta\bigl(\inf_{\vec\rho_{C_0}\not\in\cX^{\pm}_{C_0}}\cF_{\beta,C_0}^{(1)}(\vec\rho_{C_0})
-\inf_{\vec\rho_{C_0}}\cF_{\beta,C_0}^{(1)}(\vec\rho_{C_0})\bigr)\bigr\}
\end{eqnarray*}
Following the analysis of section \ref{section:largedeviation}, one gets a similar estimate for
the large deviation cost as
\begin{eqnarray}
\inf_{\vec\rho_{C_0}\not\in\cX^{\pm}_{C_0}}\cF_{\beta,C_0}^{(1)}(\vec\rho_{C_0})
-\inf_{\vec\rho_{C_0}}\cF_{\beta,C_0}^{(1)}(\vec\rho_{C_0})\ge c\ga^{2a}|C^{\ell_{\ga,-}}|
\end{eqnarray}
for some constant $c$ and $\ga$ small enough. The estimate \eqref{app2:gpm} then follows.
\end{proof}

\setcounter{equation}{0}
\section{Proof of \eqref{eq:u1-a}-\eqref{eq:u1-b}}
\label{app:u1}

In this appendix we prove that for any $\psi_B\in\cB^0$ there is
$\vec\psi^{*}_B:$
\begin{eqnarray}
\label{eq:u1-aa}
\vec \psi^*_B=
  \begin{cases}
    \vec \psi_B & \text{on}~~\delta_{\ins}^{\ell_{+,\ga}/4}[B], \\
    \vec\rho^{\pm} & \text{elsewhere}.
  \end{cases}
\end{eqnarray}
so that:
\begin{eqnarray}
\label{eq:u1-ab}
\cF^{\eff}_{\ga,B,u}(\vec\psi_B|\vw^{(\ell_0)}_{B^c})\ge
\cF^{\eff}_{\ga,B,u}(\vec\psi^*_B|\vw^{(\ell_0)}_{B^c})+c\ga^{1/4}(\ga^{1/8}|A|)
\end{eqnarray}

\vskip .5cm \noindent
The proof is analogous to the case  of the Ising model widely analyzed in \cite{errico-leip}
to which we refer for details. A sketchy version is reported here for completeness.

\vskip .5cm \noindent
Let $\Si$ as in \eqref{def: Si},
and
\begin{equation*}
\Delta=
A\sqcup\delta_{\out}^{\ell_{+,\ga}/4}[A]
\end{equation*}

Then, recalling that the interaction term appearing in the excess of free energy
is always positive, we get a lower bound by neglecting the interaction between $\Delta$ and  $B\setminus \Delta$:

\begin{eqnarray*}
\cF^{\eff}_{\ga,B,u}(\vec\psi_B|\vw^{(\ell_0)}_{B^c})\ge
\cF^{\eff}_{\ga,B\setminus \Delta,u}(\vec\psi_{B\setminus \Delta}|\vw^{(\ell_0)}_{B^c})+
 \cF^{\eff}_{\ga,\Delta, u}(\vec\psi_\Delta)
\end{eqnarray*}

where, for any sets $D, F\sqsubset \Rr^{d}$:

\begin{eqnarray*}
& &\cF^{\eff}_{\ga,D, u}(\vec\psi_D):= \int_{D} \Phi^{\eff,\pm}_u(\vec\psi_D)
+\frac{u}{4}\int_{D\times D}J_\ga(r,r')[\vec\psi_D(r)-\vec\psi_D(r')]^{2}
\\
& & \cF^{\eff}_{\ga,D, u}(\vec\psi_D| \vw_F):=
\int_{D} \Phi^{\eff,\pm}_u(\vec\psi_D)
+\frac{u}{4}\int_{D\times D}J_\ga(r,r')[\vec\psi_D(r)-\vec\psi_D(r')]^{2}
\\
& &\hskip7cm+\frac{u}{2}\int_{D\times F}
J_\ga(r,r')[\vec\psi_D(r)-\vw_F(r')]^{2}
\end{eqnarray*}

\vskip .5cm \noindent
For any $\psi_{B}\in \cB^{0}$ (\eqref{def: B0}), since
$\vec\psi_B^*(r)\equiv\vec\psi_B(r)$ on $r\in \delta_{\ins}^{\ell_{+,\ga}/4}[B]\sqcup \Si$,
and $\vec\psi_B(r)=\vec\rho^{\pm}$ on $r\in \Si$, we have that:
\begin{eqnarray*}
\cF^{\eff}_{\ga,B\setminus \Delta,u}
(\vec\psi_{B\setminus \Delta}|\vw^{(\ell_0)}_{B^c})=\cF^{\eff}_{\ga,B,u}
(\vec\psi^*_{B}|\vw^{(\ell_0)}_{B^c})
\end{eqnarray*}
in fact the distance between the sets
$\delta_{\ins}^{\ell_{+,\ga}/4}[B]$ and $\Delta$ is larger than $\ga^{-1}$ and
\linebreak ${\cF_{\ga,\Delta,u}(\vec\rho^{\pm})\equiv 0}$:

\vskip .5cm \noindent
Hence we need to prove that for any $\vec\psi_B\in\cB^{0}$:
\begin{eqnarray}
\label{eq:boundDelta}
 \cF^{\eff}_{\ga,\Delta,u}(\vec\psi_\Delta)\ge c\ga^{1/4}\left(\ga^{1/8}|A|\right)
\end{eqnarray}
\vskip .5cm \noindent
It is convenient here to fix a specific color $\p$ instead of distinguish
only disordered and ordered configurations.

 We then denote by
\begin{eqnarray*}
\cS^{\p}:=\{r\in\hat A:\| \vec\psi_{B}(r)-\vec\rho^{\p}\|_{\star}\ge \ga^{1/8} \}
\end{eqnarray*}
that can be written as the  sum of two sets $\cS_0$,  $\cS_1$:

\begin{eqnarray*}
\cS_0&:=&\{r\in \cS: \| \vec\psi_{B}(r)-\vec\rho^{\p}\|_{\star}\ge \ga^{1/8} ~~\forall \q=-1,1,\dots,Q\}
\\
\cS_1&:=&\{r\in \cS:  \exists \q\ne {\p}: \|\vec\psi_B(r)-\vec\rho^{\q}\|_{\star}\le \ga^{1/8} \}
\end{eqnarray*}

Recalling the definition of $\Phi^{\eff,\p}_u(\vec v)$ in \eqref{def:Fmfu},
 and \eqref{def:F-eff-u}-\eqref{def:fmfu},
we will prove that
there are positive constants, $c_0, c_1$, $c_2$, so that:
\begin{eqnarray}
\label{eq:s0}
& &\int_{\cS_0}    \Phi^{\eff,\p}_u(\vec\psi(r))\ge c_0\ga^{1/4}|\cS_0|
\\
\label{eq:s1}
& &\int_{\cS_1}   \Phi^{\eff,\p}_u(\vec\psi(r))\ge (c_1u+c_2(1-u))|\cS_1|
\end{eqnarray}

\vskip .5cm \noindent
\eqref{eq:s0} and \eqref{eq:s1} prove \eqref{eq:boundDelta}

\vskip 1.5cm \noindent
\begin{proof}[Proof of \eqref{eq:s0}] 
\eqref{eq:s0}  follows from the bound immediately obtained by the  explicit expression of
$\Phi^{\eff,\p}_u(\vec v)$: 

\begin{eqnarray*}
\inf_{u\in(0,1)}\inf_{\{\vec v:\|\vec v-\rho^{\q}\|_{\star}\ge \ga^{1/8} \forall \q\}}
\Phi^{\eff,\p}_u(\vec v)\ge c_0\ga^{1/4}
\end{eqnarray*}
$c_0$ a suitable constant
\end{proof}

\begin{proof}[Proof of \eqref{eq:s1}]
Suppose $\|\vec v-\vec\rho^{\q}\|_{\star}\le \ga^{1/8}$ for some  $\q\ne {\p}$.
We will prove separately two bounds:

\begin{eqnarray}
\label{eq:s1-a}
& &\int_{\cS_1}   \Phi^{\eff,\p}_u(\vec\psi(r))\ge 2c_2(1-u)|\cS_1|
\\
\label{eq:s1-b}
& &\int_{\cS_1}   \Phi^{\eff,\p}_u(\vec\psi(r))\ge 2 c_1 u|\cS_1|
\end{eqnarray}
that together give  \eqref{eq:s1}

Proof of \eqref{eq:s1-a}
\begin{eqnarray}
\label{eq:bound-s1-a} \Phi^{\eff,\p}_u(\vec v)\ge
(1-u)\bigg[-\vec\rho^{\p} \cdot(\vec
v-\vec\rho^{\p})+\frac{1}{\beta} \left[\vec v\ln\vec v-
\vec\rho^{\p}\ln\vec\rho^{\p}\right]\bigg]\ge 0
\end{eqnarray}

Since  $\vec\rho^{\p}$ is a  solution of the mean field equations
$\vec\rho^{\p}=\frac{ e^{\beta\vec\rho^{\p}}}{\sum_i e^{\beta\rho_i^{\p}}}$,
it satisfies:
\begin{eqnarray*}
\beta\rho_i^{\p}=\ln\rho_i^{\p}-\ln C
\end{eqnarray*}
with $C= \sum_i e^{\beta\rho_i^{\p}}$
and the square parenthesis in  r.h.s. of
\eqref{eq:bound-s1-a}, can be rewritten as:

\begin{eqnarray*}
&&\bigg[-\vec\rho^{\p} \cdot(\vec
v-\vec\rho^{\p})+\frac{1}{\beta} \left[\vec v\ln\vec v-
\vec\rho^{\p}\ln\vec\rho^{\p}\right]\bigg]
\\
&&\hskip1cm=-\frac{1}{\beta}\sum_i
(v_i-\rho_i^{\p})\ln\rho_i^p+\frac{1}{\beta}\sum_i
(v_i-\rho_i^{\p})\ln C^{-1}+\frac{1}{\beta} \left[\vec
v\ln\vec v- \vec\rho^{\p}\ln\vec\rho^{\p}\right]
\\
&&\hskip1cm=-\frac{1}{\beta}\sum_i
(v_i-\rho_i^{\p})\ln\rho_i^p+\frac{1}{\beta} \left[\vec
v\ln\vec v- \vec\rho^{\p}\ln\vec\rho^{\p}\right]
\end{eqnarray*}
where in the last equality we used the fact that $\sum_i
v_i=\sum_i\rho_i^{\p}=1$. We then have:

\begin{eqnarray*}
\Phi^{\eff,\p}_u(\vec v)\ge \frac{(1-u)}{\beta}~\vec v\ln
\frac{\vec v}{\vec\rho^{\p}}
\end{eqnarray*}
and by Kullback-Leibler inequality:

\begin{eqnarray*}
\Phi^{\eff,\p}_u(\vec v)&\ge &\frac{(1-u)}{2\beta}(\vec v-\vec\rho^{\p})^{2}
\\
&\ge&\frac{(1-u)}{2\beta}
\left[(\vec \rho^{\q}-\vec\rho^{\p})^{2}-\ga^{1/4}\right]
\end{eqnarray*}
We consider separately  the case when $\q=-1$, $\p>0$ (or viceversa)
and the case where both $\q,\p$ are positive.

In the first case:
\red{\begin{eqnarray*}
\Phi^{\eff,\p}_u(\vec v)
&\ge&\frac{(1-u)}{2\beta}\left[\left(\rho_A-\frac{1}{Q}\right)^{2}
+\left(\rho_B-\frac{1}{Q}\right)^{2}(Q-1)-\ga^{1/4}\right]
\\
&\ge&\frac{(1-u)}{2\beta}\left[\frac{Q(1-2/Q)^{2}}{(Q-1)}-\ga^{1/4}\right]
\end{eqnarray*}}
If both $\q,\p$ are positive
\begin{eqnarray*}
\Phi^{\eff,\p}_u(\vec v)
&\ge&\frac{(1-u)}{2\beta}\left[2\left(\rho_A-\rho_B\right)^{2}-\ga^{1/4}\right]
\\
&\ge&
\frac{(1-u)}{2\beta}\left[\left(\frac{Q(1-2/Q)}{(Q-1)}\right)^{2}-\ga^{1/4}\right]
\end{eqnarray*}

Finally, for $\ga$ small enough:
\begin{eqnarray*}
\Phi^{\eff,\p}_u(\vec v)
&\ge&\frac{(1-u)}{20\beta}
\end{eqnarray*}
We now prove \eqref{eq:s1-b}. Let $r\in \cS_1$
\begin{eqnarray*}
\int~~dr'
J_\ga(r,r')\left(\vec\psi_B(r)-\vec\psi_B(r')\right)^{2}
\ge\int~~dr'
J^{(\ell_{-,\ga})}_\ga(r,\lng r' \rng)\left(\vec\psi_B(r)-\vec\psi_B(r')\right)^{2}
-c\ga\ell_{-,\ga}
\end{eqnarray*}

where
\begin{eqnarray*}
J^{(\ell_{-,\ga})}_\ga(r,\lng r' \rng):=\frac{1}{|C^{\ell_{-,\ga}}|}
\int_{C_{r'}^{\ell_{-,\ga}}}
J_\ga(r,r'') dr''
\end{eqnarray*}
and it is constant on the cubes of the partition
$\cD^{\ell_{-,\ga}}$. By Cauchy-Schwartz inequality:
\begin{eqnarray*}
&&\frac{1}{|C^{\ell_{-,\ga}}|}\int_{C_{r'}^{\ell_{-,\ga}}}~~dr'
J^{(\ell_{-,\ga})}_\ga(r,\lng r'\rng)\left(\vec\psi_B(r)-\vec\psi_B(r')\right)^{2}
\\
&&\hskip1cm\ge
J^{(\ell_{-,\ga})}_\ga(r,\lng r' \rng)\left(\vec\psi_B(r)-\frac{1}{|C^{\ell_{-,\ga}}|}
\int_{C_{r'}^{\ell_{-,\ga}}}\vec\psi_B(r'')~~dr''\right)^{2}
\\
&&\hskip1cm\ge J^{(\ell_{-,\ga})}_\ga(r,\lng r' \rng)
\bigg(\left(\vec\rho^{\q}-\vec\rho^{\p}\right)^2-\ga^{1/4}-\ga^{2a}\bigg)
\end{eqnarray*}
if both $\q, \p>0$,
$\left(\vec\rho^{\q}-\vec\rho^{\p}\right)^2=2(\rho_A-\rho_B)^2=2
\left(\frac{Q(1-{2}/{Q})}{Q-1}\right)^2$. While if $\q$ or
$\p$ is equal to $-1$:
$\left(\vec\rho^{\q}-\vec\rho^{\p}\right)^2=2(1-2/Q)^{2}$
Then:
\begin{eqnarray*}
\int~~dr'
J_\ga(r,r')\left(\vec\psi_B(r)-\vec\psi_B(r')\right)^{2}\ge
\frac{1}{5}-c\ga\ell_{-,\ga}-\ga^{1/4}-\ga^{2a}
\end{eqnarray*}
and
\begin{eqnarray*}
\frac{u}{4}\int_{\cS_1}~~dr~dr'
J_\ga(r,r')\left(\vec\psi_B(r)-\vec\psi_B(r')\right)^{2}\ge
\left(\frac{1}{30}\right)u|\cS_1|
\end{eqnarray*}
\end{proof}

\section{Proof of Theorem \ref{thmz4.1}}
\label{app:cluster expansion}

\vskip .5cm \noindent
\eqref{z4.7b} follows from  \eqref{z4.7} and
\eqref{def:tildeH},  by
setting
      \begin{equation}
         \label{z4.13}
 e^{-\blue{\beta}\cH^{\p}_\La( \vw_\La)}=
  \sum_{\und\Ga\in \mathcal B^{+}_\La} W^{\p}(\und\Ga,\vw_\La)
     \end{equation}
To prove the remaining statements we use a cluster
expansion to express the energy $\cH^{\p}_\La(\vw_\La)$ in terms
of a sum of weights of polymers, which will then identify
the many-body potentials $U^{\p}_\Delta(\vw_\Delta)$.

Polymers are functions \blue{$I:\{\Ga\}^{\p}\to \mathbb N_+$} such that
the collection $\{\Ga: I(\Ga)>0\}$ is finite and connected,
where  two elements $\Ga$ and $\Ga'$ in $\{\Ga\}^{\p}$ are
connected if $\overline{\ssp(\Ga)}  \sqcap
\overline{\ssp(\Ga')} \ne \emptyset$. Denote by
$\mathcal P^{\p}$ the collection of all polymers and by
$\mathcal P^{\p}_\La$ those made by contours in $\{
\Ga\}^{\p}_\La$.  It then follows \blue{from Koteck\'y and Preiss},
\cite{KP}, that,  if
the Peierls constant
is large enough,
there are numbers $\varpi(I,\und s)$,
such that
      \begin{equation}
         \label{z4.14}
\ln \sum_{\und\Ga\in \,\mathcal B^{\p}_\La}
W^{\p}(\und\Ga,\vw) = \sum_{I\in\, \mathcal P^{\p}_\La}
 \varpi^{\p} (I,\vw)
    \end{equation}

Calling $\dis{\ssp_+(I)= \bigsqcup_{\Ga: I(\Ga) >0}
\ssp_+(\Ga)}$, with
$\ssp_+(\Ga):=\ssp(\Ga)\sqcup \delta_{out}^{\ell_{+,\ga}}[\ssp(\Ga)]$,
we then set:
      \begin{equation}
         \label{z4.15}
U^{\p}_\Delta( \vw_\Delta) = -\blue{\frac{1}{\beta}}
\sum_{I\in\, \mathcal P^{\p}_\La,\,
\ssp_+(I)= \Delta}
  \varpi^{\p}(I,\vw_\Delta)
    \end{equation}

\red{$\varpi^{\p} (I,\vw)$ satisfy the bound:
\begin{eqnarray}
\label{eq:kp}
\sum_{I:\ssp(I)\ni x}\|\varpi^{\p} (I,\vw)\|_{\infty}
\left\{\prod_{\Ga: I(\Ga)>0}e^{\frac{\fK_\ga}{2}  N_\Ga I(\Ga)}\right\}<1
\end{eqnarray}
\eqref{eq:kp} follows by the general theory (see \cite{KP}) after noting
that the number of contours $\#(\Delta):=\#\{\Ga:\ssp(\Ga)=\Delta\}$ is bounded by
$(Q+1)^{N_\Delta\ga^{-2\al d}}$ and for $\ga$ small enough, since $2\al\ll 1$,
$\#(\Delta)e^{-\fK_\ga N_\Delta}\ll 1$,
}\red{($\fK_\ga/2$ is not optimal)}.
\eqref{new-z4.11b} and \eqref{new-z4.11c} then follows from \eqref{z4.15}-\eqref{eq:kp}:
\begin{eqnarray*}
\beta\sum_{\Delta\ni x}|U^{\p}_\Delta( \vw_{\Delta})|&\le&
\sum_{I:
\ssp_+(I)\ni x }
  \|\varpi^{\p}(I,\vw)\|_\infty
  \\
&\le&
\sum_{I: \ssp_+(I)\ni C_x }
  \|\varpi^{\p}(I,\vw)\|_\infty \left\{\prod_{\Ga: I(\Ga)>0}
e^{\frac{\fK_\ga}{2}  (N_\Ga I(\Ga)-N_\Ga)}\right\}
\end{eqnarray*}
Where we used the fact that $N_{\Ga} I(\Ga)-N_{\Ga}\ge 0$. Since
$\min_\Ga N_\Ga= 3^{d}$, then:
\begin{eqnarray*}
\beta\sum_{\Delta\ni x}|U^{\p}_\Delta( \vw_{\Delta})|& \le&
\hskip-.3cm
e^{-\frac{\fK_\ga}{2}  3^d}\sum_{C\in [C_x\sqcup \delta_{\out}^{\ell_{+,\ga}}[C_x]]}
\sum_{I: \ssp_+(I)\sqsupset C }
  \|\varpi^{\p}(I,\vw)\|_\infty \left\{\prod_{\Ga: I(\Ga)>0}
e^{\frac{\fK_\ga}{2}  N_{\Ga} I(\Ga)}\right\}
\\
& \le& 3^d e^{-\frac{\fK_\ga}{2}  3^d}.
\end{eqnarray*}
last inequality uses the \eqref{eq:kp} and the translation invariance of $U^{\p}$,
for $\ga$ small enough \eqref{new-z4.11b} follows.
\eqref{new-z4.11c} can be  proven analogously.
The proof of \eqref{absspec} follows from \eqref{z4.6} in a similar way. Theorem
\ref{thmz4.1} is proved.
\qed

\bibliographystyle{amsalpha}

\begin{thebibliography}{A}

\bibitem{ACCN} M. Aizenman, J. T. Chayes, L Chayes, C. M. Newman:
{\it Discontinuity of the magnetization in one-dimensional
$\frac{1}{|x-y|^2}$ Ising and Potts models},
{ J. Stat. Phys.} {\bf 50}  1-40, (1988)

\vskip .5cm
\bibitem{bkmp1}F. Baffioni, T. Kuna, I. Merola, E. Presutti:
{\it A liquid vapor  phase transition in quantum statistical
mechanics},
{ Submitted to Memoirs AMS}~(2004)

\vskip.5cm
\bibitem{bkmp2}F. Baffioni, T. Kuna, I. Merola, E. Presutti:
{\it A relativized Dobrushin uniqueness condition and
applications to Pirogov-Sinai models},
{ Submitted to Memoirs AMS}~(2004)

\vskip .5cm
\bibitem{baxter73} R. J. Baxter:
{\it Potts model at the critical temperature},
{ J. Phys. C }{\bf 6}, L445 (1973) \vskip .5cm

\vskip .5cm
\bibitem{baxter78} R. J. Baxter, H. N. V. Temperley, S. E. Ashley:
{\it Triangular Potts model at its transition temperature, and related models},
{ Proc. Roy. Soc. London, Ser. A} {\bf 358}, 535 (1978)

\vskip .5cm
\bibitem{baxter} R. J. Baxter:
 { Exactly solved sodels in statistical mechanics},
Academic Press, London (1982).

\vskip .5cm
\bibitem{Biskup-Chayes} M.Biskup, L. Chayes:
 {\it Rigorous analysis of discontinous phase transitions via mean field bounds},
 { Commun. Math. Phys.} {\bf 238}, 53--93 (2003)

\vskip .5cm
\bibitem{b-chays-c} M.Biskup, L. Chayes, N.Crawford:
{\it Mean-field driven first order phase transitions in systems with long-range interactions},
{ Submitted to J. Stat. Phys.}(2005)



\vskip .5cm \bibitem{bmpz} A. Bovier, I. Merola, E. Presutti, M. Zahradn\`ik:
{\it On the Gibbs phase rule in the Pirogov-Sinai regime},
{ J. Stat. Phys.} {\bf 114}, 1235--1267 (2004)

\vskip .5cm
 \bibitem{BZ1} A. Bovier, M. Zahradn\`ik:
{\it The low-temperature phase of Kac-Ising models},
{ J. Stat. Phys.} {\bf 87}, 311--332 (1997)

\vskip .5cm
\bibitem{BZ2} A. Bovier,M. Zahradn\`ik:
{\it Cluster expansions and Pirogov-Sinai theory for long range spin systems},
{ Markov Process and Related Fields} {\bf 8}, 443--478 (2002)


\vskip .5cm
\bibitem{BKL} J. Bricmont, K. Kuroda, JL Lebowitz, {\it First order phase transitions in lattice
and continuous systems: Extension of Pirogov-Sinai theory}, Commun.
Math. Phys. 101:501 (1985).



 \vskip .5cm
\bibitem{CP} M. Cassandro, E. Presutti:
{\it Phase transitions in Ising systems with long but finite range},
{ Markov Processes and Related Fields} {\bf 2}, 241--262 (1996)



\vskip .5cm
\bibitem{FK} C. M. Fortuin, P. W. Kasteleyn:
{\it On the random cluster model I: introduction and relation to other models},
{ Physica} {\bf 57}, 536--564 (1972)

\vskip .5cm
\bibitem{GHM} H. Georgii, O. H\"aggstr\"om, C. Maes
{\it The Random Geometry of Equilibrium Phases }, Phase Transitions
and Critical Phenomena, vol 18 , Eds C. Domb and J.L. Lebowitz
(Academic Press, London) pp 1-142.


\vskip .5cm
\bibitem{Grimmett} G. R. Grimmett:
{\it The stochastic random cluster process and the uniqueness of random cluster measures},
{ Ann. Prob.} {\bf 23}, 1461--1510 (1995).


\vskip .5cm
\bibitem{Grimm-a} G. R. Grimmett:
{\it The random-cluster model},
{ Probability, Statistics and Optimization},
F.P.Kelly ed., John Wiley \& Sons, Chichester, 49--63 (1994)

\vskip .5cm
\bibitem{KS} H. Kesten, R. H. Schonmann:
 {\it Behavior in large dimension of the Potts and Heisenberg models},
{ Reviews in Math. Phys.} {\bf 1}, 147--182 (1990)

\vskip .5cm
\bibitem{KotS} R. Koteck\'y, S. B. Shlosman:
{\it First order phase transition in large entropy lattice models},
{ Commun. Math. Phys.} {\bf 83}, 493--515 (1982)

\vskip .5cm
\bibitem{KP} R. Koteck\'y, D. Preiss:  {\it Cluster
expansion for abstract polymer models} { Comm. Math.
Phys.}, {\bf 103}, 491--498 (1986)


\vskip .5cm
\bibitem{lmp} J. L. Lebowitz, A. E. Mazel and E. Presutti: {\it Liquid-vapour phase
transitions for systems with finite range interactions}, { J. Stat.
Phys.} {\bf 94}, 955--1025 (1999) \vskip .5cm

\bibitem {LMP}
J. L. Lebowitz, A. E. Mazel, E. Presutti:~ {\it Rigorous proof of a
liquid-vapour phase transition in a continuum particle
system} { Phys. Rev. Letters} {\bf 80}, 4701 (1998)

\vskip .5cm

\bibitem {LP} J. L. Lebowitz and O. Penrose:
{\it Rigorous treatment of the Van der Waals Maxwell theory of the
liquid-vapor transition}  { J. Math. Phys.} {\bf 7}, 98--113 (1966)


\vskip .5cm
\bibitem{ps} S.A. Pirogov, Ya. G. Sinai: (i) { Teor. Mat. Fiz.}
{\bf 25}, 358--369, (1975), in Russian; English
translation: {\it Phase diagrams of classical lattice
systems}, {Theor. Math. Phys.} {\bf 25} 1185--1192, (1975).
(ii) { Teor. Mat. Fiz.} {\bf 26}, 61--76, (1976), in Russian;
English translation: {\it Phase diagrams of classical
lattice systems. Continuation}, {Theor. Math. Phys.} {\bf
26} 39--49, (1976)


\vskip .5cm
\bibitem{Potts} R. Potts:
{\it Some generalyzed order-disorder transformations}
{ Proc. Camb. Phil. Soc.} {\bf48}, 106--109 (1952)
\vskip .5cm

\bibitem{errico-leip} E. Presutti: {\it Scaling limits in statistical  mechanics \&
microstructures in continuum mechanics.} ~{ In preparation} (2005)


\vskip .5cm

\bibitem{RD} S. Reynal, H. T. Diep:~
{\it Reexamination of the long-range Potts model: A multicanonical approach}
{ Phys. Rev E}{\bf 69} 026109 (2004)

\vskip .5cm

\bibitem{Sokal} A. D. Sokal: {\it Chromatic polynomials, Potts models and all that},
{ Physica A279}, 324--332 (2000)

\vskip .5cm
\bibitem{WM}D.J.A. Welsh and C. Merino, The Potts model and the
Tutte polynomial,Journal of Mathematical Physics 41(3), 2000,
1127–1149.

\vskip .5cm
\bibitem{Wu} F. Y. Wu:~ {\it The Potts model}, { R. Mod. Phys.} {\bf 54}, 235--268 (1982)


\vskip .5cm

\bibitem{Z}
 M. Zahradn\`\i k: {\it A short course on the Pirogov-Sinai theory},
 { Rend. Mat. Appl} {\bf 18}, 411--486 (1998)

\end{thebibliography}

\end{document}